# Cultural Differences and Perverse Incentives in Science Create a Bad Mix: Exploring Country-Level Publication Bias in Select ACM Conferences

Aksheytha Chelikavada[1*], Casey C. Bennett[1]

[1]Department of Computing & Digital Media, DePaul University, Chicago, IL USA

*Corresponding Author: Aksheytha Chelikavada, achelika@depaul.edu

## Abstract

In the era of big science, many national governments are helping to build well-funded teams of scientists to serve nationalistic ambitions, providing financial incentives for certain outcomes for purposes other than advancing science. That in turn can impact the behavior of scientists and create distortions in publication rates, frequency, and publication venues targeted. To that end, we provide evidence that indicates significant inequality using standard Gini Index metrics in the publication rates of individual scientists across various groupings (e.g. country, institution type, ranking-level) based on an intensive analysis of thousands of papers published in several well-known ACM conferences (HRI, IUI, KDD, CHI, SIGGRAPH, UIST, and UBICOMP) over 15 years between 2010 to 2024. Furthermore, scientists who were affiliated with the top-5 countries (in terms of research expenditure) were found to be contributing significantly more to the inequality in publication rates than others, which raises a number of questions for the scientific community. We discuss some of those questions later in the paper. We also detected several examples in the dataset of potential serious ethical problems in publications likely caused by such incentive systems. Finally, a topic modeling analysis revealed that some countries are pursuing a much narrower range of scientific topics relative to others, indicating those incentives may also be limiting genuine scientific curiosity. In summary, our findings raise awareness of systems put in place by certain national governments that may be eroding the pursuit of truth through science and gradually undermining the integrity of the global scientific community.

# 1. Introduction

## 1.1 Overview

Science, very broadly, is based off of curiosity and the pursuit of truth. Scientists are usually driven to understand how things work and find solutions to real-world problems (Holladay, 1953). Their inquisitiveness is often paired with meticulous and rigorous methodologies that prioritize evidence and logic, making sure that their findings are as accurate and reliable as possible (Popper, 2005; Voit, 2019). Yet at the same time, scientific behavior is also oftentimes driven by cultural values and more practical motivations, at least in the modern era of “big science” (Han, 2019). In other words, what was once a small “cottage industry” of scientists working alone with limited resources has, in the current day and age, transformed into well-funded scientific teams pursuing ideas that potentially may have broad economic impact and/or serve nationalistic ambitions (Adams, 2013). Those factors can make a strong impact upon the publishing patterns of scientists, given that funding typically comes from national governments who *incentivize* certain things (Franzoni et al., 2011). Research priorities are thus influenced by the goals of a nation and their political class, which arguably may lead to distorted variations in publication rates, frequency, and the kinds of publication venues targeted.

Moreover, there are some national governments, e.g. China and South Korea, that are specifically trying to achieve enhanced global recognition through technological and scientific innovation for various political reasons (Reshetnikova et al., 2021). For example, both of those aforementioned countries are aggressively pushing their scientists to publish at top international journals and conferences (based on metrics such as impact factor) to build their political reputation internationally, providing direct incentives to scientists and universities that do so (Kim & Jeong, 2023). Though there have been efforts to reform things in both those countries (Zhang & Sivertsen, 2020), the behaviors still persist as of 2024, as evidenced by our results here. Due to that kind of environment, scientists in those places face intense demands to publish frequently in particular international venues for **reasons other than purely scientific ones, creating "perverse incentives" for individual scientist behavior**.

There are of course sometimes "good incentives" for research, where high-quality research is recognized across a broad array of topics, interdisciplinary collaborations are encouraged without concern for ethnic or political affiliation, scientific ethical lapses are heavily penalized, all types of scientific results (whether it is positive or negative) are considered valuable and important, and the primary goal is advancement of human knowledge (Gärtner, Leising, & Schönbrodt, 2024). However, incentives can become harmful when they create a climate that could more broadly be thought of as a complex high-stakes "publish or perish" culture, leading scientists into compromising ethical situations where they have to produce significant results quickly (or what at least appears "significant"), often times at the expense of thoroughness and scientific validity, to meet certain productivity quotas for funding and promotions with the primary goal of benefiting one human group over another (Shin, 2019; Tian et al., 2016). It is in situations like these that the pursuit of truth can become perversely incentivized, even if an individual scientist may wish otherwise. We emphasize here that this is not intended as a criticism of individual scientists from particular countries, but rather a criticism of the "systems" put in place (e.g. by national governments) that encourage certain scientific behaviors.

We would be remiss not to note that the authors of this paper have lived and worked for years at universities in both East Asia (South Korea) and the West (United States), and have experienced/seen many of the things discussed in this paper personally.

## 1.2 Background Research on Scientific Publication Bias

The phenomenon of bias is well-studied because it can cause unfair treatment, reduce diversity in those who participate, and damage public trust in the institution in question. Scientific publications are no exception to instances of bias. Previous literature has documented several factors that lead to potential bias in a wide range of academic environments from the field of scientific bibliometrics. Many of those bias-causing factors are discussed at length later in the analysis section of this paper, which we provide examples of here in this section.

The preference for close personal relations over merit is one common issue that can be identified in many forms across the business and scientific worlds. In Italian academia, for instance, there was a well-known case that had been raised regarding nepotistic practices of hiring close relatives for academic positions, based on a retrospective analysis of last names at Italian institutions of higher education (Allesina, 2011). Additionally, in top economic journals, it was found that there is a strong possibility that in-group members, i.e. authors who share a journal's institutional affiliation, receive fewer citation counts compared to out-group members, suggesting that in-group members papers get preferential treatment (e.g. higher chance of acceptance) regardless the quality of the research (Lutmar & Reingewertz, 2021).There have also been studies on "status bias", since it is a common issue in scientific publishing. For instance, upon analysis most of the Chinese-language academic journals were found to reject the work of junior researchers and students at a higher rate than average beyond what might be expected by random chance, i.e. without considering the quality of the research itself (Tang et al., 2022). Issues like that stem from editorial department resources, editorial department culture, the scientific research peer evaluation system, and the wider academic environment. Moreover, race, ethnicity, age, religion, sex, or sexual orientation have been identified as potential factors for bias in some specialties (e.g. medical research) (Rouan et al., 2021).

On a broader scale, there have been previous comparisons of academic productivity across nations. One such comparison in particular showed how differences in language, institutional practices, and national priorities shape the publishing patterns of scientists (Bentley, 2014). Oftentimes, a country's position in human resources (e.g. researchers and their skill levels) as well as physical resources (e.g. finances) contribute to academic performance, which is often overlooked. We should also take into consideration the goal of institutions, the way they are managed, and the national government overseeing them, since they play a role in the policies towards research & development (R&D) at both national and local levels.

Of course, there is the factor of "chance" as well when it comes to publication patterns that may stem from unforeseen consequences (Harzing & Giroud, 2014). Reportedly, China leads the USA in the volume of AI-related research papers. The reason for that has been argued to be because of China's lax data protection policies and its broad diaspora allowing for research that would not be permitted in most countries due to safety concerns (Min et al., 2023). To take another example, it has been found that, ironically, private institutions and departments that grow at a moderate rate tend to have a higher publication activity than those with lower or higher rates (Jordan et al., 1989). More often than not, resources used to conduct cutting-edge research, especially in fields like AI, are only available to a few institutions that can handle the cost and manpower, which subsequently requires the gradual acquisition of both. This means there is a growing disparity when it comes to conducting influential AI research that requires sustained investment over time (Togelius & Yannakakis, 2023). This also leads to narrowing AI research as private sector researchers specialize in more computationally intensive work that offer more immediate financial rewards, averse to tackling harder theoretical problems that may not engender immediate gain (Klinger et al., 2020).

Although there is existing literature on scientific publication bias, prior research focusing on status bias, country-level bias, and the influence of well-funded research groups ***specifically*** in reputed scientific conferences is limited. However, conferences serve as a major publication venue for computer science fields, rather than the traditional journal format. That suggests more research in the area is needed, thus motivating our work here.

### 1.3 Research Aims

When scientists align their work to cultural norms and nationalistic motivations, there is a chance that it may lead to publication bias and inequality which should theoretically show up as certain scientists or certain nations publishing more scientific papers at a rate higher than expected based on chance alone. **We take the previous sentence as our central hypothesis to test in this paper**, which we investigate via several standard metrics of inequality used widely in the scientific literature (described in Section 2). We discuss some potential reasons for such inequality and why we might consider those reasons good or bad in the Discussion (Section 4).

In short, our goal here was to identify and quantify whether such inequality occurs within several well-regarded Association for Computing Machinery (ACM) conferences: Human-Robot Interaction (HRI), Intelligent User Interfaces (IUI), Knowledge Discovery and Data Mining (KDD), Conference on Human Factors in Computing Systems (CHI), Computer Graphics and Interactive Techniques (SIGGRAPH), User Interface Software and Technology (UIST), and Ubiquitous Computing (UBICOMP). Within these specific conferences, we first sought to evaluate publication patterns among individual scientists. After that, we compared the scientists by several groupings. That included comparing by institution type (top research organizations, university ranking, etc.), as well as comparing scientists from the current top-5 countries in terms of research & development expenditures relative to GDP (e.g. United States, China, Japan, Germany, and South Korea) versus scientists from several non-top-5 countries (e.g. Australia, Canada, France, Taiwan, and Turkey) based on country comparisons used in analysis by the National Science Foundation (NSF) in the US (National Science Board, 2022). Those top-5 countries are at the forefront of scientific technological advancements and show consistent, long-term growth in research output, thus serving as a useful standard to what we might consider “high” scientific productivity.

## 2. Methodology

### 2.1 Dataset Collection

To begin our study, we scraped data about authors who have published in the seven conferences: HRI, IUI, KDD, CHI, SIGGRAPH, UIST, and UBICOMP from the ACM digital library (https://dl.acm.org/conferences). To do so, we utilized Selenium, which is a Python library for web scraping. The web scraper code was written as an automated data collection tool, and functions as follows. In the ACM digital library, we first accessed the proceedings page for each conference. Then, we drilled down into each conference proceeding for each year between 2010 and 2024 by clicking on “View all proceedings”. Upon arriving at the resulting webpage, each year can be accessed sequentially where one can find “containers” of each article published that year (typically organized under various sessions at the conference). Lastly, we then navigated to the webpage of the article in order to extract the first name, last name, and the most recently associated affiliation of each author at the time of publication, which was compiled into an Excel file automatically by our web scraping Python code to comprise our “dataset” for that conference. The affiliation data point included the name of the institution and the country location.

Some further post-processing of the data was necessary before analysis, so that we would have 1 row per author and a series of columns with the number of publications in that single conference (e.g. CHI) for each year. For the first and last names, we noticed that there were inconsistencies with diacritic and middle name initials. So, we made sure to remove the diacritics and only extracted the first name and last name out of their full names and placed those in their respective columns labeled “first name” and “last name” for each Excel conference file. In order to differentiate each author, we decided to utilize unique keys. Initially, we used their first name, last name, and affiliation as a unique key, but we had to do some manual cleanup of the list when there were duplicate scientist names with different affiliated institutions, which may be due to different people with the same name or the same scientist moving from one affiliation to another. So, to streamline the process and to avoid human error, we decided to use their first and last names as a unique key. Using such a unique key is effective because each conference relates to a specific scientific field and the scientific community is generally small within each specific field.

Surprisingly, the number of duplicates was much lower than we initially feared, which came to around 21%, likely due to the relatively small size of the scientific community in each field publishing at that specific conference and the (unfortunately) limited job mobility in academia.

In order to generate the single row for each scientist, we wanted to keep a yearly count of the total number of publications for each author from 2010 to 2024 for that given conference, so we programmed the Python code to count every unique article a scientist published at each conference for each year. At the end of those processing steps, each conference had an excel file with individual columns for first name, last name, affiliation, and 15 separate columns for counts of publication for each year between 2010-2024. That dataset was then used for our analysis.

### 2.2 Possible Breach of Scientific Ethics

We found a few publications with suspiciously similar titles and the same authors (in different orders) at the ACM HRI conference between 2021-2024, which resulted in us looking into the papers more closely and discovering several potential instances of duplication publication. In other words, what appeared to be a breach of scientific ethics. Those articles were reported to the ACM digital library for further investigation. We discuss that issue more in the Results and Discussion sections below, including potential causes (e.g. perverse incentives).

### 2.3 Analysis Methods

#### 2.3.1 Methodology for Individual Scientist Comparison

To compare individual scientists, we first wanted to create a weighted publication rate that would evaluate how much an author has published and how they distributed their work over time, based on the assumption that if a conference has an acceptance rate around 20-25% (which is claimed by several of the ACM conferences analyzed here) then if an author submitted 1 paper every year for 4 years that would by random chance result in 1 publication over any 4-year period. Of course, other factors would affect that, such as the quality of submitted work, high-output labs that submit multiple papers per year, or potentially instances of publication bias. Using the above assumption, the measure we utilized is as follows:

$$R = \begin{cases} \dfrac{T}{P + 2(n-1)} \end{cases} \qquad \text{if } n \geq 1, \text{else } R = 0 \qquad (1)$$

To create a temporal version of the *R* metric over time, we defined four-year time windows starting from 2010 (2010-2013, 2011-2014, etc.). Then, we summed the total number of publications for each window, represented as T in Equation 1 and divided it by the denominator. The denominator used *P* for the number of years in the period (in this case, 4), which could be increased or decreased if one wanted to analyze a larger or smaller time period than we did. A "denominator adjustment" was added to *P* so that for each consecutive year, *n*, an author publishes a paper the denominator will increase by 50%. The idea is that scientists who publish every year at the same conference have a greater chance of publishing in future years (e.g. perhaps they are thought leaders in the field), so we accounted for that by down-weighting the resulting value (via the denominator adjustment) to ensure that those few high-output labs/scientists would not be penalized nor overshadow the rest of the dataset. The goal here after all was not to pick out specific scientists for criticism, but look for general patterns across scientists. The final output of the formula, *R*, is the publication rate for an individual scientist for each specific window at that specific ACM conference. If an author does not publish in the window (*n*=0), the publication rate is hard set to 0. **All else being equal, all authors would have an *R* value within each time window that matches the published acceptance rate of that conference (i.e. 0.2-0.25 in this case)**.

Using the above approach, we wanted to see if there were any similarities or differences in the consistency and frequency of publications based on the R values over time. To do so, we calculated the Gini index value from the R values for each author in each selected conference as a measure of publication "inequality" for that conference (Nishioka et al., 2022). **The Gini index is a widely used measure of inequality across many fields from economics to statistical science that can be applied to any underlying numerical value (e.g. money, healthcare access, voting power)**. In this case, it was used to quantitatively identify instances of long-term publication inequality in the R values over the 4-year time windows (Ceriani & Verme, 2011). We calculated a Gini index score for each individual author in that way, and we also calculated it *across* authors to create a Gini score for each conference between 2010 and 2024. Furthermore, the Gini index can be used to calculate inequality at both the individual level or group level, which can include grouping scientists by institution type, country, etc. (see Section 2.3.2 below). Such inequality may be a sign of publication bias, though other factors could contribute of course (e.g. good incentives, see Section 1).

The interpretation of Gini values varies at the individual and conference levels here, however. At the individual author level, we can theorize that the authors who have lower Gini index values, i.e. those who are publishing consistently and frequently through all the time windows, likely belong to the "in-crowd" social clique of that conference. Conversely, those with higher Gini index values would perhaps be those who publish infrequently at that conference and thus represent the out-crowd or out-group. At the conference level though, higher Gini index values signal greater inequality at that conference. **Or to put it simply, at the conference level, higher Gini index values could be considered "bad" if the goal is to increase participation and interest in science**.

To further validate our findings, we also utilized Lotka's Law as a secondary inequality measure. It is a commonly used scientific bibliometric measure to evaluate the distribution of the productivity of authors (Lotka, 1926). Though Lotka's Law is subject to a number of criticisms (Pao, 1986), it provides some further evidence here alongside the Gini index scores to support our findings.

### 2.3.2 Methodology for Grouping Comparisons

To dig deeper into the issue of publication bias and inequality, we decided to group authors in various ways, to find possible factors leading to differences across authors. For instances, such differences could be associated with their institution type (top research organizations), university ranking level, or their country location.

To do so, we used the same Gini index values we calculated in the previous section for each conference. First, we did an **institution-type analysis**, where we grouped those scientists who are affiliated with top research organizations versus those who are not, based on a previously published categorization system of the institutions in our dataset (National Science Board, 2022). Those categories were comprised of academic institutions, government organizations, healthcare organizations, NGOs, and corporate organizations. We then aggregated the Gini scores for each institutional category.

After that, we performed a **rankings-based analysis** grouping scientists who are affiliated with the world's top 10 universities in computer science versus those who are not, based on published rankings in 2024 (U.S. News & World Report, 2024). Similarly, we did a separate **country-level analysis** grouping of authors who are affiliated with the top-5 countries based on research expenditure relative to GDP (e.g. United States, China, Japan, Germany, and South Korea) versus scientists who come from non-top-5 countries (e.g. Australia, Canada, France, Taiwan, and Turkey) reported by in published NSF analyses (National Science Board, 2022). We used a subset of 5 of the 12 non-top-5 countries used in (National Science Board, 2022), so that each of our groups would have the same number of countries, choosing ones that were geographically diverse amongst the 12.

### 2.3.3 Topic Analysis by Country Group

To explore the country groups further, we conducted topic modeling using Latent Dirichlet Allocation (LDA) to find common topics within the top-5 countries mentioned in the previous section in the conferences (Klinger et al., 2020). By training on the abstracts of articles, an LDA approach can discover topics or common themes among the countries (or, vice versa, different themes) while estimating the amount of topic diversity being researched. The aim was to see if there were differences in the kinds of scientific topics scientists in those countries were researching, which perhaps might relate to any differences in publication patterns among the countries.

To do so, we first pulled out authors from our dataset who are affiliated with the top-5 countries and then extracted the abstracts of all the published articles for each country for each conference. Then, we pre-processed the abstracts in preparation for the LDA, which included text cleaning, tokenization, removal of stopwords, and lemmatization. After that, LDA was applied to the data to create a heatmap of topics for each country. For each resulting model, we visually tuned the hyperparameters with the heatmaps to optimize performance.

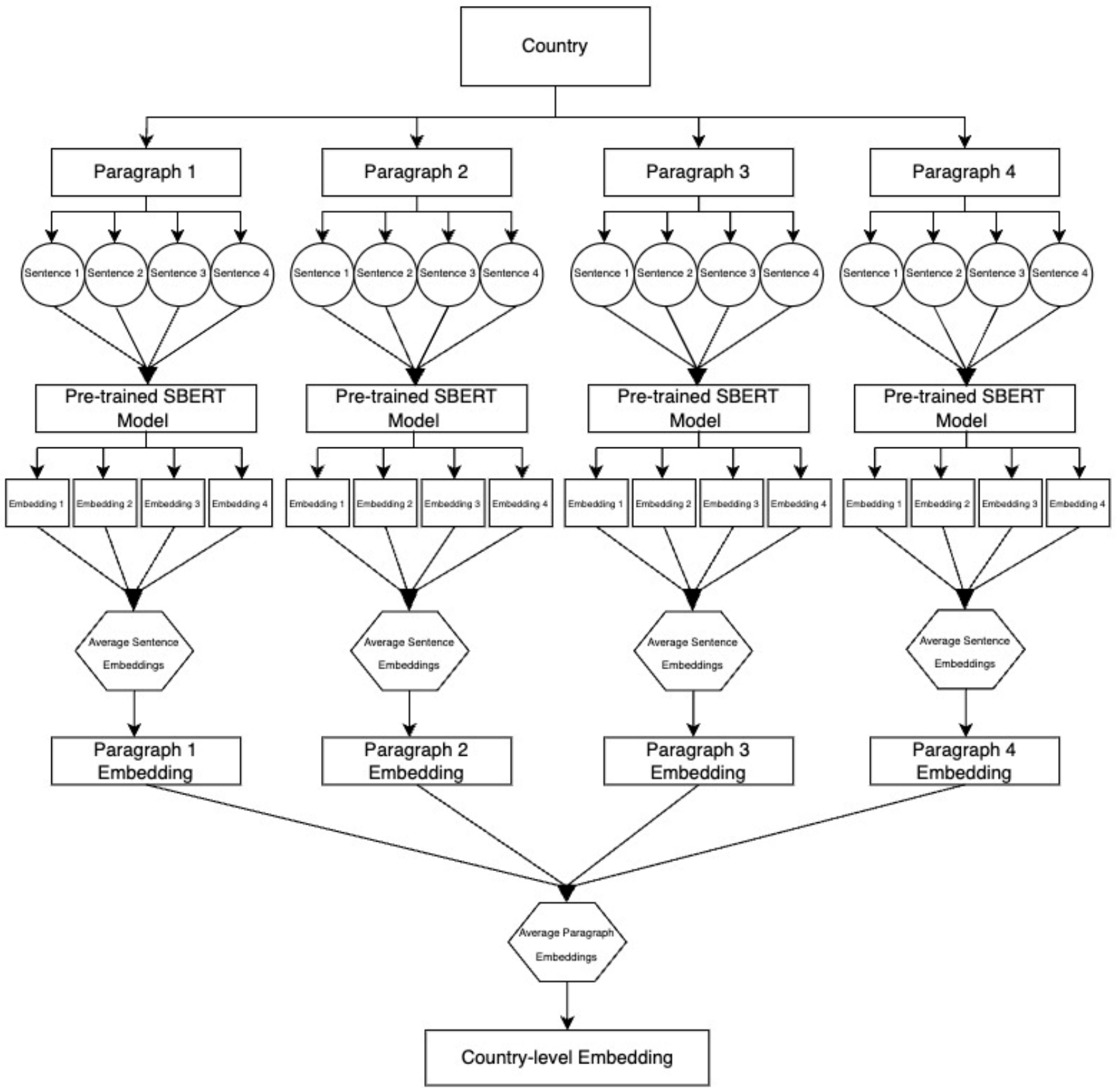

**Fig. 1** Country-level embedding – flow chart. An example of computing a country-level embedding. We are assuming there are 4 articles associated with this country and that each abstract has 4 sentences. By treating the abstract of the article as a paragraph, we performed sentence tokenization and fed the sentences to the pre-trained SBERT model to generate sentence-level embeddings. Then, we averaged the features of the sentence-level embeddings to obtain a paragraph-level embedding. Lastly, we averaged the features of the paragraph-level embeddings to obtain a single country-level embedding.

We also decided to use a more quantitative method than LDA heatmaps to identify topic preference for each country. As such, we used semantic similarity to calculate "topic similarity" among the top-5 countries for each conference (Hurtado Martín et al., 2013). That measure provides insight to how similar or different topics are across groups as a numerical score, with the groups here being the countries. Our approach to using topic similarity is as follows: if there is high similarity, then there is no topic preference that distinguishes one country from others; Adversely, if there is low similarity, then it is evidence that topic preference exists within a given country. Likewise, low similarity may also indicate reduced topic diversity compared to other countries.

To calculate semantic similarity, we used a pre-trained Sentence-BERT (SBERT) model - more specifically, all-MiniLM-L6-v2 - to process the abstracts to obtain sentence embeddings (Reimers & Gurevych, 2019). This model is hosted by the Hugging Face hub, but can be accessed through the sentence-transformers library in Python. Using the abstracts of all the published articles for each country, and treating the abstract as a paragraph, we performed sentence tokenization to get individual sentences. Then, we fed the sentences into the pre-trained Sentence-BERT (SBERT) model, which gave us associated sentence-level embedding vectors, with each of them having 384 features. Our next step was to average the features of the sentence-level embedding vectors to get a paragraph-level embedding vector. We should mention that we decided to do paragraph-level embeddings rather than sentence-level to calculate the country-level values because sentences in paragraphs are connected contextually, and we wanted to capture those relationships. Once we obtained all of our paragraph-level embedding vectors, we averaged their features to get a country-level embedding vector, which portrays the overall content of abstracts for a country (see Fig. 1). With country-level embedding vectors, we can calculate the cosine similarity among pairs of countries, which produced seven similarity matrices, each one representing one of the ACM conferences.

# 3. Results

## 3.1 Results for Individual Scientists Comparison

With our 7 datasets, each representing a specified ACM conference, we first wanted to investigate any patterns or anomalies that might point to publication bias at the individual scientist level, which would then warrant further investigation. To do so, we obtained the weighted publication rates (the "R" value from Equation 1) for all time period windows of each author between 2010 and 2024 at each conference, then calculated the mean and standard deviations of those values across authors for each conference (Table 1).

**Table 1** Mean and Standard Deviation Table

| Conference | HRI | IUI | KDD | CHI | SIGGRAPH | UIST | UBICOMP |
|---|---|---|---|---|---|---|---|
| Mean | 0.1548 | 0.1603 | 0.1692 | 0.2009 | 0.1296 | 0.1787 | 0.2029 |
| Standard Deviation | 0.2304 | 0.1914 | 0.2260 | 0.2428 | 0.2349 | 0.2126 | 0.2533 |

As can be seen in Table 1, the mean values are generally low and close in range with one another (between 0.15 and 0.20), which aligns with the low published acceptance rates of around 20-25% for those ACM conferences and matches the "all else being equal" statement we made in Section 2.3.1 above. However, we found that the standard deviation values were much larger than the mean values. This would suggest that although the *average* values match the published acceptance rates, there is a high variability (e.g. large spread) in publication rates among authors. In other words, some authors are publishing much more frequently than other ones beyond what would be expected based on chance alone.

To further analyze the spread in publication rates, we utilized the Gini index as described in Section 2 to measure publication "inequality" across authors. The overall Gini index value for each conference usually varied between 0.60 and 0.78 (Table 2). Gini index values for individual authors are provided in the online Appendix.

**Table 2** Conference-level Gini Index Value Table

| Conference | HRI | IUI | KDD | CHI | SIGGRAPH | UIST | UBICOMP |
|---|---|---|---|---|---|---|---|
| Gini Index Value | 0.7007 | 0.6116 | 0.6390 | 0.6033 | 0.7716 | 0.6038 | 0.6189 |

The values in Table 2 are considered relatively "high" for Gini index (Abramo et al., 2016) and further reinforces that there seems to be a significant inequality in the productivity among authors. In other words, there are a small number of authors who are publishing consistently across multiple years at the same conference, but there are also many authors who are publishing more sporadically or perhaps only once. While such publication inequality may be justifiable (e.g. smarter scientists, "good" incentives"), there are also many potential reasons why it may not be justifiable (e.g. researchers from the Global South who lack resources for research, personal biases of scientists, or other "perverse" incentives"). The justification of such publication inequality is, of course, a matter of debate. However, if the goal of science is to push humanity forward, then it could be argued those latter "unjustified" reasons may in fact be holding us back. That is something that warrants further research investigation and discussion amongst the scientific community.

To provide further evidence of the above point from a different angle, we utilized Lotka's Law to analyze the publication rates (Lotka, 1926). The resulting tables were very large and extensive, so we have included an example of the results for the ACM HRI conference in the online Appendix D. In those results, the observed values for the number of authors were smaller as the number of papers published increased. That is consistent with the general "power law" hypothesis basis that underlies Lotka's Law, which essentially says that the hotter the scientific field then the more unequal the distribution of scientific resources/publications will become. In other words, the bigger something gets, the narrower the opportunity to get involved becomes (e.g. professional sports is a good example of that). The growth of "big science" over the past century certainly fits such a description. All that said, the Lotka values from the conference publications appear to reinforce our findings that there is indeed a small group of authors publishing quite frequently for whatever reason, beyond what might be expected by chance.

Additionally, we calculated the Gini index value for each author, which produced lengthy lists containing the names of every single author who published at that conference over the past 15 years. Those lists are too long to include in this paper, but an example from the ACM HRI conference is shown in the online Appendix A (Table A1). The ACM HRI conference is an area that the authors of this paper are pretty familiar with, having published there ourselves in the past. Of course, analyzing such individual scientist data is difficult, but we could hypothesize that authors who have lower Gini index values in such lists of publications belong to the "in-crowd" social clique of a given conference. Data such as in Table A1 could potentially be analyzed using “social network” analysis to confirm that theory, though that remains for future research to untangle.

## 3.2 Results for Grouping Comparisons

To evaluate whether groups of authors are possibly contributing to the inequality of publication patterns found in the individual level data in the previous section, we calculated the Gini index value of various groupings of scientists: institution-type, rankings-based, country-level (see Section 2.3.2). Theoretically, if a grouping was the source of individual-level inequality shown in Section 3.1 above, then the Gini index values would differ across groups in each grouping, e.g. some countries would have higher values than others. Conversely, if the Gini index values were the same across groups, then we could theorize that the source of inequality was at the individual level. To evaluate the Gini index values against one another we used Relative Percentage Difference (RPD) (Cole & Altman, 2017; Mahinpei, 2020). We decided to use this metric as it allows us to compare the differences between values in a rigorous, standardized manner. A high RPD value would indicate a potential source of inequality for the underlying Gini index values. The results can be succinctly summarized as follows.

### 3.2.1 Institution-Type and Rankings-Based Results

With the **institution-type analysis**, the RPD values were low (generally <30%) and relatively close to one another within a limited range (shown in Table 4). This implies that institution type does not seem to contribute much to the observed publication inequality. Likewise, when we conducted **ranking-based analysis,** the RPD values were again low (<15%) and relatively close to each other within a limited range (Table 4). In short, neither the institution type nor university ranking level seem to explain much of the observed publication inequality across scientists shown in Section 3.1.

**Table 4** RPD Table for Institution-level Analysis (NTRL) and Ranking-based Analysis (USCS)

| Conference | Category | RPD |
|---|---|---|
| HRI | NTRL | 29.29% |
| | USCS | 23.18% |
| IUI | NTRL | 31.15% |
| | USCS | 11.29% |
| KDD | NTRL | 7.88% |
| | USCS | 7.42% |
| CHI | NTRL | 15.55% |
| | USCS | 3.85% |
| SIGGRAPH | NTRL | 30.94% |
| | USCS | 24.63% |
| UIST | NTRL | 9.04% |
| | USCS | 5.08% |
| UBICOMP | NTRL | 0.09% |
| | USCS | 3.84% |

### 3.2.2 Country-Level Results

However, when we conducted **country-level analysis** and compared the Gini index using RPD for the top-5 countries versus non-top-5, we found significant differences. For both the top-5 and non-top-5 countries, there was a lack of consistency in the values across conferences and countries with a less predictable pattern (see Table 5 and 6). More specifically, in the RPD values for the top-5 countries, we noticed spikes in certain countries (China and South Korea) that consistently had the highest RPD values well above other countries (sometimes by 2 or 3 times the average of all others) across all conferences, which we have highlighted in blue (see Table 5). This implies consistent, high inequality in publishing scientific papers with these two countries. Meanwhile for the non-top-5 countries, there were similar spikes but it was not consistent for the same country across all conferences (see Table 6). For the non-top-5 countries, we do note that there were a few instances where some cells had "NaN" values due to a lack of data for such country in that specific conference,

Moreover, the RPD for some of the top-5 countries in some conferences (e.g. HRI), were vastly larger than any of the non-top-5, in some cases even exceeding 100% RPD. The range of RPD scores was also larger for the top-5 countries for all conferences, except for SIGGRAPH. In short, there seems to be observable inequality at the country-level. Our interpretation of these results is that, for the top-5 countries, country-level grouping could be contributing to the inequality in publication rates among individual authors seen in Section 3.1. The fact that this phenomenon only occurred in the top-5 countries (not the non-top-5) reinforces the notion that perverse incentives at the national level may be driving this pattern among scientists.

**Table 5** RPD Table of Top-5 Countries for Select ACM Conferences

| Country | HRI | IUI | KDD | CHI | SIGGRAPH | UIST | UBICOMP |
|---|---|---|---|---|---|---|---|
| China | 122.99% | 26.39% | 11.47% | 39.33% | 39.52% | 37.58% | 3.87% |
| USA | 24.55% | 4.02% | 4.54% | 1.08% | 6.45% | 10.72% | 18.14% |
| Japan | 13.07% | 6.49% | 17.99% | 11.16% | 19.55% | 7.72% | 6.29% |
| South Korea | 100.86% | 66.49% | 29.11% | 9.99% | 13.39% | 4.82% | 35.68% |
| Germany | 41.33% | 11.84% | 2.93% | 12.48% | 12.03% | 13.38% | 4.99% |

**Table 6** RPD Table of Non-Top-5 Countries for Select ACM Conferences

| Country | HRI | IUI | KDD | CHI | SIGGRAPH | UIST | UBICOMP |
|---|---|---|---|---|---|---|---|
| Australia | 25.02% | .0024% | 5.57% | 6.98% | 97.43% | 4.36% | 11.97% |
| Canada | 9.01% | 27.79% | 7.99% | 10.75% | 4.23% | 21.56% | 2.42% |
| France | 2.19% | 27.03% | 21.04% | 20.54% | 17.73% | 13.67% | 0.46% |
| Taiwan | 4.52% | NaN | 3.21% | 14.00% | 32.63% | 18.42% | 10.15% |
| Turkey | 39.37% | NaN | NaN | 55.47% | NaN | NaN | NaN |

Here, population size should not be a concern because it does not affect the measure of inequality we used, which evaluates the *frequency distribution* of publications rather than the raw amount. Regardless, these findings should not come as a complete surprise since China and South Korea have been pushing at a national level to become global powerhouses in science through publishing scientific papers in acclaimed international journals. Currently, China outspends the United States and the European Union in research and publishes the most papers, especially in top-ranked journals, compared to the rest of the world (Hyland, 2023). Published data also suggests that South Korea has been making similar national efforts to rise in worldwide rankings through deliberate international publishing strategies (Lee & Lee, 2013).

This phenomenon could be attributed to nationalistic hyper-competitiveness that goes beyond the traditional scientific endeavor, blurring political and cultural concerns into the mix resulting in the mindset that the goal of science is primarily to outperform other countries, rather than scientific progress in and of itself (Chiang, 1990). Indeed, it has been argued previously that such a view is deeply rooted in academic communities of countries where Confucianism is a prevalent system of thought (e.g. East Asia) (Cheng, 1990). Academic success in such environments is not only seen as a personal achievement but also as a way to elevate status of one's group in society. More specifically, the pressure to excel comes from the Confucian concept of filial piety to uphold family honor and fulfill societal expectations, i.e. the group's success is the individual's success, whatever the means (Kim, 2009). As such, we could interpret these above results as the interplay of Confucianist power hierarchies upon traditional Western scientific systems, leading to those countries to create group-level systems that encourage certain individual level behaviors (aggressive publishing practices aimed at particular international journals and conferences) to serve national interests, rather than scientific ones.

To encourage such publishing behaviors, it is well-known that some governments and institutions are establishing what we might consider "perverse incentives". For instance, a key factor in China's rise as a publishing powerhouse is its government implementing policies and targeted funding for research with a focus on quantity over quality, including direct financial "rewards" to scientists for publishing in certain scientific venues (Hyland, 2023), despite efforts to reform the system in recent years (Zhang & Sivertsen, 2020). In South Korea, its government too has been implementing various policies and financial incentives such as income tax reductions for successful scientists (Jung & Mah, 2014), with a similar focus on quantity over quality (Kim & Bak, 2016). For publishing in leading journals such as Nature and Science, a South Korean researcher could in the past earn up to $17,000 USD for a single publication (Lee & Lee, 2013). Moreover, to compete for government research funding and university-level incentives (e.g. promotions), scientists at institutions in both those countries are placed under great pressure to publish in certain top-ranked venues with high frequency, using a government-curated list of internationally renowned journals & conferences, while at the same time absconding other traditional scientific activities such as peer reviewing other scientists' papers (Tian et al., 2016). That of course has had the unfortunate side-effect of a number of scientific publishing scandals in China and South Korea over the past couple decades, including a number of paper retractions and instances of plagiarism (Do, 2021). In the new era of AI (e.g. large-language models, or LLMs) being used to write scientific papers, there is renewed concern over those issues in recent years (e.g. paper mills), which we return to discussing more later in this paper.

**Table 7** Topic Semantic Similarity Matrices (via SBERT model)

| | | China | Germany | Japan | South Korea | USA |
|---|---|---|---|---|---|---|
| **HRI Matrix** | **China** | 1.000 | 0.844 | 0.887 | 0.872 | 0.860 |
| | **Germany** | 0.844 | 1.000 | 0.975 | 0.930 | 0.987 |
| | **Japan** | 0.887 | 0.975 | 1.000 | 0.945 | 0.976 |
| | **South Korea** | 0.872 | 0.930 | 0.945 | 1.000 | 0.936 |
| | **USA** | 0.860 | 0.987 | 0.976 | 0.936 | 1.000 |
| | | **China** | **Germany** | **Japan** | **South Korea** | **USA** |
| **IUI Matrix** | **China** | 1.000 | 0.858 | 0.873 | 0.791 | 0.881 |
| | **Germany** | 0.858 | 1.000 | 0.878 | 0.798 | 0.924 |
| | **Japan** | 0.873 | 0.878 | 1.000 | 0.872 | 0.919 |
| | **South Korea** | 0.791 | 0.798 | 0.872 | 1.000 | 0.889 |
| | **USA** | 0.881 | 0.924 | 0.919 | 0.889 | 1.000 |
| | | **China** | **Germany** | **Japan** | **South Korea** | **USA** |
| **KDD Matrix** | **China** | 1.000 | 0.940 | 0.961 | 0.946 | 0.993 |
| | **Germany** | 0.940 | 1.000 | 0.951 | 0.935 | 0.960 |
| | **Japan** | 0.961 | 0.951 | 1.000 | 0.955 | 0.970 |
| | **South Korea** | 0.946 | 0.935 | 0.955 | 1.000 | 0.952 |
| | **USA** | 0.993 | 0.960 | 0.970 | 0.952 | 1.000 |
| | | **China** | **Germany** | **Japan** | **South Korea** | **USA** |
| **CHI Matrix** | **China** | 1.000 | 0.966 | 0.958 | 0.984 | 0.965 |
| | **Germany** | 0.966 | 1.000 | 0.972 | 0.978 | 0.940 |
| | **Japan** | 0.958 | 0.972 | 1.000 | 0.964 | 0.900 |
| | **South Korea** | 0.984 | 0.978 | 0.964 | 1.000 | 0.961 |
| | **USA** | 0.965 | 0.940 | 0.900 | 0.961 | 1.000 |
| | | **China** | **Germany** | **Japan** | **South Korea** | **USA** |
| **SIGGRAPH Matrix** | **China** | 1.000 | 0.952 | 0.733 | 0.882 | 0.973 |
| | **Germany** | 0.952 | 1.000 | 0.799 | 0.908 | 0.970 |
| | **Japan** | 0.733 | 0.799 | 1.000 | 0.887 | 0.804 |
| | **South Korea** | 0.882 | 0.908 | 0.887 | 1.000 | 0.914 |
| | **USA** | 0.973 | 0.970 | 0.804 | 0.914 | 1.000 |
| | | **China** | **Germany** | **Japan** | **South Korea** | **USA** |
| **UIST Matrix** | **China** | 1.000 | 0.954 | 0.942 | 0.938 | 0.960 |
| | **Germany** | 0.954 | 1.000 | 0.958 | 0.947 | 0.937 |
| | **Japan** | 0.942 | 0.958 | 1.000 | 0.919 | 0.914 |
| | **South Korea** | 0.938 | 0.947 | 0.919 | 1.000 | 0.923 |
| | **USA** | 0.960 | 0.937 | 0.914 | 0.923 | 1.000 |
| | | **China** | **Germany** | **Japan** | **South Korea** | **USA** |
| **UBICOMP Matrix** | **China** | 1.000 | 0.904 | 0.939 | 0.902 | 0.956 |
| | **Germany** | 0.904 | 1.000 | 0.948 | 0.971 | 0.966 |
| | **Japan** | 0.939 | 0.948 | 1.000 | 0.923 | 0.956 |
| | **South Korea** | 0.902 | 0.971 | 0.923 | 1.000 | 0.972 |
| | **USA** | 0.956 | 0.966 | 0.956 | 0.972 | 1.000 |

## 3.3 Results for Topic Analysis by Country Group

We conducted topic modeling using LDA for the conferences, which identified the most common published topics by conference and by country. However, that was not as informative as first hoped because no clear topic preferences emerged from the analysis at the country-level using the heatmaps from LDA, either by individual country or by country group. As such, there was no clear "theme" for any particular country based on LDA. For brevity, those results can be found in the online Appendix B (Table B1).

As a more quantitative alternative to LDA, we attempted to use semantic similarity to calculate "topic similarity" among the countries for each conference to gain further insight (see Section 2.3.3). Via that method, we could obtain a quantified degree of similarity which would help us better understand the relationship between topics of the countries. We calculated the similarity scores and formatted it into a matrix for easier analysis. At the end, we had a total of 7 matrices, each one representing a conference (Table 7). For all the conferences, we found that the topics of the top-5 countries were very similar to one another, with the similarity scores approaching 1.0 in most cases and rarely dipping below 0.9. That is not entirely surprising, since most of these conferences have a defined set of topics in their call for papers (CFPs). Nevertheless, the results tell us that the topics being studied/published in these countries are quite identical to one another semantically-speaking and that there are not any notable differences in topic diversity. This leads us to infer that observed publication inequality across countries (see Section 3.2) as well as across individuals (see Section 3.1) is not being driven by topic diversity, based on these results.

## 3.4 Evidence of Ethical Problems due to Perverse Incentives

In this section, we provide some further evidence of potential ethical concerns that we accidentally discovered during our analysis related to the country-level grouping analysis of publication analysis in Section 3.2.2. Those concerns relate to two countries (China and South Korea) and likely can be tied the "perverse incentives" we mention in Section 3.2.2. We emphasize here again that this is not intended as a criticism of individual scientists from particular countries, but rather a criticism of the "systems" put in place (e.g. by national governments) that encourage certain scientific behaviors.

### 3.4.1 China

In China, there have been growing number of reports of researchers utilizing paper mills to quickly produce papers for publication, sometimes multiple versions with slightly altered content for duplicate publication. Paper mills deliberately mass produce and sell papers to scientists, who can then publish them in journals and conferences (Candal-Pedreira, 2022). Those papers can sometimes make it through peer review even when using fabricated data and results, which is extremely concerning as it can lead to misleading theories and results (Christopher, 2021). Moreover, in recent years, there is growing alarm over the use of AI and LLMs, such as ChatGPT, to further fuel paper mills and potentially jeopardize the entire scientific ecosystem with a flood of spurious papers that the peer review system cannot appropriately handle (Gray 2024; Kendall & da Silva, 2024).

Within our results, we noticed that the RPD value for the HRI conference seemed unusually high for the top-5 countries (see Table 5 in Section 3.2.2), so we investigated that abnormality further. Keeping in mind the main characteristics of papers produced by paper mills (e.g. duplicate content, AI-generated content), we manually reviewed each paper from the HRI conference between 2010 and 2024 by scanning paper titles and abstracts for semantic similarity (Parker et al., 2024; Patel, 2022). Via that process, we discovered 19 total articles (all from institutions in China), within which we identified 8 pairs of published papers (16 total papers) that were exceedingly similar in terms of content and appeared to be duplicate publications. Screenshots of those papers in the ACM digital library can be seen in the online Appendix C. Upon closer review, the other three papers did not seem to breach scientific publishing ethics.

Those 8 pairs had similar titles and abstracts and the same exact authors, except the author order was shuffled and slight alterations were made to the titles/abstracts. Within the body of the papers themselves, the sections were sometimes reorganized and split differently, but they had nearly identical content with many of the images being reused. Obviously, those papers appear to violate ethical scientific publishing guidelines, so the issue was reported to the ACM Digital Library. However, we do stress that this is indicative of a system-level problem, which in these cases create incentives that individual scientists are simply responding to. That is human nature, and if the same systems existed in another country, the outcomes would likely be similar, unfortunately.

### 3.4.2 South Korea

In South Korea, there has been a higher-than-average global average of influx of academics from abroad, whether they are relocating from elsewhere or Korean nationals returning back home after studying/working overseas. For instance, nearly 5% of scientists taking new positions in academia had moved from abroad to South Korea between 2017-2019. This is higher than the global average of 3.7% (Woolston, 2020). These researchers come with access to resources (such as fluency in English) to publish in international journals, which is advantageous to institutions and the country as a whole because of the visibility that is gained (Lee & Lee, 2013). As mentioned in Section 3.2.2, the South Korean government and individual institutions provide further financial incentives for publishing in such venues.

Interestingly, this phenomenon can be found in the ACM conferences analyzed in our dataset too. For instance, we identified that for HRI, around 72% of articles had been written by at least one author who is affiliated with a South Korean institution that had either overseas training or overseas research work experience. That is higher than the total percentage of academics in South Korea who had such overseas training/experience (roughly 50%) as per published estimates (Kim, 2010). In the IUI conference, the percentage is even higher, with all of the articles in our dataset (100%) having been written by at least one Korean institution affiliated author who has such overseas training/experience. One possible explanation for those results is the interplay of those researchers' English language and international research skills with the incentives from the Korean government, resulting in such outcomes. However, it begs an ethical question about whether a local scientist who studied and trained within South Korea (perhaps due to financial limitations) and as a result has limited English fluency would be penalized in such a system averse to their peers. Ideally, science should reward scientists for doing great science regardless of their backgrounds.

# 4 Discussion

## 4.1 Summary

In summary, we analyzed publications from several prestigious ACM conferences (HRI, IUI, KDD, CHI, SIGGRAPH, UIST, and UBICOMP) over 15 years between 2010 to 2024, and we discovered indications of potential publication bias and inquality at those conferences, where some individual scientists appear to be publishing scientific papers at a higher rate than chance alone given the published acceptance rates at those conferences, even after taking into account factors such as high output labs/scientists. **In other words, there appears to be behavior amongst some scientists that is intended to "game" the publication system**. To support our claim, we present a range of evidence that stems from the scraped data about authors who have published their work in the chosen conferences during that time period.

Such evidence includes the significant inequality among the publication rates of the individual authors, that may indicate in-group "social cliques" within some ACM conferences (Section 3.1). We also evaluated whether such publishing inequality at the individual-level was influenced by group-level factors based on author affiliation, including institution-type, rankings-based, and country-level. We found that institution-type and rankings-based grouping were not able to explain the publication inequality, but country-level did appear to have a significant contribution to the publication inequality at the individual-level (Section 3.2). In short, scientists who were affiliated with the top-5 countries based on research expenditure (United States, China, Japan, Germany, and South Korea) seem to be behaving in ways that are contributing more to publication frequency patterns that diverge from what we might expect from chance alone, i.e. what one might call "publication bias". RPD values (which account for population differences) were much higher than expected, particularly at some individual conferences.

We further investigated ethical problems that we discovered in two particular countries (China, South Korea) that both showed up in our country-level analysis results above and where governments are thought to be providing "perverse incentives" to scientists for reasons other than advancing science itself. There is notable previously published evidence indicating that China and South Korea at the national-level are consistently pushing for their scientists to publish aggressively at certain venues (even at the expense of ethical considerations), which we argue here can be attributed to nationalistic hyper-competitiveness intertwined with cultural factors all intending to serve national goals. We present several examples within our publication dataset of potential recent ethical issues (Section 3.4). To further back our claims, we conducted topic modeling using LDA as well as semantic similarity to find "topic similarity" and found that publication inequality is not being driven by scientists from some countries covering more diverse scientific topics, so that can be ruled out as a factor (Section 3.3).

These results suggest a closer investigation of the practices of some national governments on science in general may be warranted, as the systems those governments are putting in place may be incentivizing individual behaviors that are ultimately damaging to the scientific community. We discuss some broader implications of these results below.

## 4.2 Implications

We hope that our findings will raise awareness in various issues related to the influence of national governments' policies on modern science, while possibly even challenging pre-existing standards in academia and scientific publishing.

First and foremost, we want to bring attention to the mental health and well-being of researchers, especially early-career researchers who are affiliated with institutions in some of the countries discussed in this paper (e.g. China and South Korea) where national governments are known to be providing perverse incentives to scientists.

This specific group of people are most at-risk in such environments and pressured to have a certain number of publications within a very small time-frame, which causes many to have substantial "work-related anxiety" as their tenure and promotions are on the table (Tian et al., 2016; Zhong & Liu, 2022). Oftentimes, those who exhibit such anxiety will also have depressive symptoms, possible substance-use issues, sleeping difficulties, and severe fatigue. Unfortunately, these symptoms can end up being a precursor to suicide in some cases (Doering et al., 2024). Getting help for such conditions is seen with much dismay because in many Asian countries there is severe shame and social ostracism associated with mental illnesses (Yoo, 2018). This stigma makes it even harder for them to access support and leaves them in a very vulnerable state with their livelihood on the line.

Another issue is that, with paper mills becoming increasingly more popular, many fake articles are appearing in different journals across various disciplines (Brundy & Thornton, 2024). These fake articles often include fabricated data, which compromises the integrity of scientific theories and creates concerning foundations for future work that may be based on them (Christopher, 2021). Moreover, if such fraud is discovered, it can also unfairly damage the credibility of all scientists from that particular country associated with such shoddy work (Horta & Li, 2023; Mckie, 2024). When such articles are used as foundations for future work, many resources such as research funding and the valuable time of other researchers are wasted. This causes many, even those within the scientific community, to have a difficult time in trusting scientific literature (Chambers, 2024). Our paper reveals more potential evidence for such activity at ACM conferences, which goes to show that it is prevalent and common, unfortunately, and something that we as a global scientific community must more seriously address together in a concerted effort rather than piecemeal (Else & Van Noorden, 2021).

National hyper-competitiveness urges countries to compete with one another to become leaders in science and innovation. However, we must not neglect global scientific interests that are geared toward advancing human knowledge for all mankind, not just any one affiliation (ethnic, racial, tribal, national, etc.), but for all of us. Although competition is important, collaboration is the driving factor for new questions and ideas (Van den Besselaar et al., 2012). It is important to foster international collaborations and to build such networks as global science is often driven by those relationships, which rely on mutual trust (Gui et al., 2019). A common standard to compare researchers is through their h-index and the impact factor of the journals they publish in (Boell & Wilson, 2010), but it is debatable whether such metrics help foster a collaborative spirit. In this paper, we have seen the distress and pressure this creates for researchers often forcing them down the path of quantity over quality, as well as the consequences of "perverse incentives" on the practice of science. Instead, a better approach may be to perhaps focus on the overall impact an individual scientist's work has in the specific field and the rigor of their studies using new measures. For example, this might mean that we would look into how a researcher's work helped advance the field based on the size of their collaborative network or how they helped solve real-world issues based on social development indexes or other non-academic metrics (Ravenscroft et al., 2017). Those are just a couple potential ideas, but there is a plethora of ways the scientific community could attempt to address this issue.

## 4.3 Scientific "Economy" and the Breakdown of Peer Review

There has been a rapid growth in the overall number of scientific publications in recent years (Bornmann & Mutz, 2015; Drodz & Ladomery, 2024; McCook, 2006; Park et al., 2023), which may be partially due to the "perverse incentives" for scientists discussed in this paper. At the same time, there is growing awareness of an overall "crisis" in the peer review process for scientific publications, with a lack of available qualified peer reviewers. To put it bluntly, if we assume that any random scientific paper has a 25% chance of getting accepted, and all publications need at least 2 reviewers, then any scientist should technically be reviewing at least 2 papers for each paper they publish every year, though some published estimates have stated as high as 3-4 (Raoult, 2020). So, if college professor X published 5 papers last year, then they should have peer reviewed 10-20 papers. However, there is no current incentive for scientists to do so, outside of some personal honor code. Universities typically do not factor peer reviewing into decisions on promotion or tenure, for that matter (Day, 2022; Malcolm, 2018). Combined with the aforementioned "perverse incentives" for aggressive publication, this lack of incentives for peer reviewing is a very plausible part of the peer review crisis.

Technically speaking, scientific publication has always operated as an *informal economy*, where scientists submit papers to be peer-reviewed by other scientists, with the general understanding that they return the favor-in-kind and careers on-the-line (Ioannidis, 2005; Partha & David, 1994; Stephen, 1996; Van Dalen, 2021). **However, perhaps the time has come to *formalize* that scientific economy for publications**, to counter the peer review crisis and perverse incentives of national governments. We propose that could be created rather simply by instituting "tokens" (e.g. via blockchain), which scientists would receive after conducting a peer review. This is similar to other suggestions in recent years (Mohan, 2019; Spearpoint, 2017). Those tokens could then be used as credits when they submit their own papers for publication, with each submission requiring some number of credits. To make it fairer, early-stage researchers could receive a pool of tokens to start with, and perhaps tokens could be partially refunded if a paper was rejected after peer review. Furthermore, peer reviewers could be rated on the quality of their peer review by the editors of the journal/conference, which might affect how many tokens they receive and thus ensure they make a reasonable effort. In essence, this would create a more formal scientific economy that is less likely to be gamed and more resistant to behaviors engendered through perverse incentives.

The difficult part of the above would be the administration of such a system. In real economies, that is of course handled by governing bodies at a national level. However, science is an international effort at this stage, with many different actors. Thus, the challenge is to ensure the integrity of such a formal science economy, and whether the token system would be managed by one large journal publisher, or some group of publishers, or perhaps a non-profit organization on behalf of scientists. Also, there is an open question of whether all publishers (journals and/or conferences) would be included, or whether a publication venue's participation in the token system would be restricted somehow. The problem is if it is not restricted, that may lead to abuse whereas small groups of scientists could setup "sham" conferences or journals to inappropriately obtain tokens. All that said, there is already some precedent for what we are suggesting, in the form of international bodies like Science Citation Index (SCI) or Web-of-Science (WOS) or ORC-ID that could perhaps be used to restrict access to the token system based on some sort of quality-control, in order to prevent any abuse.

Regardless, we think the above concept is something that needs to be seriously discussed across the scientific community, as this paper and other recent papers on the topic have shown we are likely reaching a critical point in the history of science, and that scientific practices we have traditionally used in the past are no longer suitable for current conditions (Park et al., 2023).

## 4.4 Limitations

This study has several potential limitations. For example, the size of the datasets we scraped varied from conference to conference. There were some conferences where the datasets were very large (CHI had 17,279 rows) and some that had much smaller datasets (HRI had 3,787 rows). Variations in sample size could potentially impact any analysis results, or produce bias in the form of under-representation. The same issue occurred with the countries as well, since some countries had a larger dataset of authors and publications averse to others. We attempted to deal with that issue by using metrics that focused on ratios (Gini Index, RPD) rather than raw absolute counts.

In terms of the data within the datasets, we attempted to scrape all the available papers on the ACM digital library for a defined set of well-known ACM conferences, but there could be other conferences or papers of interest for analysis of publication bias that were not included. That might include conferences outside the ACM realm, e.g. IEEE conferences, as well as potentially journals. However, for our purposes here, the ACM Digital Library provided a high-quality source of data. Overall, though we have looked through a small subset of ACM conferences, we could get a more comprehensive analysis of publication bias by adding additional conferences to any future analysis.

Finally, another limitation is regarding the author affiliations, as we did not take into consideration of the author's earlier affiliations but only their affiliation at time of publication in the ACM conference. This has the possibility of skewing our results when evaluating various groupings of scientists. With the authors, we also deliberately overlooked varying levels of contribution such as 1$^{st}$ author, 2$^{nd}$ author, 3$^{rd}$ author, and so on. The results may be different if only the 1$^{st}$ author or corresponding author was considered, for instance. Regardless, author affiliations may have affected some parts of our analysis, such as the ranking-based groupings. Though we can also not ignore the reality that there is the possibility of the ranking system used for university ranking-level to be flawed. Oftentimes, there are minute differences in the scoring and ranking of universities, but ranking agencies will capitalize on the first decimal value differences and sensationalize rankings (Soh, 2015).

## 4.5 Future Work

Our work has the potential to open up future exploration to find the prevalence and influence of publication bias and inequality in the scientific world, potentially creating a better "scientific ecosystem" for all scientists to produce better science for the **ultimate benefit of all humanity** rather than one group over another. However, there is still much research to be done in this area, and many unanswered questions.

At the moment, we are in the new era of AI where many are utilizing LLMs to write scientific papers. This has caused a sudden increase in scientific publications from around 2019, which raises concerns in the originality of such work but also adds pressure to peer-reviewers (as mentioned in Section 4.3). We ourselves too have noticed such increase in publications at around the same time in this study, given the number of reviews invites we receive every week in recent years. This raises the question of whether there is any correlation between the prevalence of AI and the rise in publications. That topic is one that needs quantitative scientific investigation to verify, similar to the work presented in this paper.

On a broader scale, there are of course both "unjustified" and "justified" reasons for publication inequality, like the inequality evidenced in this paper. Yet, this raises the deeper philosophical question of what makes a reason unjustified or justified. That question takes us back to the work of Latour and Popper and others in asking what "science" really is (Dunbar, 2000; Latour, 1987; Popper, 2005). As such, further investigations would be necessary to consider the global scientific interests and whether the "unjustified" reasons are advancing science (e.g. through competition) or instead holding us back from reaching higher heights in science. Likewise, there could be further analysis in the "social network" of scientists using bibliometric techniques to identify those who belong to the so called scientific "in-crowd" clique of the ACM conferences. That could be followed by an analysis of their publication citation patterns to evaluate whether such in-group/out-group behavior in science is related to the pace of scientific advancement. For instance, it would be a relatively simple matter to look at whether a smaller, more-defined in-crowd in a field leads to a quicker pace of discovery in that field based on some evaluation metric derived from scraped Google Scholar results. The harder part would be deciding what would make for a good evaluation metric (e.g. growth in impact factor), taking into account differences across fields (Coccia, 2022).

There are a few other issues that still need to be considered in future research, such as automated tools to detect plagiarism. In order to maintain integrity of publication ethics, ACM has introduced a software named Crossref iThenticate, which can be used to check for similarity to other literature and web content (Association for Computing Machinery, n.d.). Although this is a very important tool, we must take into consideration that there is a possibility for flagged papers to create a "first impression", and this would incline the reviewers to possibly reject them unfairly. Research could investigate more sophisticated ways of looking for plagiarism or unethical publication practices. Along the same lines, for semantic similarity across publications we used SBERT in our analysis here. However, to catch the even subtler differences among the topics across various conferences/journals, a fine-tuned model could be used, especially if those conferences/journals have defined set of topics for their CFPs. For instance, one can fine-tune a base model, such as SciBERT, on the datasets first to make the model more sensitive (Beltagy & Cohan, 2019). That, however, will require significant resources such as GPUs and large storage spaces, so future research could explore the tradeoff between sensitivity and computational resources for that use case.

## Acknowledgements

We would like to thank our many research collaborators for their thoughtful discussions on this topic prior to this publication, as well as some of our colleagues in South Korea who contributed to the initial idea behind this research.

# Appendix A.

## Table A1

| Author First Name | Author Last Name | Affiliation | Gini Index |
|---|---|---|---|
| Andi | Peng | MIT, Cambridge, MA, USA | 0.9167 |
| Cloe | Emnett | Colorado School of Mines, Golden, CO, USA | 0.9167 |
| Philip | Stafford | Indiana University, Bloomington, IN, USA | 0.9167 |
| Mark | Higger | Colorado School of Mines, Golden, CO, USA | 0.9167 |
| Theing | Oo | Collaborative Robotics and Intelligent Systems Institute, Oregon State University, Corvallis, OR, USA | 0.9167 |
| Gregory | LeMasurier | University of Massachusetts Lowell, Lowell, MA, USA | 0.9167 |
| Geronimo | Gorostiaga Zubizarreta | Georgia Institute of Technology, Atlanta, GA, USA | 0.9167 |
| Lisa | Scherf | Technische Universität Darmstadt, Darmstadt, Germany | 0.9167 |
| Dorothea | Koert | Technische Universität Darmstadt, Darmstadt, Germany | 0.9167 |
| Eileen | Roesler | George Mason University, Fairfax, VA, USA | 0.9167 |
| Arsha | Ali | University of Michigan, Ann Arbor, MI, USA | 0.9167 |
| Fares | Abawi | University of Hamburg, Hamburg, Germany | 0.9167 |
| Philipp | Allgeuer | University of Hamburg, Hamburg, Germany | 0.9167 |
| Di | Fu | University of Hamburg, Hamburg, Germany | 0.9167 |
| Stefan | Wermter | University of Hamburg, Hamburg, Germany | 0.9167 |
| Alex | Chow | University of California, San Diego, La Jolla, CA, USA | 0.9167 |
| Valeria | Villani | Department of Sciences and Methods for Engineering, University of Modena and Reggio Emilia, Reggio Emilia, Italy | 0.9167 |
| Lorenzo | Sabattini | Department of Sciences and Methods for Engineering, University of Modena and Reggio Emilia, Reggio Emilia, Italy | 0.9167 |
| Raphaelle | Roy | Fédération ENAC ISAE-SUPAERO ONERA, Université de Toulouse, Toulouse, France | 0.9167 |
| Anke | Brock | Fédération ENAC ISAE-SUPAERO ONERA, Université de Toulouse, Toulouse, France | 0.9167 |
| Liubove | Orlov Savko | Rice University, Houston, TX, USA | 0.9167 |
| Zhiqin | Qian | Rice University, Houston, TX, USA | 0.9167 |
| Gregory | Gremillion | Army Research Laboratory, Aberdeen Proving Ground, MD, USA | 0.9167 |
| Catherine | Neubauer | Army Research Laboratory, Aberdeen Proving Ground, MD, USA | 0.9167 |
| Andres | Ramirez Duque | University of Glasgow, Glasgow, United Kingdom | 0.9167 |
| Josh | Bhagat Smith | Oregon State University, Corvallis, OR, USA | 0.9167 |
| Prakash | Baskaran | Oregon State University, Corvallis, OR, USA | 0.9167 |
| Kota | Nieda | Nara Institute of Science and Technology, Ikoma, Japan | 0.9167 |
| Dawn | Tilbury | University of Michigan, Ann Arbor, MI, USA | 0.9167 |
| Dante | Arroyo | Faculty of Science and Engineering, Pontificia Universidad Catolica del Peru, Lima, Peru | 0.9167 |
| Cristian | Barrue | Institut de Robòtica i Informàtica Industrial, CSIC-UPC, Barcelona, Spain | 0.9167 |
| Guillem | Alenya | Institut de Robòtica i Informàtica Industrial, CSIC-UPC, Barcelona, Spain | 0.9167 |
| Adel | Baselizadeh | Department of Informatics, University of Oslo, OSLO, Norway | 0.9167 |
| Diana | Lindblom | Department of Informatics, University of Oslo, OSLO, Norway | 0.9167 |
| Jim | Torresen | Department of Informatics, University of Oslo & RITMO Centrer for Interdisciplinary Studies in Rhythm, Time and Motion, University of Oslo, OSLO, Norway | 0.9167 |
| Artem | Bazhenov | Skolkovo Institute of Science and Technology, Moscow, Russian Federation | 0.9167 |
| Vladimir | Berman | Skolkovo Institute of Science and Technology, Moscow, Russian Federation | 0.9167 |
| Artem | Lykov | Skolkovo Institute of Science and Technology, Moscow, Russian Federation | 0.9167 |
| Dzmitry | Tsetserukou | Skolkovo Institute of Science and Technology, Moscow, Russian Federation | 0.9167 |
| Bram | Van Deurzen | Expertise Centre for Digital Media, UHasselt - Flanders Make, Diepenbeek, Belgium | 0.9167 |
| Kris | Luyten | Expertise Centre for Digital Media, UHasselt - Flanders Make, Diepenbeek, Belgium | 0.9167 |
| Adina | Panchea | Université de Sherbrooke, Research Center of Aging & Université de Sherbrooke, Interdisciplinary Institute for Technological Innovation, Sherbrooke, QC, Canada | 0.9167 |
| Yao-Cheng | Chan | School of Information, University of Texas at Austin, Austin, Texas, USA | 0.9167 |
| Cristina | Gena | Department of Computer Science, University of Turin, Torino, Italy | 0.9167 |
| Arthi | Haripriyan | University of California, San Diego, La Jolla, CA, USA | 0.9167 |
| Vignesh | Prasad | Technische Universität Darmstadt, Darmstadt, Germany | 0.9167 |
| Jan | Peters | Technische Universität Darmstadt & German Research Center for AI, Darmstadt, Germany | 0.9167 |
| Georgia | Chalvatzaki | Technische Universität Darmstadt & Hessian Center for AI, Darmstadt, Germany | 0.9167 |
| Xiaoxuan | Hei | Autonomous Systems and Robotics Lab, U2IS, ENSTA Paris, Institut Polytechnique de Paris, Palaiseau, France | 0.9167 |
| Heng | Zhang | Autonomous Systems and Robotics Lab, U2IS, ENSTA Paris, Institut Polytechnique de Paris, Palaiseau, France | 0.9167 |
| Eric | Nichols | Honda Research Institute Japan, Wako, Saitama, Japan | 0.9167 |
| Ayse | Dogan | Sabanci University, Istanbul, Turkey | 0.9167 |
| Shogo | Okada | Japan Advanced Institute of Science and Technology, Nomi, Japan | 0.9167 |
| Marieke | Van Otterdijk | ROBIN, RITMO, University of Oslo, Oslo, Norway | 0.9167 |
| Xuezhu | Wang | Tsinghua University, Beijing, China | 0.9167 |
| Ruilin | Xiong | Tsinghua University, Beijing, China | 0.9167 |
| Xun | Cui | Tsinghua University, Beijing, China | 0.9167 |
| Bernardo | Marques | IEETA, DETI, LASI, University of Aveiro, Aveiro, Portugal | 0.9167 |
| Eurico | Pedrosa | IEETA, DETI, LASI, University of Aveiro, Aveiro, Portugal | 0.9167 |
| Tsvetomila | Mihaylova | Aalto University, Espoo, Finland | 0.9167 |
| Max | Pascher | TU Dortmund University, Dortmund, Germany | 0.9167 |
| Jens | Gerken | TU Dortmund University, Dortmund, Germany | 0.9167 |
| Stefan | Reitmann | Lund University & Zittau/Görlitz University of Applied Sciences, Lund, Sweden | 0.9167 |
| Anna | Kim | Indiana University, Bloomington, Indiana, USA | 0.9167 |
| Changdan | Cao | Tsinghua University, Beijing, China | 0.9167 |
| Xiangyu | Sun | Harbin Institute of Technology, Harbin, China | 0.9167 |
| Yilin | Wang | Tsinghua University, Beijing, Beijing, China | 0.9167 |
| Qingwei | Wang | Tsinghua University, Beijing, China | 0.9167 |
| Ruisi | Sun | Tsinghua University, Beijing, China | 0.9167 |
| Qi | Xin | Tsinghua University, Beijing, China | 0.9167 |
| Xinyu | Wang | Tsinghua University, Beijing, China | 0.9167 |
| Angela | Pan Ding | Tsinghua University, Beijing, China | 0.9167 |
| Zipeng | Zhang | Tsinghua University, Beijing, China | 0.9167 |
| Zihui | Chen | Tsinghua University, Beijing, China | 0.9167 |
| Aakash | Yadav | University of Wisconsin-Madison, Madison, WI, USA | 0.9167 |
| Ranjana | Mehta | University of Wisconsin-Madison, Madison, WI, USA | 0.9167 |
| Ningning | Zhang | Tsinghua University, Beijing, China | 0.9167 |
| Ziwei | Chi | Tsinghua University, Beijing, China | 0.9167 |
| Zhiling | Xu | Tsinghua University, Beijing, China | 0.9167 |
| Qi | Chen | Tsinghua University & Wuhan Textile University, Beijing, China | 0.9167 |
| Valentina | Campo | Tsinghua University, Beijing, China | 0.9167 |
| Sarah | Kriz | University of Washington, Seattle, WA, USA | 0.9167 |
| Dilip | Limbu | Institute for Infocomm Research, Singapore, Singapore | 0.9167 |
| Lawrence | Por | Institute for Infocomm Research, Singapore, Singapore | 0.9167 |
| Takafumi | Matsumaru | Shizuoka University, Hamamatsu, Japan | 0.9167 |
| Yuichi | Ito | Shizuoka University, Hamamatsu, Japan | 0.9167 |
| Wataru | Saitou | Shizuoka University, Hamamatsu, Japan | 0.9167 |
| Weslie | Khoo | Indiana University, Bloomington, IN, USA | 0.8712 |
| Manasi | Swaminathan | Indiana University, Bloomington, IN, USA | 0.8712 |
| David | Crandall | Indiana University, Bloomington, IN, USA | 0.8712 |
| Mark-Robin | Giolando | Oregon State University, Corvallis, OR, USA | 0.8712 |
| Nathaniel | Dennler | University of Southern California, Los Angeles, CA, USA | 0.8712 |
| Lynne | Baillie | Heriot-Watt University, Edinburgh, United Kingdom | 0.8712 |
| Kyrie | Amon | Indiana University, Bloomington, IN, USA | 0.8611 |
| Reiden | Webber | Georgia Institute of Technology, Atlanta, GA, USA | 0.8611 |
| Alan | Lindsay | Heriot-Watt University, Edinburgh, United Kingdom | 0.8611 |
| Ronald | Petrick | Heriot-Watt University, Edinburgh, United Kingdom | 0.8611 |
| Carlos | Granados | Faculty of Science and Engineering, Pontificia Universidad Catolica del Peru, Lima, Peru | 0.8611 |
| Jouh | Chew | Honda Research Institute Japan, Saitama, Japan | 0.8611 |
| Melissa | Donnermann | Julius-Maximilians-University Würzburg, Würzburg, Germany | 0.8611 |
| Yao | Lu | Tsinghua University, Beijing, China | 0.8611 |
| Haipeng | Mi | Tsinghua University, Beijing, China | 0.8611 |
| Mary | Foster | University of Glasgow, Glasgow, United Kingdom | 0.8514 |
| Cong | Shi | University of Miami, Miami, FL, USA | 0.8452 |
| Yanheng | Li | City University of Hong Kong, Hong Kong, Hong Kong | 0.8452 |
| Sawyer | Collins | Indiana University Bloomington, Bloomington, IN, USA | 0.8452 |
| Kenna | Baugus Henkel | Mississippi State University, Mississippi State, MS, USA | 0.8452 |
| Patricia | Piedade | ITI, LARSYS, Instituto Superior Técnico, Universidade de Lisboa, Lisbon, Portugal | 0.8452 |
| Alyssa | Hanson | Colorado School of Mines, Golden, CO, USA | 0.8452 |
| Hiroki | Sato | Indiana University, Bloomington, IN, USA | 0.8452 |
| Ulas | Karli | Yale University, New Haven, CT, USA | 0.8452 |
| Victor | Antony | Johns Hopkins University, Baltimore, MD, USA | 0.8452 |
| Marianne | Bossema | Amsterdam University of Applied Sciences, Amsterdam, Netherlands | 0.8452 |
| Lamia | Elloumi | Amsterdam University of Applied Sciences, Amsterdam, Netherlands | 0.8452 |
| Matthijs | Smakman | HU University of Applied Sciences Utrecht, Utrecht, Netherlands | 0.8452 |
| Muhammad | Mohammed Zaffir | Nara Institute of Science and Technology, Ikoma, Nara, Japan | 0.8452 |
| Takahiro | Wada | Nara Institute of Science and Technology, Ikoma, Nara, Japan | 0.8452 |
| Filipa | Rocha | LASIGE, Faculdade de Ciências, Universidade de Lisboa, Lisboa, Portugal | 0.8452 |
| Stuart | Reeves | Mixed Reality Lab, School of Computer Science, University of Nottingham, Nottingham, Nottinghamshire, United Kingdom | 0.8452 |
| Zulfiqar | Zaidi | Georgia Institute of Technology, Atlanta, GA, USA | 0.8452 |
| Arjun | Krishna | Georgia Institute of Technology, Atlanta, GA, USA | 0.8452 |
| Samia | Bhatti | University of Michigan, Ann Arbor, MI, USA | 0.8452 |
| Yifei | Zhu | Colorado School of Mines, Golden, CO, USA | 0.8452 |
| Helen | Zhou | Yale University, New Haven, CT, USA | 0.8452 |
| Qiping | Zhang | Yale University, New Haven, CT, USA | 0.8452 |
| Lux | Miranda | Department of Information Technology, Uppsala University, Uppsala, Sweden | 0.8452 |
| Joel | Fischer | School of Computer Science, University of Nottingham, Nottingham, United Kingdom | 0.8452 |
| Takuto | Akiyoshi | Nara Institute of Science and Technology, Ikoma, Japan | 0.8452 |
| Adrian | Anhuaman | Faculty of Science and Engineering, Pontificia Universidad Catolica del Peru, Lima, Peru | 0.8452 |
| Itala | Latorre | Faculty of Science and Engineering, Pontificia Universidad Catolica del Peru, Lima, Peru | 0.8452 |
| Sebastian | Chion | Faculty of Science and Engineering, Pontificia Universidad Catolica del Peru, Lima, Peru | 0.8452 |
| William | Meza | Faculty of Science and Engineering, Pontificia Universidad Catolica del Peru, Lima, Peru | 0.8452 |
| Sao | Nguyen | U2IS, ENSTA Paris, IP Paris & IMT Atlantique, Lab-STICC, UMR CNRS 6285, Palaiseau, 91120, France | 0.8452 |
| Jane | Stuart-Smith | University of Glasgow, Glasgow, United Kingdom | 0.8452 |
| Katharina | Brunnmayr | Institute of Visual Computing and Human-Centered Technology, HCI Group, Technische Universität Wien, Vienna, Austria | 0.8452 |
| Andres | Ramirez-Duque | University of Glasgow, Glasgow, United Kingdom | 0.8452 |
| Daniel | Sidobre | CNRS, UPS, LAAS-CNRS, Université de Toulouse, Toulouse, France | 0.8452 |
| Pratyusha | Ghosh | University of California, San Diego, La Jolla, CA, USA | 0.8452 |
| Rabeya | Jamshad | University of California San Diego, La Jolla, CA, USA | 0.8452 |
| Carl | Bettosi | Heriot-Watt University, Edinburgh, United Kingdom | 0.8452 |
| Maike | Paetzel-Prusmann | Disney Research, Zurich, Switzerland | 0.8452 |
| Toshihiro | Hiraoka | Mobility Research Division, Japan Automobile Research Institute, Tokyo, Japan | 0.8452 |
| Niko | Kleer | DFKI, Saarland Informatics Campus, Saarbrücken, Germany | 0.8452 |
| Michael | Feld | DFKI, Saarland Informatics Campus, Saarbrücken, Germany | 0.8452 |
| Kurima | Sakai | Advanced Telecommunications Research Institute International, Keihanna Science City, Kyoto, Japan | 0.8452 |
| Rosalind | Picard | MIT Media Laboratory, Cambridge, MA, USA | 0.8452 |
| Zachary | Kaufman | Indiana University Bloomington, Bloomington, Indiana, USA | 0.8452 |
| Arinah | Karim | Indiana University Bloomington, Bloomington, Indiana, USA | 0.8452 |
| Yijie | Guo | The Future Laboratory, Tsinghua University, Beijing, China | 0.8452 |
| Sehoon | Ha | School of Interactive Computing, Georgia Institute of Technology, Atlanta, GA, USA | 0.8452 |
| Tugce | Pekcetin | Department of Cognitive Science, Graduate School of Informatics, Middle East Technical University, Ankara, Turkey | 0.8452 |
| Cengiz | Acarturk | Department of Cognitive Science, Jagiellonian University, Krakow, Poland | 0.8452 |
| Linda | Onnasch | Technische Universität Berlin, Berlin, Germany | 0.8452 |
| Barbara | Bruno | Karlsruhe Institute of Technology, Karlsruhe, Germany | 0.8452 |
| Fabian | Schirmer | Institute Digital Engineering, Technical University of Applied Sciences, Würzburg-Schweinfurt, Germany | 0.8452 |
| Giulia | Scorza Azzara | DIBRIS, University of Genoa & RBCS, Italian Institute of Technology, Genoa, Italy | 0.8452 |
| Joo-Hyun | Song | Cognitive, Linguistic, and Psychological Sciences, Brown University, Providence, RI, USA | 0.8452 |
| Shenando | Stals | Heriot-Watt University, Edinburgh, United Kingdom | 0.8452 |
| Isobel | Voysey | University of Edinburgh, Edinburgh, United Kingdom | 0.8452 |
| Anita | Vrins | Tilburg University, Tilburg, Netherlands | 0.8452 |
| Ethel | Pruss | Tilburg University, Tilburg, Netherlands | 0.8452 |
| Caterina | Ceccato | Tilburg University, Tilburg, Netherlands | 0.8452 |
| Jos | Prinsen | Tilburg University, Tilburg, Netherlands | 0.8452 |
| Leimin | Tian | Faculty of Engineering, Monash University, Melbourne, Australia | 0.8452 |
| Jose | Cortazar | Art and Design Department, Pontifical Catholic University of Peru, Lima, Peru | 0.8452 |
| Chacharin | Lertyosbordin | King's Mongkut University of Technology Thonburi, Bangkok, Thailand | 0.8452 |
| Teeratas | Asavareongchai | Shrewsbury International School Bangkok, Bangkok, Thailand | 0.8452 |
| Nichaput | Khurukitwanit | Shrewsbury International School Bangkok, Bangkok, Thailand | 0.8452 |
| Sirin | Liukasemsarn | Darunsikkhalai School, bangkok, Bangkok, Thailand | 0.8452 |
| Frederic | Delaunay | University of Plymouth, Plymouth, United Kingdom | 0.8452 |
| Joachim | de Greeff | University of Plymouth, Plymouth, United Kingdom | 0.8452 |
| P. | De Silva | Toyohashi University of Technology, Toyohashi, Japan | 0.8452 |
| Yuta | Yoshiike | Toyohashi University of Technology, Toyohashi, Japan | 0.8452 |
| Christian | Kroos | University of Western Sydney, Sydney, Australia | 0.8452 |
| Greg | Mori | Simon Fraser University, Burnaby, BC, Canada | 0.8452 |
| Richard | Vaughan | Simon Fraser University, Burnaby, BC, Canada | 0.8452 |
| Kate | Candon | Yale University, New Haven, CT, USA | 0.8377 |
| Minja | Axelsson | University of Cambridge, Cambridge, United Kingdom | 0.8333 |
| Mirjam | De Haas | HU University of Applied Sciences, Utrecht, Netherlands | 0.8333 |
| Joshua | Zonca | CONTACT Unit, Italian Institute of Technology, Genoa, Italy | 0.8333 |
| Nialah | Wilson-Small | New York University, Brooklyn, NY, USA | 0.8333 |
| Kirstin | Petersen | Cornell University, Ithaca, NY, USA | 0.8333 |
| Qiaoqiao | Ren | Ghent University - imec, Ghent, Belgium | 0.8333 |
| Natasha | Randall | Indiana University, Bloomington, IN, USA | 0.8333 |
| Patric | Jensfelt | KTH Royal Institute of Technology, Stockholm, Sweden | 0.8333 |
| Sjoerd | Hendriks | Chalmers University of Technology, Gothenburg, Sweden | 0.8333 |
| Wei-Chu | Chen | Indiana University Bloomington, Bloomington, IN, USA | 0.8333 |
| Nguyen | Tuyen | King's College London, London, United Kingdom | 0.8333 |
| Tibor | Bosse | Radboud University, Nijmegen, Netherlands | 0.8333 |
| Hanyang | Hu | Tsinghua University, Beijing, China | 0.8333 |
| Mengyu | Chen | Tsinghua University, Beijing, China | 0.8333 |
| Ruhan | Wang | Tsinghua University, Beijing, China | 0.8333 |
| Zhilong | Zhao | Tsinghua University, Beijing, China | 0.8333 |
| Yanran | Chen | Tsinghua University, Beijing, China | 0.8333 |
| Qingyu | Hu | Tsinghua University, Beijing, China | 0.8333 |
| Siran | Ma | Tsinghua University, Beijing, China | 0.8333 |
| Houze | Li | Tsinghua University, Beijing, China | 0.8333 |
| Aryaman | Pandya | Motional, Boston, MA, USA | 0.8333 |
| Paul | Schmitt | MassRobotics, Boston, MA, USA | 0.8333 |
| Dian | Lv | Tsinghua University, Beijing, China | 0.8333 |
| Jirui | Liu | Tsinghua University, Beijing, China | 0.8333 |
| Jiancheng | Zhong | Tsinghua University, Beijing, China | 0.8333 |
| Zhiyao | Ma | Tsinghua University, Beijing, China | 0.8333 |
| Seong | Lee | Cornell University, Ithaca, NY, USA | 0.8333 |
| Nicholas | Britten | Virginia Tech, Blacksburg, VA, USA | 0.8333 |
| Aksel | Holmgren | Linköping University, Linköping, Sweden | 0.8333 |
| Max | Pettersson | Linköping University, Linköping, Sweden | 0.8333 |
| Arna | Aimysheva | Nazarbayev University, Astana, Kazakhstan | 0.8333 |
| Maria | Insuasty | Universidad del Rosario, Bogota, Colombia | 0.8333 |
| Yahan | Xie | Tsinghua University, Beijing, China | 0.8333 |
| Hanhui | Yang | Tsinghua University, Beijing, China | 0.8333 |
| Jing | Huang | Tsinghua University, Beijing, China | 0.8333 |
| Jing | Wang | Tsinghua University, Beijing, China | 0.8333 |
| Hanxuan | Li | Tsinghua University, Beijing, China | 0.8333 |
| Sebastian | Caballa | Pontificia Universidad Católica del Perú, Lima, Peru | 0.8333 |
| Eleanor | Avrunin | Yale University, New Haven, CT, USA | 0.8333 |
| Eric | Sauser | Swiss Federal Institute of Technology Lausanne, Lausanne, Switzerland | 0.8333 |
| Aude | Billard | Swiss Federal Institute of Technology Lausanne, Lausanne, Switzerland | 0.8333 |
| Antonio | Andriella | Pal Robotics, Barcelona, Spain | 0.8141 |
| Leigh | Levinson | Indiana University Bloomington, Bloomington, IN, USA | 0.8095 |
| Long-Jing | Hsu | Indiana University, Bloomington, IN, USA | 0.8081 |
| Isabel | Neto | INESC-ID, Instituto Superior Técnico, Universidade de Lisboa, Lisbon, Portugal | 0.8017 |
| Aida | Amirova | Nazarbayev University, Astana, Kazakhstan | 0.8007 |
| Alberto | Sanfeliu | Universitat Politècnica de Catalunya - BarcelonaTech (UPC) & Institut de Robòtica i Informàtica Industrial (CSIC-UPC), Barcelona, Spain | 0.7955 |
| Chuang | Yu | UCL Interaction Centre, Computer Science Department, University College London, London, United Kingdom | 0.7955 |
| Nurziya | Oralbayeva | Nazarbayev University, Astana, Kazakhstan | 0.7939 |
| Ayaka | Fujii | National Institute of Advanced Industrial Science and Technology, Tokyo, Japan | 0.7917 |
| Laura | Kunold | Ruhr University Bochum, Bochum, Germany | 0.7917 |
| Selen | Akay | Sabanci University, Istanbul, Turkey | 0.7917 |
| Paul | Schermerhorn | Indiana University, Bloomington, IN, USA | 0.7917 |
| Hugo | Nicolau | ITI, LARSYS, Instituto Superior Técnico, Universidade de Lisboa, Lisbon, Portugal | 0.7885 |
| Andrew | Schoen | University of Wisconsin - Madison, Madison, WI, USA | 0.7885 |
| Ruben | Janssens | IDLab - AIRO, Ghent University - imec, Ghent, Belgium | 0.7885 |
| Connor | Esterwood | University of Michigan, Ann Arbor, MI, USA | 0.7833 |
| Lionel | Robert | University of Michigan, Ann Arbor, MI, USA | 0.7833 |
| Feiran | Zhang | Norwegian University of Science and Technology, Trondheim, Norway | 0.7833 |
| Oriana | Ferrari | Eindhoven University of Technology, Eindhoven, Netherlands | 0.7833 |
| Hidenobu | Sumioka | ATR, Kyoto, Japan | 0.7833 |
| Kim | Baraka | Free University Amsterdam, Amsterdam, Netherlands | 0.7803 |
| Frank | Broz | Intelligent Systems, Delft University of Technology, Delft, Netherlands | 0.7767 |
| Bengisu | Cagiltay | Computer Sciences Department, University of Wisconsin - Madison, Madison, WI, USA | 0.7717 |
| Dakota | Sullivan | University of Wisconsin - Madison, Madison, WI, USA | 0.7717 |
| Mudit | Verma | School of Computing and Augmented Intelligence, Arizona State University, Tempe, AZ, USA | 0.7717 |
| Yu-Wen | Chang | Department of Management Information Systems, National Chengchi University, Taipei, Taiwan | 0.7717 |
| Ching-Chih | Tsao | Cornell Tech, New York, New York, USA | 0.7717 |
| J. | Dominguez-Vidal | Institut de Robòtica i Informàtica Industrial (CSIC-UPC), Barcelona, Spain | 0.7717 |
| Takashi | Minato | RIKEN, Seika-cho, Japan | 0.7717 |
| Hirokazu | Kato | Nara Institute of Science and Technology, Ikoma, Japan | 0.7717 |
| Saurav | Singh | Rochester Institute of Technology, Rochester, New York, USA | 0.7717 |
| Jamison | Heard | Rochester Institute of Technology, Rochester, New York, USA | 0.7717 |
| Kumar | Akash | Honda Research Institute USA, Inc., San Jose, CA, USA | 0.7717 |
| Teruhisa | Misu | Honda Research Institute USA, Inc., San Jose, CA, USA | 0.7717 |
| Keiichi | Yamazaki | Saitama University, Saitama, Japan | 0.7717 |
| Elizabeth | Twamley | University of California, San Diego & VA San Diego Healthcare System, San Diego, CA, USA | 0.7667 |
| Elior | Carsenti | milab, Reichman University, Herzliya, Israel | 0.7667 |
| Nathan | White | University of Wisconsin - Madison, Madison, WI, USA | 0.7667 |
| Angelique | Taylor | Cornell University, New York City, NY, USA | 0.7667 |
| Hannen | Wolfe | Colby College, Waterville, ME, USA | 0.7667 |
| Chao | Zhang | Eindhoven University of Technology, Eindhoven, Netherlands | 0.7667 |
| Mohammad | Yasar | University of Virginia, Charlottesville, VA, USA | 0.7667 |
| Md | Islam | AWS Generative AI Innovation Center, Seattle, WA, USA | 0.7667 |
| Adam | Norton | New England Robotics Validation and Experimentation (NERVE) Center, University of Massachusetts Lowell, Lowell, MA, USA | 0.7667 |
| Lorenzo | Ferrini | PAL Robotics & Technische Universität Wien, Barcelona, Spain | 0.7667 |
| Matthias | Rehm | Technical Faculty of IT and Design, Aalborg University, Aalborg, Denmark | 0.7667 |
| Sara | Cooper | Honda Research Institute Japan, Wako, Japan | 0.7667 |
| Ruth | Stock-Homburg | Technische Universität Darmstadt, Darmstadt, Germany | 0.7667 |
| Kristiina | Jokinen | AI Research Center, National Institute of Advanced Industrial Science and Technology, Tokyo, Japan | 0.7667 |
| Graham | Wilcock | CDM Interact & University of Helsinki, Helsinki, Finland | 0.7667 |
| Nadine | Reissner | KUKA Deutschland GmbH, Augsburg, Germany | 0.7667 |
| Matthias | Kraus | Chair for Human-Centered Artificial Intelligence, University of Augsburg, Augsburg, Germany | 0.7667 |
| Sabahat | Bagci | Sabanci University, Istanbul, Türkiye | 0.7667 |
| Eleonore | Lumer | Digital Linguistics Lab, Bielefeld University, Bielefeld, Germany | 0.7667 |
| Hendrik | Buschmeier | Digital Linguistics Lab, Bielefeld University, Bielefeld, Germany | 0.7667 |
| Yi | Zhao | Sydney School of Architecture, Design and Planning, The University of Sydney, Sydney, Australia | 0.7667 |
| Akihiro | Maehigashi | Shizuoka University, Shizuoka, Japan | 0.7667 |
| Verena | Hafner | Adaptive Systems Group, Humboldt-Universitaet zu Berlin, Berlin, Germany | 0.7667 |
| Jiayuan | Dong | Grado Department of Industrial and Systems Engineering, Virginia Tech, Blacksburg, VA, USA | 0.7667 |
| Shuqi | Yu | Department of Human Development and Family Science, Virginia Tech, Blacksburg, VA, USA | 0.7667 |
| Koeun | Choi | Department of Human Development and Family Science, Virginia Tech, Blacksburg, VA, USA | 0.7667 |
| Federica | Nenna | University of Padova, Padova, Italy | 0.7667 |
| Luciano | Gamberini | University of Padova, Padova, Italy | 0.7667 |
| Yuichiro | Fujimoto | Nara Institute of Science and Technology, Ikoma, Japan | 0.7667 |
| Marc-Antoine | Maheux | Interdisciplinary Institute for Technological Innovation, Université de Sherbrooke, Sherbrooke, QC, Canada | 0.7667 |
| Nils | Tolksdorf | Faculty of Arts and Humanities, Psycholinguistics, Paderborn University, Paderborn, Germany | 0.7667 |
| Katharina | Rohlfing | Faculty of Arts and Humanities, Psycholinguistics, Paderborn University, Paderborn, Germany | 0.7667 |
| Elliott | Hauser | School of Information, University of Texas at Austin, Austin, TX, USA | 0.7667 |
| Thomas | Demeester | IDLab, Ghent University - imec, Ghent, Belgium | 0.7667 |
| Christopher | Peters | KTH Royal Institute of Technology, Stockholm, Sweden | 0.7667 |
| Benjamin | Dossett | University of Denver, Denver, CO, USA | 0.7667 |
| Paul | Rybski | Carnegie Mellon University, Pittsburgh, PA, USA | 0.7667 |
| Bryan | Morse | Brigham Young University, Provo, UT, USA | 0.7667 |
| Rehj | Cantrell | Indiana University, Bloomington, IN, USA | 0.7667 |
| Yuichiro | Yoshikawa | Osaka University, Toyonaka, Osaka, Japan | 0.7661 |
| Micol | Spitale | Politecnico di Milano, Milan, Italy | 0.7655 |
| Soyon | Kim | University of California, San Diego, La Jolla, CA, USA | 0.7652 |
| Danilo | Gallo | Naver Labs Europe, Grenoble, France | 0.7652 |
| Tommaso | Colombino | Naver Labs Europe, Grenoble, France | 0.7652 |
| Shreepriya | Gonzalez-Jimenez | Naver Labs Europe, Meylan, France | 0.7652 |
| Cecile | Boulard | Naver Labs Europe, Grenoble, France | 0.7652 |
| Marius | Hoggenmuller | The University of Sydney, Sydney, NSW, Australia | 0.7652 |
| Erin | Hedlund-Botti | Georgia Institute of Technology, Atlanta, GA, USA | 0.7652 |
| Kaisa | Vaananen | Tampere University, Tampere, Finland | 0.7652 |
| Stefano | Guidi | University of Siena, Siena, Italy | 0.7652 |
| Christian | Pek | KTH Royal Institute of Technology, Stockholm, Sweden | 0.7652 |
| Daniel | Weber | University of Tübingen, Tübingen, Germany | 0.7652 |
| Andreas | Zell | University of Tübingen, Tübingen, Germany | 0.7652 |
| Jin | Ryu | Cornell University, Ithaca, NY, USA | 0.7652 |
| Rachel | Au | New Mexico State University, Las Cruces, NM, USA | 0.7652 |
| Raida | Karim | The University of Washington, Seattle, WA, USA | 0.7652 |
| Hanna | Lee | The University of Washington, Seattle, WA, USA | 0.7652 |
| Sara | Mongile | Italian Institute of Technology & University of Genoa, Genoa, Italy | 0.7652 |
| Ana | Tanevska | Uppsala University, Uppsala, Sweden | 0.7652 |
| James | Berzuk | University of Manitoba, Winnipeg, Canada | 0.7652 |
| Seonghee | Lee | Cornell University, Ithaca, NY, USA | 0.7652 |
| Leon | Bodenhagen | University of Southern Denmark, Odense, Denmark | 0.7652 |
| Javier | Laplaza | Universitat Politècnica de Catalunya, Barcelona, Spain | 0.7652 |
| Tobiaz | Paulsson | Uppsala University, Uppsala, Sweden | 0.7652 |
| Mengyu | Zhong | Uppsala University, Uppsala, Sweden | 0.7652 |
| Daniella | DiPaola | Massachusetts Institute of Technology, Cambridge, MA, USA | 0.7652 |
| Nicolas | Rodriguez | Institut de Robòtica i Informàtica industrial (CSIC-UPC), Barcelona, Spain | 0.7652 |
| Jonas | Jorgensen | University of Southern Denmark, Odense, Denmark | 0.7652 |
| Ali | Ayub | University of Waterloo, Waterloo, ON, Canada | 0.7652 |
| Cheng-Yi | Tang | National Chengchi University, Taipei, Taiwan Roc | 0.7652 |
| Shamil | Sarmonov | Nazarbayev University, Astana, Kazakhstan | 0.7652 |
| Aidar | Shakerimov | Nazarbayev University, Astana, Kazakhstan | 0.7652 |
| Jules | van Gurp | Eindhoven University of Technology, Eindhoven, Netherlands | 0.7652 |
| Johan | Michalove | Cornell University, Ithaca, NY, USA | 0.7652 |
| Roberto | Raez | Pontificia Universidad Católica del Perú, Lima, Peru | 0.7652 |
| Andreas | Kolling | University of Pittsburgh, Pittsburgh, PA, USA | 0.7652 |
| Yoshinori | Kobayashi | Saitama University & JST, PRESTO, Saitama, Japan | 0.7652 |
| Yoshinori | Kuno | Saitama University, Saitama, Japan | 0.7652 |
| Michael | Gielniak | Georgia Institute of Technology, Atlanta, GA, USA | 0.7652 |
| Maria | Parreira | KTH Royal Institute of Technology, Stockholm, Sweden | 0.7567 |
| Alexandra | Bejarano | Colorado School of Mines, Golden, CO, USA | 0.7564 |
| Vivienne | Chi | Brown University, Providence, RI, USA | 0.7500 |
| Bryce | Ikeda | University of North Carolina at Chapel Hill, Chapel Hill, NC, USA | 0.7500 |
| Tariq | Iqbal | University of Virginia, Charlottesville, VA, USA | 0.7500 |
| Antonio | Paolillo | Dalle Molle Institute for Artificial Intelligence (IDSIA) USI-SUPSI, Lugano, Switzerland | 0.7500 |
| Eike | Schneiders | School of Computer Science, University of Nottingham, Nottingham, United Kingdom | 0.7500 |
| Anais | Garrell | Universitat Politècnica de Catalunya - BarcelonaTech (UPC) & Institut de Robòtica i Informàtica Industrial (CSIC-UPC), Barcelona, Spain | 0.7500 |
| Robin | Murphy | Texas A&M University, College Station, TX, USA | 0.7500 |
| Eva | Wiese | Technische Universität Berlin, Berlin, Germany | 0.7500 |
| Nazerke | Rakhymbayeva | Nazarbayev University, Astana, Kazakhstan | 0.7500 |
| Kolja | Kuhnlenz | Coburg University of Applied Sciences and Arts, Coburg, Germany | 0.7500 |
| Lucia | Sarmiento Calderon | Pontifical Catholic University of Peru, Lima, Peru | 0.7500 |
| Leonardo | Gomez Hormaza | Pontifical Catholic University of Peru, Lima, Peru | 0.7500 |
| Ornnalin | Phaijit | University of New South Wales, Sydney, NSW, Australia | 0.7500 |
| Claude | Sammut | University of New South Wales, Sydney, NSW, Australia | 0.7500 |
| Katrina | Ling | New Mexico State University, Las Cruces, NM, USA | 0.7500 |
| Katie | Seaborn | Tokyo Institute of Technology, Tokyo, Japan | 0.7500 |
| Peter | Pennefather | gDial, Inc., Toronto, ON, Canada | 0.7500 |
| Vinh | Nguyen | U.S. National Institute of Standards and Technology, Gaithersburg, MD, USA | 0.7500 |
| Yuga | Adachi | University of Tsukuba, Tsukuba, Japan | 0.7500 |
| Carmen | Espinoza | Pontifical Catholic University of Peru, Lima, Peru | 0.7500 |
| Andres | Alamo | Pontifical Catholic University of Peru, Lima, Peru | 0.7500 |
| Daniel | Bambusek | Brno University of Technology, Brno, Czech Rep | 0.7500 |
| Zdenek | Materna | Brno University of Technology, Brno, Czech Rep | 0.7500 |
| Michal | Kapinus | Brno University of Technology, Brno, Czech Rep | 0.7500 |
| Vitezslav | Beran | Brno University of Technology, Brno, Czech Rep | 0.7500 |
| Pavel | Smrz | Brno University of Technology, Brno, Czech Rep | 0.7500 |
| Phani | Singamaneni | LAAS-CNRS, Toulouse, France | 0.7500 |
| Emily | Cross | University of Glasgow & Macquarie University, Glasgow, United Kingdom | 0.7500 |
| Ayumu | Kawahara | University of Tsukuba, Tsukuba, Japan | 0.7500 |
| Wing-ting | Law | Hong Kong Productivity Council, Hong Kong, Hong Kong | 0.7500 |
| Ki-sing | Li | Hong Kong Productivity Council, Hong Kong, Hong Kong | 0.7500 |
| Kam-wah | Fan | Hong Kong Productivity Council, Hong Kong, Hong Kong | 0.7500 |
| Tiande | Mo | Hong Kong Productivity Council, Hong Kong, Hong Kong | 0.7500 |
| Chi-kin | Poon | Hong Kong Productivity Council, Hong Kong, Hong Kong | 0.7500 |
| Meriam | Moujahid | Heriot-Watt University, Edinburgh, United Kingdom | 0.7500 |
| Oliver | Lemon | Heriot-Watt University, Edinburgh, United Kingdom | 0.7500 |
| Jose | Aguas Lopes | Semasio, Edinburgh, United Kingdom | 0.7500 |
| Guy | Laban | University of Glasgow, Glasgow, United Kingdom | 0.7500 |
| Roberto | Raez Pereyra | Pontificia Universidad Católica del Perú, Lima, Peru | 0.7500 |
| Catholijn | Jonker | Delft University of Technology & Leiden University, Delft & Leiden, Netherlands | 0.7500 |
| Bruce | Wilson | Heriot-Watt University, Edinburgh, United Kingdom | 0.7500 |
| Antonia | Krummheuer | Aalborg University, Aalborg, Denmark | 0.7500 |
| Nina | Riether | Bielefeld University, Bielefeld, Germany | 0.7500 |
| Britta | Wrede | Bielefeld University, Bielefeld, Germany | 0.7500 |
| Waki | Kamino | Cornell University, Ithaca, NY, USA | 0.7422 |
| Sam | Thellman | IDA, Linköping University, Linköping, Sweden | 0.7422 |
| Terran | Mott | Colorado School of Mines, Golden, CO, USA | 0.7418 |
| Yuyi | Liu | Kyoto University & ATR, Kyoto, Japan | 0.7402 |
| Norihiro | Hagita | ATR, Kyoto, Japan | 0.7271 |
| Hadas | Erel | miLab, Reichman University, Herzliya, Israel | 0.7250 |
| Franziska | Babel | Dept.of Computer Science, Linköping University, Linköping, Sweden | 0.7226 |
| Sharifa | Alghowinem | MIT Media Lab, Cambridge, MA, USA | 0.7188 |
| Kantwon | Rogers | Georgia Institute of Technology, Atlanta, GA, USA | 0.7188 |
| Mohammad | Obaid | Chalmers University of Technology, Gothenburg, Sweden | 0.7188 |
| Carolin | Strassmann | Institute of Computer Science, University of Applied Sciences Ruhr West, Bottrop, Germany | 0.7167 |
| Sabrina | Eimler | Institute of Computer Science, University of Applied Sciences Ruhr West, Bottrop, Germany | 0.7167 |
| Maryam | Alimardani | Tilburg University, Tilburg, Netherlands | 0.7149 |
| Johannes | Kraus | Ulm University, Ulm, Germany | 0.7147 |
| Anastasia | Ostrowski | Massachusetts Institute of Technology, Cambridge, MA, USA | 0.7130 |
| Martin | Baumann | Ulm University, Ulm, Germany | 0.7130 |
| Tom | Ziemke | Dept. of Computer Science, Linköping University, Linköping, Sweden | 0.7115 |
| Taishi | Sawabe | Nara Institute of Science and Technology, Ikoma, Japan | 0.7108 |
| Philipp | Hock | Dept.of Computer Science, Linköping University, Linköping, Sweden | 0.7024 |
| Maria | Lupetti | Human Centred Design, TU Delft, Delft, Netherlands | 0.7024 |
| Saad | Elbeleidy | Colorado School of Mines & Peerbots, Golden, CO, USA | 0.7024 |
| Silvia | Rossi | University of Naples Federico II, Naples, Italy | 0.7024 |
| Elin | Bjorling | The University of Washington, Seattle, WA, USA | 0.7024 |
| Alessandra | Rossi | University of Naples Federico II, Naples, Italy | 0.7024 |
| Eric | Rosen | Brown University, Providence, RI, USA | 0.7024 |
| Burcu | Urgen | Department of Psychology, Department of Neuroscience, Bilkent University, UMRAM, Ankara, Turkey | 0.6983 |
| Philipp | Kranz | Institute Digital Engineering, Technical University of Applied Sciences, Würzburg-Schweinfurt, Germany | 0.6983 |
| Jan | Schmitt | Institute Digital Engineering, Technical University of Applied Sciences, Würzburg-Schweinfurt, Germany | 0.6983 |
| Tobias | Kaupp | Institute Digital Engineering, Technical University of Applied Sciences, Würzburg-Schweinfurt, Germany | 0.6983 |
| Tiffany | Chen | Georgia Institute of Technology, Atlanta, GA, USA | 0.6968 |
| Charles | Kemp | Georgia Institute of Technology, Atlanta, GA, USA | 0.6968 |
| Birgit | Lugrin | University of Würzburg, Würzburg, Germany | 0.6930 |
| Sarah | Gillet | KTH Royal Inst. of Tech., Stockholm, Sweden | 0.6927 |
| Nina | Moorman | Georgia Institute of Technology, Atlanta, GA, USA | 0.6927 |
| Ray | LC | City University of Hong Kong, Hong Kong, Hong Kong | 0.6927 |
| Tal | Oron-Gilad | Ben Gurion University of the Negev, Be'er Sheva, Israel | 0.6927 |
| Theresa | Schmiedel | University of Applied Sciences and Arts Northwestern Switzerland, Basel, Switzerland | 0.6927 |
| Vivienne | Zhong | University of Applied Sciences and Arts Northwestern Switzerland, Basel, Switzerland | 0.6927 |
| Chrystopher | Nehaniv | University of Waterloo, Waterloo, ON, Canada | 0.6927 |
| Munjal | Desai | University of Massachusetts Lowell, Lowell, USA | 0.6927 |
| Sara | Kiesler | Carnegie Mellon University, Pittsburgh, PA, USA | 0.6927 |
| Jenay | Beer | Georgia Institute of Technology, Atlanta, GA, USA | 0.6927 |
| Nazli | Cila | Human-Centered Design, Delft University of Technology, Delft, Netherlands | 0.6909 |
| Masayuki | Kanbara | Nara Institute of Science and Technology, Ikoma, Japan | 0.6909 |
| Donald | McMillan | Stockholm University, Stockholm, Sweden | 0.6905 |
| Sooyeon | Jeong | Northwestern University, Chicago, IL, USA | 0.6905 |
| Debasmita | Ghose | Yale University, New Haven, CT, USA | 0.6905 |
| Thomas | Beelen | University of Twente, Enschede, Netherlands | 0.6905 |
| Khiet | Truong | University of Twente, Enschede, Netherlands | 0.6905 |
| Anna-Lisa | Vollmer | Bielefeld University, Bielefeld, Germany | 0.6905 |
| Kuanhao | Zheng | ATR, Kyoto, Japan | 0.6905 |
| Oskar | Palinko | University of Southern Denmark, Odense, Denmark | 0.6861 |
| Alisha | Bevins | University of Nebraska-Lincoln, Lincoln, NE, USA | 0.6859 |
| Pia | Dautzenberg | RWTH Aachen University, Aachen, Germany | 0.6859 |
| Stefan | Ladwig | RWTH Aachen University, Aachen, NRW, Germany | 0.6859 |
| Adam | Stogsdill | RAISonance, Greenwood Village, CO, USA | 0.6859 |
| Grace | Clark | Colorado School of Mines, Golden, CO, USA | 0.6859 |
| Jindan | Huang | Tufts University, Medford, MA, USA | 0.6859 |
| Isaac | Sheidlower | Tufts University, Medford, MA, USA | 0.6859 |
| Minjung | Park | Carnegie Mellon University, Pittsburgh, PA, USA | 0.6859 |
| Olov | Engwall | KTH Royal Institute of Technology, Stockholm, Sweden | 0.6859 |
| Mariacarla | Staffa | University of Naples Parthenope, Naples, Italy | 0.6859 |
| Kei | Okada | The University of Tokyo, Tokyo, Japan | 0.6859 |
| Takanori | Komatsu | Meiji University, Tokyo, Japan | 0.6859 |
| Alexander | Arntz | University of Applied Sciences Ruhr West, Oberhausen, Germany | 0.6859 |
| Nicola | Webb | Bristol Robotics Laboratory, University of the West of England, Bristol, United Kingdom | 0.6859 |
| Rahatul | Ananto | McGill University, Montreal, QC, Canada | 0.6859 |
| Rebecca | Ramnauth | Yale University, New Haven, CT, USA | 0.6833 |
| Emmanuel | Adeniran | Yale University, New Haven, CT, USA | 0.6833 |
| Michal | Lewkowicz | Yale University, New Haven, CT, USA | 0.6833 |
| Zhao | Zhao | University of Toronto Mississauga, Mississauga, ON, Canada | 0.6833 |
| Rhonda | McEwen | University of Toronto Mississauga, Mississauga, ON, Canada | 0.6833 |
| Gaspar | Melsion | KTH Royal Institute of Technology, Stockholm, Sweden | 0.6833 |
| Kaitlynn | Pineda | Yale University, New Haven, CT, USA | 0.6833 |
| Jana | Tumova | KTH Royal Institute of Technology, Stockholm, Sweden | 0.6833 |
| Kazuo | Hiraki | The University of Tokyo, Tokyo, Japan | 0.6833 |
| Victoria | Chen | Oregon State University, Corvallis, OR, USA | 0.6833 |
| Kleio | Baxevani | University of Delaware, Newark, DE, USA | 0.6833 |
| Herbert | Tanner | University of Delaware, Newark, DE, USA | 0.6833 |
| Elena | Kokkoni | University of California, Riverside, Riverside, CA, USA | 0.6833 |
| Ela | Liberman-Pincu | Ben Gurion University of the Negev, Beer Sheva, Israel | 0.6833 |
| Birthe | Nesset | Heriot-Watt University, Edinburgh, United Kingdom | 0.6833 |
| Juhyun | Kim | Korea Institute of Robotics and Technology Convergence, Pohang, Republic of Korea | 0.6833 |
| Min-Gyu | Kim | Korea Institute of Robotics and Technology Convergence, Pohang, Republic of Korea | 0.6833 |
| Heejin | Jeong | University of Illinois at Chicago, Chicago, IL, USA | 0.6833 |
| Jimin | Rhim | McGill University, Montreal, Canada | 0.6833 |
| Eduard | Fosch-Villaronga | Leiden University, Leiden, Netherlands | 0.6833 |
| Nicholas | Rabb | Tufts University, Medford, MA, USA | 0.6833 |
| Theresa | Law | Tufts University, Medford, MA, USA | 0.6833 |
| Carlos | Martinho | INESC-ID and IST, Technical University of Lisbon, Lisbon, Portugal | 0.6833 |
| Mikhail | Medvedev | University of Massachusetts Lowell, Lowell, USA | 0.6833 |
| Akanksha | Prakash | Georgia Institute of Technology, Atlanta, GA, USA | 0.6833 |
| Cory-Ann | Smarr | Georgia Institute of Technology, Atlanta, GA, USA | 0.6833 |
| Tracy | Mitzner | Georgia Institute of Technology, Atlanta, GA, USA | 0.6833 |
| Wendy | Rogers | Georgia Institute of Technology, Atlanta, GA, USA | 0.6833 |
| Reuben | Aronson | Tufts University, Medford, MA, USA | 0.6786 |
| Wonse | Jo | University of Michigan, Ann Arbor, MI, USA | 0.6786 |
| Lionel | Robert Jr. | University of Michigan, Ann Arbor, MI, USA | 0.6786 |
| Aida | Zhanatkyzy | Nazarbayev University, Astana, Kazakhstan | 0.6781 |
| Zhansaule | Telisheva | Nazarbayev University, Astana, Kazakhstan | 0.6781 |
| Sophia | Steinhaeusser | University of Würzburg, Würzburg, Germany | 0.6779 |
| Giulia | Perugia | Eindhoven University of Technology, Eindhoven, Netherlands | 0.6681 |
| Ronald | Cumbal | KTH Royal Institute of Technology, Stockholm, Sweden | 0.6667 |
| Matous | Jelinek | University of Southern Denmark, Sønderborg, Denmark | 0.6667 |
| Anouk | Neerincx | Utrecht University, Utrecht, Netherlands | 0.6667 |
| Youssef | Mohamed | KTH The Royal Institute of Technology, Stockholm, Sweden | 0.6667 |
| Angelica | Lim | Simon Fraser University, Burnaby, Canada | 0.6667 |
| Cynthia | Matuszek | University of Maryland, Baltimore County, Baltimore, MD, USA | 0.6667 |
| Thao | Phung | Colorado School of Mines, Golden, CO, USA | 0.6667 |
| Joakim | Gustafson | KTH Royal Institute of Technology, Stockholm, Sweden | 0.6667 |
| Rebecca | Currano | Stanford University, Stanford, CA, USA | 0.6667 |
| Muyleng | Ghuy | Georgia Institute of Technology, Atlanta, GA, USA | 0.6667 |
| Diljot | Garcha | University of Manitoba | 0.6667 |
| Michelle | Park | Stanford University | 0.6667 |
| Aaquib | Tabrez | University of Colorado Boulder | 0.6667 |
| Tathagata | Chakraborti | IBM Research AI | 0.6667 |
| Elie | Saad | Delft University of Technology, Delft, The Netherlands | 0.6667 |
| Amit | Pandey | SoftBank Robotics Europe, Paris, France | 0.6667 |
| Nak | Chong | JAIST, Kanazawa, Japan | 0.6667 |
| Xingkun | Liu | Heriot-Watt University, Edinburgh, UK | 0.6667 |
| Laura | Hoffman | Bielefeld University, Bielefeld, Germany | 0.6667 |
| Ezgi | Mamus | Koç University, Istanbul, Turkey | 0.6667 |
| Jean-Marc | Montanier | SoftBank Robotics, Paris, France | 0.6667 |
| Cansu | Oranc | Koç University, Istanbul, Turkey | 0.6667 |
| Ora | Oudgenoeg-Paz | Utrecht University, Utrecht, the Netherlands | 0.6667 |
| Daniel | Garcia | University of Plymouth, Plymouth, United Kingdom | 0.6667 |
| Josje | Verhagen | Utrecht University, Utrecht, the Netherlands | 0.6667 |
| Christopher | Wallbridge | University of Plymouth, Plymouth, United Kingdom | 0.6667 |
| Tilbe | Goksun | Koç University, Istanbul, Turkey | 0.6667 |
| Aylin | Kuntay | Koç University, Istanbul, Turkey | 0.6667 |
| Paul | Leseman | Utrecht University, Utrecht, the Netherlands | 0.6667 |
| Michael | Novitzky | Massachusetts Institute of Technology | 0.6667 |
| Michael | Benjamin | Massachusetts Institute of Technology | 0.6667 |
| Caileigh | Fitzgerald | Massachusetts Institute of Technology | 0.6667 |
| Henrik | Schmidt | Massachusetts Institute of Technology | 0.6667 |
| Patric | Spence | University of Central Florida | 0.6667 |
| Tatsuya | Nomura | Ryukoku University, Otsu, Japan | 0.6667 |
| Alessandro | Di Nuovo | Sheffield Hallam University, Sheffield, United Kingdom | 0.6667 |
| Daniela | Conti | Sheffield Hallam University, Sheffield, United Kingdom | 0.6667 |
| Shirley | Elprama | imec-SMIT-Vrije Universiteit Brussel, Brussels, Belgium | 0.6667 |
| An | Jacobs | imec-SMIT-Vrije Universiteit Brussel, Brussels, Belgium | 0.6667 |
| Pablo | Esteban | Vrije Universiteit Brussel, Brussels, Belgium | 0.6667 |
| Kheng | Koay | University of Hertfordshire, Hatfield, England | 0.6667 |
| Keunwook | Kim | Seoul National University, Seoul, Korea | 0.6667 |
| Minkyung | Kim | KAIST, Daejeon, Korea | 0.6667 |
| Hotae | Lee | Seoul National University, Seoul, Korea | 0.6667 |
| Jaehoon | Kim | Seoul National University, Seoul, Korea | 0.6667 |
| Jungwook | Moon | Seoul National University, Seoul, Korea | 0.6667 |
| Scott | Niekum | University of Texas at Austin | 0.6667 |
| Meg | Tonkin | University of Technology Sydney, Sydney, Australia | 0.6667 |
| Xun | Wang | Commonwealth Bank of Australia, Sydney, Australia | 0.6667 |
| William | Judge | Commonwealth Bank of Australia, Sydney, Australia | 0.6667 |
| Siddharta | Srinivasa | Carnegie Mellon University, Pittsburgh, USA | 0.6667 |
| Fernando | Garcia | École Polytechnique Fédérale de Lausanne (EPFL), Lausanne, Switzerland | 0.6667 |
| Alexis | Jacq | École Polytechnique Fédérale de Lausanne (EPFL), Lausanne, Switzerland | 0.6667 |
| S. | Sundar | Penn State University, University Park, PA, USA | 0.6667 |
| Marissa | McCoy | Yale University, New Haven, CT, USA | 0.6667 |
| Rosemarijn | Looije | TNO Human Factors, Soesterberg, Netherlands | 0.6667 |
| Sofia | Thunberg | Linköping University, Linköping, Sweden | 0.6667 |
| Takamasa | Iio | University of Tsukuba, Tsukuba, Japan | 0.6667 |
| Yu | Zhang | Arizona State University, Tempe, AZ, USA | 0.6667 |
| Philipp | Schaper | University of Wuerzburg, Wuerzburg, Germany | 0.6667 |
| Santosh | Banisetty | Colorado School of Mines, Golden, CO, USA | 0.6667 |
| Zhao | Han | University of South Florida, Tampa, FL, USA | 0.6657 |
| Junya | Nakanishi | Osaka University, Osaka, Japan | 0.6613 |
| Hadas | Kress-Gazit | Cornell University, Ithaca, NY, USA | 0.6548 |
| Katherine | Tsui | Toyota Research Institute, Cambridge, MA, USA | 0.6528 |
| Francesco | Rea | CONTACT Unit, Italian Institute of Technology, Genoa, Italy | 0.6525 |
| Jun | Baba | CyberAgent, Inc., Tokyo, Japan | 0.6523 |
| Alessandra | Sciutti | CONTACT Unit, Italian Institute of Technology, Genoa, Italy | 0.6510 |
| Dorsa | Sadigh | Google Deepmind, Mountain View, CA, USA | 0.6458 |
| Erdem | Biyik | Stanford University, Stanford, CA, USA | 0.6458 |
| Anna | Dobrosovestnova | Human-Computer Interaction Group, TU Wien, Vienna, Austria | 0.6447 |
| Omer | Gvirsman | Curiosity Lab, Department of Industrial Engineering, Tel-Aviv University, Tel Aviv, Israel, Israel | 0.6389 |
| Manuel | Giuliani | Kempten University of Applied Sciences, Kempten, Germany | 0.6389 |
| Bram | Willemsen | Tilburg University, Tilburg, the Netherlands | 0.6389 |
| Justin | Hart | The University of Texas at Austin, Austin, TX, USA | 0.6367 |
| Alyssa | Kubota | San Francisco State University, San Francisco, CA, USA | 0.6342 |
| Ruchen | Wen | Univ. of Maryland Baltimore County, Baltimore, MD, USA | 0.6342 |
| Andreea | Bobu | University of California, Berkeley, Berkeley, CA, USA | 0.6337 |
| Avram | Block | Motional, Boston, MA, USA | 0.6333 |
| Nathan | Tsoi | Yale University, New Haven, CT, USA | 0.6204 |
| Timothy | Adamson | Yale University, New Haven, CT, USA | 0.6204 |
| Michael | Johnson | Georgia Institute of Technology, Atlanta, GA, USA | 0.6204 |
| Joe | Connolly | Yale University, New Haven, CT, USA | 0.6204 |
| Sachie | Yamada | Tokai University & ATR, Kanagawa & Kyoto, Japan | 0.6204 |
| Raquel | Oliveira | Instituto Universitário de Lisboa (ISCTE-IUL), CIS-IUL, Portugal and INESC-ID, Portugal | 0.6204 |
| Rui | Prada | Universidade de Lisboa | 0.6204 |
| Ayberk | Ozgur | École Polytechnique Fédérale de Lausanne, Lausanne, Switzerland | 0.6204 |
| Paul | Baxter | Plymouth University, Plymouth, United Kingdom | 0.6204 |
| Amal | Nanavati | University of Washington, Seattle, WA, USA | 0.6194 |
| Manisha | Natarajan | Georgia Institute of Technology, Atlanta, GA, USA | 0.6194 |
| Angelo | Cangelosi | University of Manchester, Manchester, United Kingdom | 0.6194 |
| Luis | Morales Saiki | Advanced Telecommunications Research Institute International, Kyoto, Japan | 0.6194 |
| Dagoberto | Cruz-Sandoval | University of California, San Diego, San Diego, CA, USA | 0.6181 |
| Mirjam | de Haas | HU University of Applied Sciences Utrecht, Utrecht, Netherlands | 0.6146 |
| Letian | Chen | Georgia Institute of Technology, Atlanta, GA, USA | 0.6127 |
| Rebecca | Stower | KTH Royal Institute of Technology, Stockholm, Sweden | 0.6127 |
| Stephanie | Rosenthal | Carnegie Mellon University, Pittsburgh, PA, USA | 0.6127 |
| Yeping | Wang | University of Wisconsin-Madison, Madison, WI, USA | 0.6127 |
| Yunus | Terzioglu | Northeastern University, Boston, MA, USA | 0.6127 |
| Aditi | Talati | Stanford University, Stanford, CA, USA | 0.6127 |
| Thomas | Groechel | University of Southern California, Los Angeles, CA, USA | 0.6127 |
| Anouk | van Maris | University of the West of England, Bristol, United Kingdom | 0.6127 |
| Hannah | Pelikan | Department of Culture and Society, Linköping University, Linköping, Östergötland, Sweden | 0.6125 |
| Glenda | Hannibal | TU Wien, Vienna, Austria | 0.6125 |
| Nick | Walker | University of Washington, Seattle, WA, USA | 0.6111 |
| Houston | Claure | Yale University, New Haven, CT, USA | 0.6111 |
| Dimosthenis | Kontogiorgos | Adaptive Systems Group, Humboldt-Universität zu Berlin, Berlin, Germany | 0.6111 |
| Chad | Edwards | Western Michigan University, Kalamazoo, MI, USA | 0.6111 |
| Autumn | Edwards | Western Michigan University, Kalamazoo, MI, USA | 0.6111 |
| Ella | Velner | University of Twente, Enschede, Netherlands | 0.6111 |
| Nicolas | Spatola | Istituto Italiano di Tecnologia, Genova, Italy | 0.6111 |
| De'Aira | Bryant | Georgia Institute of Technology, Atlanta, GA, USA | 0.6111 |
| Muneeb | Ahmad | Heriot-Watt University, Edinburgh, United Kingdom | 0.6111 |
| Eric | Pairet | Heriot-Watt University & University of Edinburgh, Edinburgh, United Kingdom | 0.6111 |
| Jaap | Ham | Eindhoven University of Technology, Eindhoven, Netherlands | 0.6111 |
| Panos | Markopoulos | Eindhoven University of Technology, Eindhoven, Netherlands | 0.6111 |
| Yaacov | Koren | Tel Aviv University, Tel Aviv, Israel | 0.6111 |
| Michael | Suguitan | Cornell University, Ithaca, NY, USA | 0.6111 |
| Andrew | Murtagh | Akara Robotics, Dublin, Ireland | 0.6111 |
| JongSuk | Choi | Korea Institute of Science and Technology (KIST), Seoul, Republic of Korea | 0.6111 |
| Sonja | Stange | Bielefeld University, Bielefeld, Germany | 0.6111 |
| Samuel | Mascarenhas | Universidade de Lisboa | 0.6111 |
| Anthony | Harrison | U.S. Naval Research Laboratory | 0.6111 |
| Hae-Sung | Lee | Yonsei University, Seoul, Korea | 0.6111 |
| JeeHang | Lee | KAIST, Daejeon, Korea | 0.6111 |
| Jinwoo | Kim | Yonsei University, Seoul, Korea | 0.6111 |
| Mitsuhiro | Goto | NIPPON TELEGRAPH AND TELEPHONE CORPORATION, Kanagawa, Japan | 0.6111 |
| Laurie | Santos | Yale University | 0.6111 |
| Kirsten | Bergmann | Bielefeld University, Bielefeld, Germany | 0.6111 |
| Chandrayee | Basu | University of California Merced, Merced, CA, USA | 0.6111 |
| Mukesh | Singhal | University of California Merced, Merced, CA, USA | 0.6111 |
| Tomohiro | Yamada | NTT Service Evolution Laboratories, Yokosuka-shi, Japan | 0.6111 |
| Nikolai | Bock | University of Duisburg-Essen, Duisburg, Germany | 0.6111 |
| Jurgen | Brandstetter | University of Canterbury, Christchurch, New Zealand | 0.6111 |
| Denise | Hebesberger | Akademie für Altersforschung, Vienna, Austria | 0.6111 |
| John | Voiklis | Brown University, Providence, RI, USA | 0.6111 |
| Alexandru | Litoiu | Yale University, New Haven, CT, USA | 0.6111 |
| Corey | Cusimano | University of Pennsylvania, Philadelphia, PA, USA | 0.6111 |
| Maha | Salem | University of Hertfordshire, Hatfield, United Kingdom | 0.6111 |
| Rui | Fang | Michigan State University, East Lansing, MI, USA | 0.6111 |
| Joyce | Chai | Michigan State University, East Lansing, MI, USA | 0.6111 |
| Keinan | Vanunu | IDC Herzliya, Herzliya, Israel | 0.6111 |
| Jakub | Zlotowski | University of Canterbury, Christchurch, New Zealand | 0.6111 |
| Brian | Gleeson | University of British Columbia, Vancouver, BC, Canada | 0.6111 |
| Elizabeth | Jochum | Aalborg University, Aalborg, Denmark | 0.6111 |
| Janie | Grant | University of Canberra, Canberra, ACT, Australia | 0.6111 |
| David | Goedicke | Cornell Tech, New York, NY, USA | 0.6090 |
| Karon | MacLean | University of British Columbia, Vancouver, BC, Canada | 0.6090 |
| Gabriel | Skantze | KTH Royal Institute of Technology, Stockholm, Sweden | 0.6090 |
| Ethan | Gordon | University of Washington, Seattle, WA, USA | 0.6042 |
| Xiajie | Zhang | Massachusetts Institute of Technology, Cambridge, MA, USA | 0.6042 |
| Felix | Lindner | Ulm University, Ulm, Germany | 0.6042 |
| Dieta | Kuchenbrandt | Center of Excellence in Cognitive Interaction Technology (CITEC), Bielefeld, Germany | 0.6042 |
| Qin | Zhu | Department of Engineering Education, Virginia Tech, Blacksburg, VA, USA | 0.6009 |
| Luca | Gambardella | USI-SUPSI, Lugano, Switzerland | 0.5996 |
| Julie | Adams | Oregon State University, Corvallis, OR, USA | 0.5967 |
| Jan | de Wit | Tilburg University, Tilburg, Noord-Brabant, Netherlands | 0.5946 |
| Emiel | Krahmer | Tilburg University, Tilburg, Noord-Brabant, Netherlands | 0.5901 |
| Matthew | Gombolay | Georgia Institute of Technology, Atlanta, GA, USA | 0.5858 |
| Mariah | Schrum | School of Interactive Computing, Georgia Institute of Technology & Robotics, University of California, Berkeley, Atlanta, GA, USA | 0.5839 |
| Huili | Chen | MIT Media Lab, Cambridge, MA, USA | 0.5833 |
| Yoyo | Hou | Cornell University, Ithaca, NY, USA | 0.5833 |
| Jacob | Crandall | Brigham Young University, Provo, UT, USA | 0.5833 |
| David | Porfirio | U.S. Naval Research Laboratory, NRC Postdoctoral Research Associate, Washington, DC, USA | 0.5833 |
| Oya | Celiktutan | King's College London, London, United Kingdom | 0.5833 |
| Gabriele | Abbate | Dalle Molle Institute for Artificial Intelligence (IDSIA) USI-SUPSI, Lugano, Switzerland | 0.5833 |
| Francois | Ferland | Université de Sherbrooke, Interdisciplinary Institute for Technological Innovation & Université de Sherbrooke, Research Center of Aging, Sherbrooke, QC, Canada | 0.5833 |
| Deborah | Szapiro | University of Technology Sydney, Sydney, Australia | 0.5833 |
| Elizabeth | Broadbent | Psychological Medicine, University of Auckland, Auckland, New Zealand | 0.5833 |
| Ho | Ahn | Electrical, Computer and Software Engineering, The University of Auckland, Auckland, New Zealand | 0.5833 |
| Elisabeth | Andre | Chair for Human-Centered Artificial Intelligence, University of Augsburg, Augsburg, Germany | 0.5833 |
| Ho | Siu | Lincoln Laboratory, Massachusetts Institute of Technology, Lexington, MA, USA | 0.5833 |
| Kotaro | Funakoshi | Tokyo Institute of Technology, Tokyo, Japan | 0.5833 |
| Cindy | Grimm | Oregon State University, Corvallis, OR, USA | 0.5833 |
| Jason | Borenstein | Georgia Institute of Technology, Atlanta, GA, USA | 0.5833 |
| Sayanti | Roy | Purdue University Northwest, Hammond, Indiana, USA | 0.5833 |
| Heqiu | Song | RWTH Aachen University, Aachen, Germany | 0.5833 |
| AJung | Moon | McGill University, Montreal, QC, Canada | 0.5833 |
| Patrick | Holthaus | University of Hertfordshire, Hatfield, United Kingdom | 0.5833 |
| Shiri | Azenkot | Jacobs Technion-Cornell Institute, Cornell Tech, New York, NY, USA | 0.5833 |
| Aws | Albarghouthi | University of Wisconsin-Madison, Madison, WI, USA | 0.5833 |
| Shelly | Bagchi | National Institute of Standards and Technology, Gaithersburg, MD, USA | 0.5833 |
| Yoonseob | Lim | Korea Institute of Science and Technology, Seoul, Republic of Korea | 0.5833 |
| Megan | Zimmerman | U.S. National Institute of Standards and Technology, Gaithersburg, MD, USA | 0.5833 |
| Jeremy | Marvel | U.S. National Institute of Standards and Technology, Gaithersburg, MD, USA | 0.5833 |
| Franca | Garzotto | Politecnico di Milano, Milan, Italy | 0.5833 |
| Paul | Vogt | Tilburg University, Tilburg, Noord-Brabant, Netherlands | 0.5833 |
| Matthew | Walter | Toyota Technological Institute at Chicago, Chicago, IL, USA | 0.5833 |
| Michio | Okada | Toyohashi University of Technology, Toyohashi, Japan | 0.5833 |
| Pamela | Hinds | Stanford University, Stanford, CA, USA | 0.5833 |
| Charles | Rich | Worcester Polytechnic Institute, Worcester, MA, USA | 0.5833 |
| Candace | Sidner | Worcester Polytechnic Institute, Worcester, MA, USA | 0.5833 |
| Stefanie | Tellex | Massachusettes Institute of Technology, Cambridge, MA, USA | 0.5833 |
| Tom | Williams | Colorado School of Mines, Golden, CO, USA | 0.5781 |
| Ilaria | Torre | Computer Science and Engineering, Chalmers University of Technology, Gothenburg, Sweden | 0.5774 |
| Cristina | Zaga | University of Twente, Enschede, Netherlands | 0.5774 |
| Pieter | Wolfert | Donders Institute for Brain, Cognition and Behaviour, Radboud University, Nijmegen, Netherlands | 0.5770 |
| Sarah | Sebo | The University of Chicago, Chicago, USA | 0.5682 |
| Alap | Kshirsagar | Technische Universität Darmstadt, Darmstadt, Germany | 0.5682 |
| Min | Lee | University of Texas at Austin, Austin, TX, USA | 0.5682 |
| Conor | McGinn | Trinity College Dublin, Dublin, Ireland | 0.5680 |
| Vanessa | Evers | University of Twente & Nanyang Technological University, Enschede, Netherlands | 0.5667 |
| Raquel | Thiessen | University of Manitoba, Winnipeg, MAN, Canada | 0.5667 |
| Boris | Gromov | USI-SUPSI, Lugano, Switzerland | 0.5667 |
| Michael | Koller | Technische Universität Wien, Vienna, Austria | 0.5667 |
| Sebastian | Schneider | University of Applied Sciences Cologne, Cologne, Germany | 0.5667 |
| Nhan | Tran | Colorado School of Mines, Golden, CO, USA | 0.5667 |
| Sarah | Strohkorb Sebo | Yale University, New Haven, CT, USA | 0.5667 |
| Dylan | Hadfield-Menell | UC Berkeley | 0.5667 |
| Andrew | Williams | University of Kansas | 0.5667 |
| Dahyun | Kang | Center for Intelligent and Interactive Robotics, Korea Institute of Science and Technology, Seoul, Korea, Republic of | 0.5661 |
| Sanne | van Waveren | Georgia Institute of Technology, Atlanta, GA, USA | 0.5661 |
| Hatice | Gunes | University of Cambridge, Cambridge, United Kingdom | 0.5643 |
| Mike | Ligthart | Vrije Universiteit Amsterdam, Amsterdam, Netherlands | 0.5627 |
| Emilia | Barakova | Eindhoven University of Technology, Eindhoven, Netherlands | 0.5627 |
| Tapomayukh | Bhattacharjee | Cornell University, Ithaca, NY, USA | 0.5625 |
| Helen | Hastie | Heriot-Watt University, Edinburgh, United Kingdom | 0.5604 |
| X. | Yang | University of Michigan, Ann Arbor, MI, USA | 0.5595 |
| Brittany | Duncan | University of Nebraska-Lincoln, Lincoln, NE, USA | 0.5565 |
| Myounghoon | Jeon | Grado Department of Industrial and Systems Engineering, Virginia Tech, Blacksburg, VA, USA | 0.5565 |
| Serge | Thill | Radboud University, Nijmegen, Netherlands | 0.5565 |
| Ryan | Jackson | Colorado School of Mines, Golden, CO, USA | 0.5565 |
| Fotios | Papadopoulos | University of Plymouth, Plymouth, United Kingdom | 0.5565 |
| Thorsten | Schodde | Bielefeld University, Bielefeld, Germany | 0.5565 |
| Irvin | Cardenas | Kent State University, Kent | 0.5565 |
| Jong-Hoon | Kim | Kent State University, Kent | 0.5565 |
| Alessandro | Giusti | Dalle Molle Institute for Artificial Intelligence (IDSIA) USI-SUPSI, Lugano, Switzerland | 0.5493 |
| Jerome | Guzzi | USI-SUPSI, Lugano, Switzerland | 0.5493 |
| Elizabeth | Phillips | Psychology, George Mason University, Fairfax, VA, USA | 0.5449 |
| Hiroyuki | Umemuro | Tokyo Institute of Technology, Tokyo, Japan | 0.5443 |
| Katherine | Kuchenbecker | MPI for Intelligent Systems, Stuttgart, Germany | 0.5435 |
| Mattia | Racca | Aalto University, Espoo, Finland | 0.5435 |
| Ross | Knepper | Cornell University | 0.5435 |
| Brian | Mok | Stanford University, Stanford, CA, USA | 0.5435 |
| Justin | Huang | University of Washington, Seattle, WA, USA | 0.5435 |
| Micheline | Ziadee | American University of Science and Technology, Beirut, Lebanon | 0.5435 |
| Majd | Sakr | Carnegie Mellon University, Pittsburgh, PA, USA | 0.5435 |
| Sarath | Sreedharan | Colorado State University, Fort Collins, CO, USA | 0.5417 |
| Rianne | van den Berghe | Windesheim University of Applied Sciences, Almere, Netherlands | 0.5417 |
| Marcela | Munera | University of Bristol, Bristol, United Kingdom | 0.5417 |
| Carlos | Cifuentes | Bristol Robotics Laboratory, UWE, Bristol, United Kingdom | 0.5417 |
| Meia | Chita-Tegmark | Tufts University, Medford, MA, USA | 0.5389 |
| Pragathi | Praveena | University of Wisconsin-Madison, Madison, WI, USA | 0.5389 |
| Ross | Mead | Semio AI, Inc., Los Angeles, CA, USA | 0.5389 |
| Junko | Kanero | Sabanci University, Istanbul, Turkey | 0.5379 |
| David | Robb | Heriot-Watt University, Edinburgh, United Kingdom | 0.5370 |
| Heather | Knight | Collaborative Robotics and Intelligent Systems Institute, Oregon State University, Corvallis, OR, USA | 0.5362 |
| Agnieszka | Wykowska | Istituto Italiano di Tecnologia, Genova, Italy | 0.5357 |
| Chris | Crawford | Department of Computer Science, The University of Alabama, Tuscaloosa, AL, USA | 0.5357 |
| William | Smart | Oregon State University, Corvallis, OR, USA | 0.5357 |
| Ewart | de Visser | U.S. Air Force Academy, Air Force Academy, CO, USA | 0.5357 |
| Zahra | Zahedi | Arizona State University, Tempe, AZ, USA | 0.5357 |
| Yuki | Okafuji | CyberAgent, Inc., Tokyo, Japan | 0.5357 |
| Bob | Schadenberg | University of Twente, Enschede, Netherlands | 0.5357 |
| Michael | Lewis | University of Pittsburgh, Pittsburgh, PA, USA | 0.5357 |
| Mayumi | Mohan | MPI for Intelligent Systems, Stuttgart, Germany | 0.5333 |
| Kerstin | Haring | University of Denver, DENVER, CO, USA | 0.5333 |
| Akiko | Yamazaki | Tokyo University of Technology, Hachioji, Japan | 0.5333 |
| Jose | Lopes | Heriot Watt University, Edinburgh, United Kingdom | 0.5333 |
| Minae | Kwon | Stanford University, Stanford, CA, USA | 0.5333 |
| Dylan | Losey | Stanford University, Stanford, CA, USA | 0.5333 |
| Diana | Loffler | University of Siegen, Siegen, Germany | 0.5333 |
| Swapna | Joshi | Indiana University | 0.5333 |
| Jacob | Nielsen | University of Southern Denmark | 0.5333 |
| Ronald | Moore | University of Kansas | 0.5333 |
| Mirko | Gelsomini | Politecnico di Milano, Milan, Italy | 0.5333 |
| Andre | Cleaver | Tufts University | 0.5333 |
| Jivko | Sinapov | Tufts University | 0.5333 |
| Abhijeet | Agnihotri | Oregon State University | 0.5333 |
| Joseph | Daly | UWE, Frenchay Campus, Bristol, UK | 0.5333 |
| Xuan | Zhao | Brown University, Providence, RI, USA | 0.5333 |
| Sofia | Petisca | Instituto Universitário de Lisboa (ISCTE-IUL), CIS-IUL & INESC-ID, Lisbon, Portugal | 0.5333 |
| Norman | Su | Indiana University, Bloomington, IN, USA | 0.5333 |
| Jean-Claude | Martin | Université Paris-Saclay, Orsay, France | 0.5333 |
| Wan-Ling | Chang | Indiana University, Bloomington, IN, USA | 0.5333 |
| Gurit | Birnbaum | IDC Herzliya, Herzliya, Israel | 0.5333 |
| Harry | Reis | University of Rochester, Rochester, NY, USA | 0.5333 |
| Omri | Sass | Cornell Tech, New York, NY, USA | 0.5333 |
| Hiroyuki | Kidokoro | ATR & Osaka University, Kyoto, Japan | 0.5333 |
| Rohan | Paleja | Georgia Institute of Technology, Atlanta, GA, USA | 0.5323 |
| Koen | Hindriks | Vrije Universiteit Amsterdam, Amsterdam, Netherlands | 0.5307 |
| Meiying | Qin | Yale University, New Haven, CT, USA | 0.5303 |
| Vicky | Charisi | Joint Research Centre, Seville, Spain | 0.5303 |
| Hooman | Hedayati | University of North Carolina at Chapel Hill, Chapel Hill, NC, USA | 0.5238 |
| Filipa | Correia | ITI, LARSYS, Instituto Superior Técnico, Universidade de Lisboa, Lisbon, Portugal | 0.5230 |
| Casey | Bennett | Depaul University, Chicago, IL, USA | 0.5219 |
| Jennifer | Piatt | Indiana University Bloomington, Bloomington, IN, USA | 0.5219 |
| Somaya | Ben Allouch | Amsterdam University of Applied Sciences, Amsterdam, Netherlands | 0.5219 |
| Yuhan | Hu | Cornell University, Ithaca, NY, USA | 0.5219 |
| Daniel | Brown | University of Utah, Salt Lake City, UT, USA | 0.5219 |
| Alessandro | Roncone | University of Colorado Boulder, Boulder, CO, USA | 0.5219 |
| Anders | Sorensen | The Maersk M. Moller Institute, University of Southern Denmark, Odense, Denmark | 0.5219 |
| Yugo | Hayashi | College of Comprehensive Psychology, Ritsumeikan University, Ibaraki, Osaka, Japan | 0.5219 |
| Brenna | Argall | Northwestern University, Shirley Ryan Ability Lab, Chicago, IL, USA | 0.5219 |
| Timothy | Bickmore | Northeastern University, Boston, MA, USA | 0.5219 |
| Markus | Vincze | Technische Universität Wien, Vienna, Austria | 0.5219 |
| Yasuto | Nakanishi | Keio University, Fujisawa, Japan | 0.5219 |
| Kate | Loveys | The University of Auckland, Auckland, New Zealand | 0.5219 |
| Kazuki | Mizumaru | Hokkaido University, Sapporo, Hokkaido, Japan | 0.5219 |
| Tetsuo | Ono | Hokkaido University, Sapporo, Hokkaido, Japan | 0.5219 |
| Murat | Aksu | U.S. National Institute of Standards and Technology, Gaithersburg, MD, USA | 0.5219 |
| Brian | Antonishek | U.S. National Institute of Standards and Technology, Gaithersburg, MD, USA | 0.5219 |
| Yue | Wang | Clemson University, Clemson, SC, USA | 0.5219 |
| Jeonghye | Han | Cheongju National University of Education, Republic of Korea | 0.5219 |
| John | Antanitis | Carnegie Mellon University, Pittsburgh, PA, USA | 0.5219 |
| Chaoran | Liu | ATR, Keihanna, Kyoto, Japan | 0.5219 |
| Naomi | Fitter | Collaborative Robotics and Intelligent Systems (CoRIS) Institute, Oregon State University, Corvallis, OR, USA | 0.5192 |
| Katie | Winkle | Uppsala University, Uppsala, Sweden | 0.5109 |
| Anara | Sandygulova | Nazarbayev University, Astana, Kazakhstan | 0.5109 |
| Sarita | Herse | University of Technology Sydney, Sydney, NSW, Australia | 0.5076 |
| Jonathan | Vitale | University of Technology Sydney, Sydney, NSW, Australia | 0.5076 |
| Mary-Anne | Williams | University of New South Wales, Sydney, NSW, Australia | 0.5076 |
| Lars | Jensen | University of Southern Denmark Kolding, Denmark | 0.5076 |
| Sonya | Kwak | Center for Intelligent and Interactive Robotics, Korea Institute of Science and Technology, Seoul, Korea, Republic of | 0.5052 |
| Samantha | Reig | Carnegie Mellon University, Pittsburgh, PA, USA | 0.5048 |
| Trenton | Schulz | Norwegian Computing Center, Oslo, Norway | 0.5000 |
| Gabriella | Lakatos | University of Hertfordshire, Hatfield, United Kingdom | 0.5000 |
| Andrea | Bajcsy | University of California, Berkeley, Berkeley, CA, USA | 0.5000 |
| Jesse | de Pagter | TU Wien, Vienna, Austria | 0.5000 |
| Minsu | Jang | Electronics and Telecommunications Research Institute, Daejeon, Republic of Korea | 0.5000 |
| Roel | Boumans | Radboud University, Nijmegen, Netherlands | 0.5000 |
| Kosuke | Wakabayashi | Ritsumeikan University, Ibaraki, Japan | 0.5000 |
| Karolina | Zawieska | Aarhus University, Aarhus, Denmark | 0.5000 |
| Marta | Romeo | University of Manchester, Manchester, United Kingdom | 0.5000 |
| Claire | Liang | Cornell University, Ithaca, NY, USA | 0.5000 |
| Tatsuya | Kawahara | Kyoto University, Kyoto, Japan | 0.5000 |
| Patricia | Arriaga | Iscte-iul, Lisboa, Portugal | 0.5000 |
| Francisco | Melo | Universidade de Lisboa | 0.5000 |
| Hirotaka | Osawa | Univeristy of Tsukuba, Tsukuba, Japan | 0.5000 |
| Nathan | Kirchner | University of Technology, Sydney, Sydney, NSW, Australia | 0.5000 |
| Katia | Sycara | Carnegie Mellon University, Pittsburgh, PA, USA | 0.5000 |
| Peter | McOwan | Queen Mary University of London, London, United Kingdom | 0.5000 |
| David | St-Onge | Laval University, Québec, PQ, Canada | 0.5000 |
| Nicolas | Reeves | University of Québec in Montreal, Montreal, PQ, Canada | 0.5000 |
| Fumihide | Tanaka | University of Tsukuba, Tsukuba, Japan | 0.4890 |
| Michael | Walker | University of North Carolina at Chapel Hill, Chapel Hill, NC, USA | 0.4881 |
| Masahiro | Shiomi | ATR, Kyoto, Japan | 0.4833 |
| Elaine | Short | Tufts University, Medford, MA, USA | 0.4801 |
| Paul | Robinette | Massachusetts Institute of Technology | 0.4781 |
| David | Hsu | National University of Singapore, Singapore | 0.4700 |
| Thomas | Arnold | Tufts University, Medford, MA, USA | 0.4700 |
| Ailie | Turton | Bristol Robotics Laboratory, Bristol, United Kingdom | 0.4657 |
| Hamish | Tennent | Cornell University, Ithaca, NY, USA | 0.4657 |
| Jin | Lee | MIT Media Lab | 0.4657 |
| Maartje | de Graaf | Utrecht University, Utrecht, Netherlands | 0.4634 |
| Peter | Kahn | University of Washington, Seattle, WA, USA | 0.4567 |
| Brian | Gill | Seattle Pacific University, Seattle, WA, USA | 0.4567 |
| Jolina | Ruckert | University of Washington, Seattle, WA, USA | 0.4567 |
| Heather | Gary | University of Washington, Seattle, WA, USA | 0.4567 |
| Damith | Herath | University of Western Sydney, Sydney, Australia | 0.4567 |
| Masayuki | Inaba | The University of Tokyo, Tokyo, Japan | 0.4529 |
| Enkelejda | Kasneci | Technical University of Munich, Munich, Germany | 0.4529 |
| Daniel | Rea | University of New Brunswick, Fredericton, NB, Canada | 0.4516 |
| Harold | Soh | National University of Singapore, Singapore, Singapore | 0.4494 |
| Alan | Wagner | The Pennsylvania State University, State College, PA, USA | 0.4444 |
| Katrin | Lohan | Heriot-Watt University & NTB University of Applied Sciences in Technology, Buchs, CH, Edinburgh, United Kingdom | 0.4444 |
| Francesco | Mondada | École Polytechnique Fédérale de Lausanne, Lausanne, Switzerland | 0.4444 |
| Laura | Hiatt | U.S. Naval Research Laboratory, Washington, DC, USA | 0.4394 |
| Dominic | Letourneau | Interdisciplinary Institute for Technological Innovation, Université de Sherbrooke, Sherbrooke, QC, Canada | 0.4394 |
| Francois | Michaud | Interdisciplinary Institute for Technological Innovation, Université de Sherbrooke, Sherbrooke, QC, Canada | 0.4394 |
| Marc | Hanheide | University of Lincoln, Lincoln, United Kingdom | 0.4394 |
| Dylan | Moore | Stanford University, Stanford, CA, USA | 0.4394 |
| James | Landay | Stanford University | 0.4394 |
| Momotaz | Begum | University of New Hampshire, Durham, NH, USA | 0.4394 |
| Scott | Hudson | Carnegie Mellon University, Pittsburgh, PA, USA | 0.4394 |
| Kazuhiro | Sasabuchi | Microsoft, Tokyo, Japan | 0.4394 |
| Katsushi | Ikeuchi | Microsoft, Redmond, WA, USA | 0.4394 |
| Daniel | Rakita | Yale University, New Haven, CT, USA | 0.4386 |
| Malte | Jung | Cornell University, Ithaca, NY, USA | 0.4366 |
| Elizabeth | Carter | Robotics Institute, Carnegie Mellon University, Pittsburgh, PA, USA | 0.4349 |
| Chung | Park | Department of Biomedical Engineering, School of Engineering and Applied Science, George Washington University, Washington, District of Columbia, USA | 0.4312 |
| Sichao | Song | CyberAgent, Inc., Tokyo, Japan | 0.4306 |
| David | Feil-Seifer | Yale University, New Haven, CT, USA | 0.4298 |
| Wafa | Johal | University of Melbourne, Melbourne, VIC, Australia | 0.4202 |
| Serena | Booth | Massachusetts Institute of Technology, Cambridge, MA, USA | 0.4167 |
| Sanja | Dogramadzi | University of Sheffield, Sheffield, United Kingdom | 0.4167 |
| Pauline | Chevalier | Italian Institute of Technology, Genoa, Italy | 0.4167 |
| Mai | Chang | University of Texas Austin, Austin, TX, USA | 0.4167 |
| Hiroko | Kamide | Nagoya University, Nagoya, Japan | 0.4167 |
| Takahiro | Matsumoto | Keio University, Yokohama-shi, Japan | 0.4167 |
| Tiago | Ribeiro | INESC-ID and Instituto Superior Técnico, Universidade de Lisboa, Lisboa, Portugal | 0.4167 |
| Christoforos | Mavrogiannis | University of Washington, Seattle, WA, USA | 0.4145 |
| Stefan | Kopp | CITEC, Bielefeld University, Bielefeld, Germany | 0.4056 |
| Michael | Goodrich | Brigham Young University, Provo, UT, USA | 0.4000 |
| Zachary | Henkel | Mississippi State University, Mississippi State, MS, USA | 0.3981 |
| Eduardo | Sandoval | University of New South Wales, Sydney, NSW, Australia | 0.3958 |
| Kerstin | Dautenhahn | University of Waterloo, Waterloo, ON, Canada | 0.3958 |
| Bahar | Irfan | Evinoks Service Equipment Industry and Commerce Inc., Bursa, Turkey | 0.3917 |
| Patricia | Alves-Oliveira | University of Washington, Seattle, WA, USA | 0.3854 |
| Randy | Gomez | Honda Research Institute Japan, Wako, Japan | 0.3821 |
| Eliot | Smith | Indiana University | 0.3810 |
| Stela | Seo | University of Manitoba, Canada | 0.3810 |
| Seiji | Yamada | National Institute of Informatics & The Graduate University for Advanced Studies (SOKENDAI), Tokyo, Japan | 0.3796 |
| Elin | Topp | Lund University, Lund, Sweden | 0.3796 |
| Oren | Zuckerman | Interdisciplinary Center (IDC), Herzliya, Israel | 0.3796 |
| Rinat | Rosenberg-Kima | Technion - Israel Institute of Technology, Haifa, Israel | 0.3796 |
| Reid | Simmons | Carnegie Mellon University, Pittsburgh, PA, USA | 0.3796 |
| Michal | Luria | Carnegie Mellon University, Pittsburgh, PA, USA | 0.3782 |
| Jessica | Cauchard | Magic Lab, Industrial Engineering and Management, Ben Gurion University of the Negev & CNRS, UPS, LAAS-CNRS, Université de Toulouse, Be'er Sheva, Israel | 0.3750 |
| James | Kennedy | Disney Research, Glendale, CA, USA | 0.3679 |
| Subbarao | Kambhampati | School of Computing and Augmented Intelligence, Arizona State University, Tempe, AZ, USA | 0.3624 |
| Hae | Park | MIT Media Lab, Cambridge, MA, USA | 0.3571 |
| Astrid | Rosenthal-von der Putten | RWTH Aachen University, Aachen, Germany | 0.3567 |
| Daniel | Szafir | University of North Carolina at Chapel Hill, Chapel Hill, NC, USA | 0.3454 |
| Pierre | Dillenbourg | École Polytechnique Fédérale de Lausanne, Lausanne, Switzerland | 0.3444 |
| Astrid | Weiss | Institute of Visual Computing and Human-Centered Technology, HCI Group, Technische Universität Wien, Vienna, Austria | 0.3417 |
| Kerstin | Fischer | University of Southern Denmark, Sonderborg, Denmark | 0.3410 |
| Christopher | Crick | Oklahoma State University, Stillwater, OK, USA | 0.3333 |
| Christian | Dondrup | Heriot-Watt University, Edinburgh, United Kingdom | 0.3333 |
| Yoren | Gaffary | Univ Rennes, INSA Rennes, Inria, CNRS & Irisa-UMR6074, Rennes, France | 0.3333 |
| Steven | Sherrin | Indiana University | 0.3333 |
| Monika | Lohani | University of Utah | 0.3333 |
| Greg | Trafton | U.S. Naval Research Laboratory | 0.3333 |
| Susan | Fussell | Cornell University, Ithaca, NY, USA | 0.3333 |
| Matthew | Pan | University of British Columbia, Vancouver, BC, Canada | 0.3333 |
| James | Tompkin | Brown University, Providence, RI, USA | 0.3333 |
| Shih-Yi | Chien | Department of Management Information Systems, National Chengchi University, Taipei, Taiwan | 0.3306 |
| Marlena | Fraune | Department of Psychology, New Mexico State University, Las Cruces, NM, USA | 0.3234 |
| Aditi | Ramachandran | Vän Robotics, Columbia, SC, USA | 0.3204 |
| Malcolm | Doering | Kyoto University, Kyoto, Japan | 0.3167 |
| J. | Trafton | Naval Research Laboratory, Washington, DC, USA | 0.3167 |
| Cindy | Bethel | Mississsippi State University, Mississippi State, MS, USA | 0.3114 |
| Iolanda | Leite | KTH Royal Institute of Technology, Stockholm, Sweden | 0.3103 |
| Ute | Leonards | University of Bristol, Bristol, United Kingdom | 0.3095 |
| Dylan | Glas | Futurewei Technologies | 0.3081 |
| Andrea | Thomaz | University of Texas at Austin, Austin, TX, USA | 0.3078 |
| EunJeong | Cheon | Syracuse University, Syracuse, NY, USA | 0.3047 |
| Hee | Lee | Michigan State University, East Lansing, MI, USA | 0.3039 |
| Jamy | Li | University of Twente, Enschede, Netherlands | 0.3000 |
| Elizabeth | Cha | University of Southern California, Los Angeles, CA, USA | 0.3000 |
| Nikolas | Martelaro | Stanford University, Stanford, CA, USA | 0.3000 |
| Selma | Sabanovic | Indiana University Bloomington, Bloomington, IN, USA | 0.2939 |
| Carlos | Ishi | RIKEN, Sorakugun, Seika-cho, Kyoto, Japan | 0.2902 |
| Myrthe | Tielman | Delft University of Technology, Delft, Netherlands | 0.2902 |
| James | Young | University of Manitoba, Winnipeg, Canada | 0.2874 |
| Friederike | Eyssel | Bielefeld University, Center for Cognitive Interaction Technology (CITEC), Bielefeld, Germany | 0.2870 |
| Emmanuel | Senft | Idiap Research Institute, Martigny, Switzerland | 0.2869 |
| Adriana | Tapus | Autonomous Systems and Robotics Lab, U2IS, ENSTA Paris, Institut Polytechnique de Paris, Palaiseau, France | 0.2827 |
| Nicole | Salomons | Yale University, New Haven, CT, USA | 0.2826 |
| Sean | Andrist | Microsoft Research, Redmond | 0.2801 |
| Leila | Takayama | Hoku Labs, Santa Cruz, CA, USA | 0.2791 |
| Allison | Sauppe | University of Wisconsin-La Crosse, La Crosse, WI, USA | 0.2778 |
| Bertram | Malle | Brown University, Providence, RI, USA | 0.2717 |
| Rachid | Alami | LAAS-CNRS, Toulouse, France | 0.2696 |
| Laurel | Riek | University of California, San Diego, San Diego, CA, USA | 0.2526 |
| Marynel | Vazquez | Yale University, New Haven, CT, USA | 0.2509 |
| Goren | Gordon | Curiosity Lab, Department of Industrial Engineering, Tel Aviv Univeristy & Curiosity Robotics, Tel Aviv, Israel | 0.2500 |
| Michael | Gleicher | University of Wisconsin - Madison, Madison, WI, USA | 0.2458 |
| Elizabeth | Croft | University of Victoria & Monash University, Victoria, British Columbia, Canada | 0.2429 |
| Megan | Strait | University of Texas Rio Grande Valley, Edinburg, TX, USA | 0.2429 |
| Mark | Neerincx | Delft University of Technology, Delft, Netherlands | 0.2372 |
| Samuel | Spaulding | Massachusetts Institute of Technology, Cambridge, MA, USA | 0.2358 |
| Vaibhav | Unhelkar | Rice University, Houston, TX, USA | 0.2337 |
| Daniel | Ullman | Brown University, Providence, RI, USA | 0.2337 |
| Ginevra | Castellano | Department of Information Technology, Uppsala University, Uppsala, Sweden | 0.2332 |
| Solace | Shen | Cornell University | 0.2303 |
| Andre | Pereira | KTH Royal Institute of Technology, Stockholm, Sweden | 0.2294 |
| Denise | Geiskkovitch | Dept. of Computing and Software, McMaster University, Hamilton, ON, Canada | 0.2262 |
| Keisuke | Nakamura | Honda Research Institute Japan, Saitama, Japan | 0.2197 |
| Hiroshi | Ishiguro | Osaka University, Osaka, Japan | 0.2188 |
| Severin | Lemaignan | PAL Robotics, Barcelona, Spain | 0.2164 |
| Christoph | Bartneck | University of Canterbury, Christchurch, New Zealand | 0.2125 |
| Cynthia | Breazeal | MIT Media Lab, Cambridge, MA, USA | 0.2092 |
| Paul | Bremner | University of the West of England, Bristol, United Kingdom | 0.2051 |
| Ville | Kyrki | Aalto University, Espoo, Finland | 0.1990 |
| Amir | Aly | University of Plymouth, Plymouth, United Kingdom | 0.1909 |
| Terry | Fong | U.S. National Aeronautics and Space Administration, Moffett Field, CA, USA | 0.1909 |
| Stefanos | Nikolaidis | University of Southern California, Los Angeles, CA, USA | 0.1791 |
| Xiang | Tan | Carnegie Mellon University, Pittsburgh, PA, USA | 0.1776 |
| Wendy | Ju | Cornell Tech, New York, NY, USA | 0.1766 |
| Drazen | Brscic | Kyoto University & ATR, Kyoto, Japan | 0.1765 |
| Praminda | Caleb-Solly | University of Nottingham, Nottingham, United Kingdom | 0.1728 |
| Ana | Paiva | INESC-ID, Instituto Superior Técnico, Universidade de Lisboa & Örebro University & Umeå University, Lisbon, Portugal | 0.1699 |
| Maja | Mataric | University of Southern California, Los Angeles, CA, USA | 0.1687 |
| Ayanna | Howard | The Ohio State University, Columbus, OH, USA | 0.1667 |
| Farshid | Amirabdollahian | University of Hertfordshire, Hatfield, United Kingdom | 0.1667 |
| Sonia | Chernova | Georgia Institute of Technology, Atlanta, GA, USA | 0.1667 |
| Bilge | Mutlu | Computer Sciences Department, University of Wisconsin - Madison, Madison, WI, USA | 0.1600 |
| Maya | Cakmak | University of Washington, Seattle, WA, USA | 0.1596 |
| Bradley | Hayes | University of Colorado Boulder, Boulder, CO, USA | 0.1538 |
| Terrence | Fong | NASA Ames Research Center, Moffett Field, CA, USA | 0.1464 |
| David | Sirkin | Stanford University, Stanford, CA, USA | 0.1431 |
| Guy | Hoffman | Cornell University, Ithaca, NY, USA | 0.1425 |
| Anca | Dragan | University of California, Berkeley, Berkeley, CA, USA | 0.1393 |
| Aaron | Steinfeld | Carnegie Mellon University, Pittsburgh, PA, USA | 0.1375 |
| Siddhartha | Srinivasa | University of Washington, Seattle, WA, USA | 0.1316 |
| Chien-Ming | Huang | Johns Hopkins University, Baltimore, MD, USA | 0.1305 |
| Tony | Belpaeme | IDLab - AIRO, Ghent University - imec, Ghent, Belgium | 0.1299 |
| Takayuki | Kanda | Kyoto University, Kyoto, Japan | 0.1235 |
| Matthias | Scheutz | Tufts University, Medford, MA, USA | 0.1225 |
| Brian | Scassellati | Yale University, New Haven, CT, USA | 0.1202 |
| Holly | Yanco | University of Massachusetts Lowell, Lowell, MA, USA | 0.1146 |
| Satoru | Satake | ATR, Kyoto, Japan | 0.1136 |
| Jodi | Forlizzi | Carnegie Mellon University, Pittsburgh, PA, USA | 0.1017 |
| Julie | Shah | MIT, Cambridge, MA, USA | 0.1013 |
| Henny | Admoni | Carnegie Mellon University, Pittsburgh, USA | 0.0588 |
| Michita | Imai | Keio University, Kouhoku-ku, Yokohama-shi, Japan | 0.0227 |

# Appendix B.

**Table B1**

| Conference | Country | Keywords |
|---|---|---|
| HRI | China | [robot, interaction, behavior, orbo]<br>[child, interaction, robot, design]<br>[robot, navigation, people, human]<br>[robot, interaction, study, people]<br>[child, participant, toy, interaction] |
| | USA | [robot, human, learning, interaction]<br>[robot, human, interaction, child]<br>[robot, user, task, hri]<br>[robot, human, user, task]<br>[robot, human, participant, interaction] |
| | Japan | [robot, human, interaction, social]<br>[robot, human, agent, participant]<br>[robot, people, interaction, human]<br>[robot, touch, hand, behavior]<br>[robot, human, interaction, people] |
| | South Korea | [service, robot, restaurant, touch]<br>[robot, medical, staff, label]<br>[robot, design, user, communication]<br>[robot, human, interaction, social]<br>[robot, social, human, tactility] |
| | Germany | [robot, human, social, research]<br>[human, robot, uncertainty, task]<br>[robot, study, social, perception]<br>[robot, human, interaction, user]<br>[robot, human, social, risk] |

| Conference | Country | Keywords |
|---|---|---|
| IUI | China | [user, gesture, touch, multi]<br>[user, system, item, based]<br>[system, user, interaction, language]<br>[system, manual, marker, interaction]<br>[user, context, explanation, style] |
| | USA | [user, task, system, interface]<br>[user, paper, data, screen]<br>[user, model, system, explanation]<br>[user, model, system, tool]<br>[user, system, model, based] |
| | Japan | [data, query, user, transformation]<br>[user, system, model, text]<br>[user, viewpoint, system, object]<br>[object, user, animation, hand]<br>[user, system, sensor, interface] |
| | South Korea | [tool, model, tag, error]<br>[user, video, dissatisfaction, label]<br>[e, ltgms, visual, work]<br>[user, step, interaction, video]<br>[description, group, image, text] |
| | Germany | [user, device, based, result]<br>[user, model, interaction, driving]<br>[user, system, interaction, intelligent]<br>[user, display, gesture, warning]<br>[user, image, human, password] |

| Conference | Country | Keywords |
|---|---|---|
| KDD | China | [data, method, model, real]<br>[learning, model, data, task]<br>[model, data, graph, method]<br>[user, graph, network, node]<br>[model, data, learning, feature] |
| | USA | [user, social, network, algorithm]<br>[system, data, result, time]<br>[data, model, method, prediction]<br>[model, data, method, algorithm]<br>[graph, algorithm, clustering, network] |
| | Japan | [model, data, recommendation, method]<br>[method, data, learning, model]<br>[method, data, time, pattern]<br>[data, method, model, performance]<br>[method, data, graph, time] |
| | South Korea | [data, model, method, feature]<br>[data, based, algorithm, two]<br>[graph, time, data, network]<br>[aspect, node, change, network]<br>[data, graph, tensor, real] |
| | Germany | [model, data, pattern, algorithm]<br>[time, data, model, learning]<br>[data, clustering, graph, cluster]<br>[graph, data, model, method]<br>[data, user, model, graph] |

| Conference | Country | Keywords |
|---|---|---|
| CHI | China | [ai, user, study, human]<br>[user, design, study, based]<br>[design, user, study, based]<br>[user, study, based, system]<br>[user, design, data, visualization] |
| | USA | [user, design, study, system] |
| | Japan | [user, design, study, speech]<br>[user, system, study, feedback]<br>[study, design, user, tool]<br>[user, system, design, device]<br>[paper, user, system, magnetic] |
| | South Korea | [user, study, design, virtual]<br>[user, touch, study, design]<br>[user, study, design, interaction]<br>[user, design, study, based]<br>[user, study, based, video] |
| | Germany | [user, touch, input, interaction]<br>[user, study, data, design]<br>[user, task, time, display]<br>[user, study, virtual, experience]<br>[user, design, vr, study] |

| **Conference** | **Country** | **Keywords** |
|---|---|---|
| SIGGRAPH | China | [motion, method, model, expression]<br>[image, view, method, representation]<br>[method, network, model, structure]<br>[image, model, motion, method]<br>[method, model, point, based] |
| | USA | [method, based, model, image]<br>[time, real, approach, surface]<br>[material, light, display, field]<br>[motion, image, method, model]<br>[model, image, method, based] |
| | Japan | [method, using, display, user]<br>[motion, object, user, used] |
| | South Korea | [motion, based, model, method]<br>[image, supervised, lens, object] |
| | Germany | [image, field, model, method]<br>[model, based, approach, image]<br>[shape, algorithm, based, surface]<br>[object, light, image, based]<br>[surface, approach, neural, method] |

| **Conference** | **Country** | **Keywords** |
|---|---|---|
| UIST | China | [user, based, image, hand]<br>[user, design, device, paper]<br>[user, e, interaction, video]<br>[user, target, ring, based]<br>[design, user, color, based] |
| | USA | [user, technique, system, design]<br>[user, design, tool, model]<br>[user, system, hand, model]<br>[user, system, design, video]<br>[user, device, object, data] |
| | Japan | [user, system, design, method]<br>[ring, virtual, taste, garment]<br>[user, system, device, haptic]<br>[user, method, system, object]<br>[user, gesture, design, technique] |
| | South Korea | [touch, finger, gesture, user]<br>[table, palmrest, push, reader]<br>[user, feedback, interaction, design]<br>[user, using, study, device] |
| | Germany | [touch, user, surface, object]<br>[display, user, gaze, position]<br>[user, system, adaptation, object]<br>[user, object, device, present]<br>[user, system, physical, present] |

| Conference | Country | Keywords |
|---|---|---|
| UBICOMP | China | [user, based, system, device]<br>[signal, system, based, gesture]<br>[user, data, based, model]<br>[system, sensing, signal, device]<br>[system, time, sensing, data] |
| | USA | [user, device, system, based]<br>[data, sensor, system, feature]<br>[data, system, user, model]<br>[sensing, system, based, signal]<br>[user, system, based, device] |
| | Japan | [method, sensor, using, data]<br>[user, system, method, data]<br>[user, data, using, device]<br>[data, using, sensor, method]<br>[data, activity, recognition, method] |
| | South Korea | [user, system, stress, activity]<br>[user, phone, data, based]<br>[user, data, system, study]<br>[study, user, system, based]<br>[user, device, system, study] |
| | Germany | [user, device, interaction, human]<br>[system, activity, based, sensor]<br>[user, data, system, model]<br>[data, user, based, eye]<br>[system, recognition, based, using] |

# Appendix C.

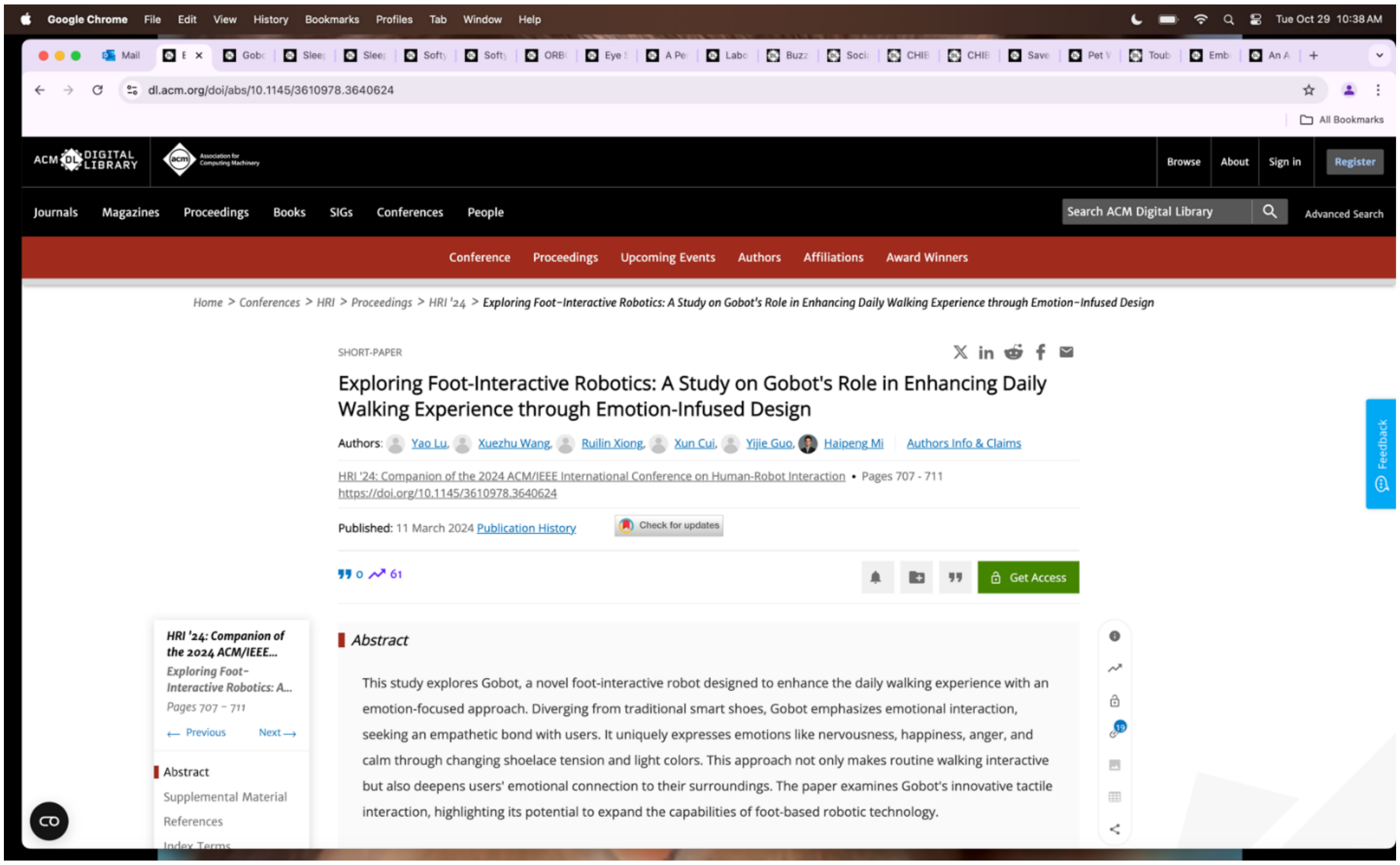
dl.acm.org/doi/abs/10.1145/3610978.3640624
Home > Conferences > HRI > Proceedings > HRI '24 > Exploring Foot-Interactive Robotics: A Study on Gobot's Role in Enhancing Daily Walking Experience through Emotion-Infused Design
SHORT-PAPER
Exploring Foot-Interactive Robotics: A Study on Gobot's Role in Enhancing Daily Walking Experience through Emotion-Infused Design
Authors: Yao Lu, Xuezhu Wang, Ruilin Xiong, Xun Cui, Yijie Guo, Haipeng Mi
Authors Info & Claims
HRI '24: Companion of the 2024 ACM/IEEE International Conference on Human-Robot Interaction • Pages 707 - 711
https://doi.org/10.1145/3610978.3640624
Published: 11 March 2024 Publication History
Check for updates
Get Access
Abstract
This study explores Gobot, a novel foot-interactive robot designed to enhance the daily walking experience with an emotion-focused approach. Diverging from traditional smart shoes, Gobot emphasizes emotional interaction, seeking an empathetic bond with users. It uniquely expresses emotions like nervousness, happiness, anger, and calm through changing shoelace tension and light colors. This approach not only makes routine walking interactive but also deepens users' emotional connection to their surroundings. The paper examines Gobot's innovative tactile interaction, highlighting its potential to expand the capabilities of foot-based robotic technology.

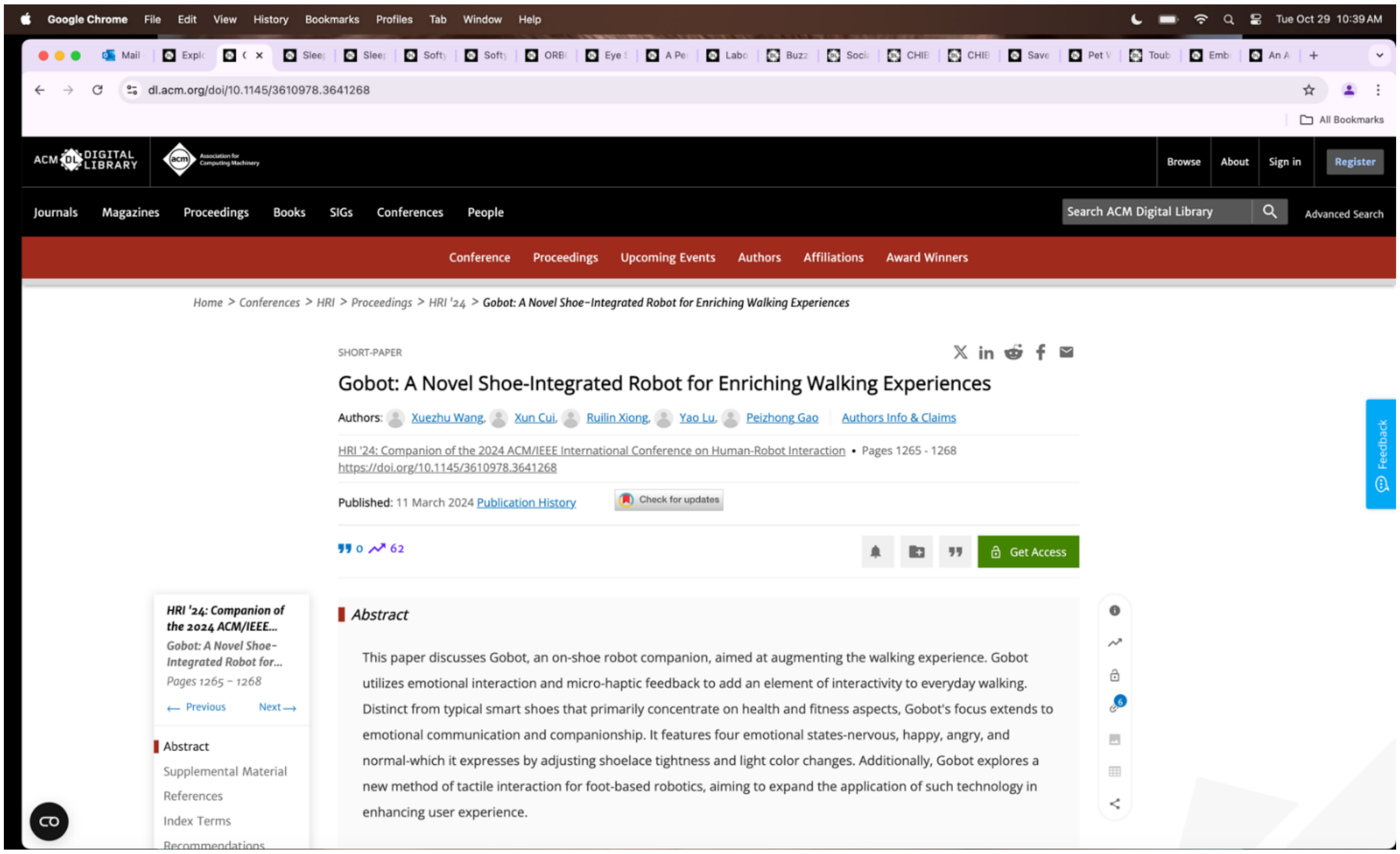
dl.acm.org/doi/10.1145/3610978.3641268
Home > Conferences > HRI > Proceedings > HRI '24 > Gobot: A Novel Shoe-Integrated Robot for Enriching Walking Experiences
SHORT-PAPER
Gobot: A Novel Shoe-Integrated Robot for Enriching Walking Experiences
Authors: Xuezhu Wang, Xun Cui, Ruilin Xiong, Yao Lu, Peizhong Gao
Authors Info & Claims
HRI '24: Companion of the 2024 ACM/IEEE International Conference on Human-Robot Interaction • Pages 1265 - 1268
https://doi.org/10.1145/3610978.3641268
Published: 11 March 2024 Publication History
Check for updates
Get Access
Abstract
This paper discusses Gobot, an on-shoe robot companion, aimed at augmenting the walking experience. Gobot utilizes emotional interaction and micro-haptic feedback to add an element of interactivity to everyday walking. Distinct from typical smart shoes that primarily concentrate on health and fitness aspects, Gobot's focus extends to emotional communication and companionship. It features four emotional states-nervous, happy, angry, and normal-which it expresses by adjusting shoelace tightness and light color changes. Additionally, Gobot explores a new method of tactile interaction for foot-based robotics, aiming to expand the application of such technology in enhancing user experience.

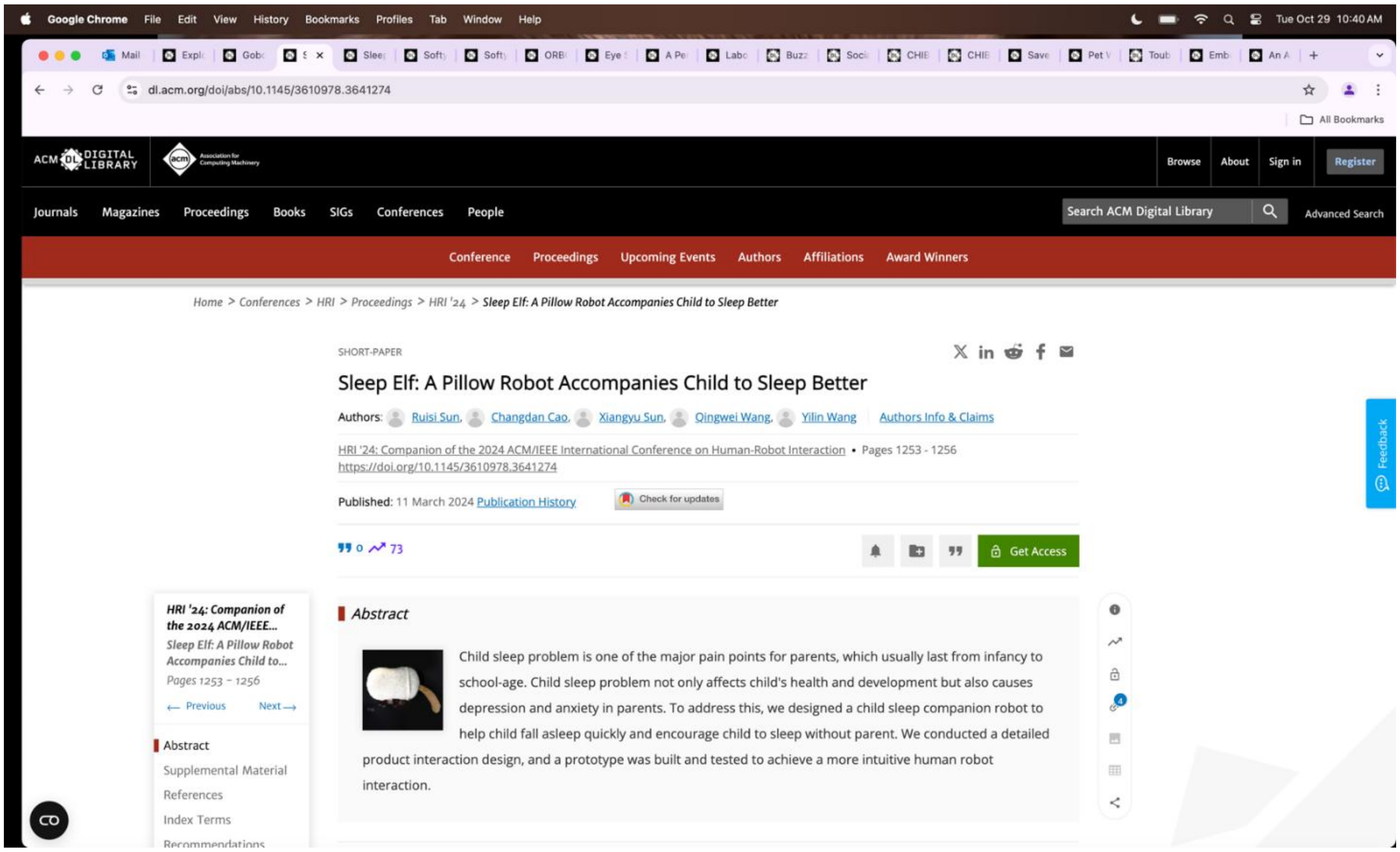
dl.acm.org/doi/abs/10.1145/3610978.3641274
Home > Conferences > HRI > Proceedings > HRI '24 > Sleep Elf: A Pillow Robot Accompanies Child to Sleep Better
SHORT-PAPER
Sleep Elf: A Pillow Robot Accompanies Child to Sleep Better
Authors: Ruisi Sun, Changdan Cao, Xiangyu Sun, Qingwei Wang, Yilin Wang
HRI '24: Companion of the 2024 ACM/IEEE International Conference on Human-Robot Interaction • Pages 1253 - 1256
https://doi.org/10.1145/3610978.3641274
Published: 11 March 2024
Abstract
Child sleep problem is one of the major pain points for parents, which usually last from infancy to school-age. Child sleep problem not only affects child's health and development but also causes depression and anxiety in parents. To address this, we designed a child sleep companion robot to help child fall asleep quickly and encourage child to sleep without parent. We conducted a detailed product interaction design, and a prototype was built and tested to achieve a more intuitive human robot interaction.

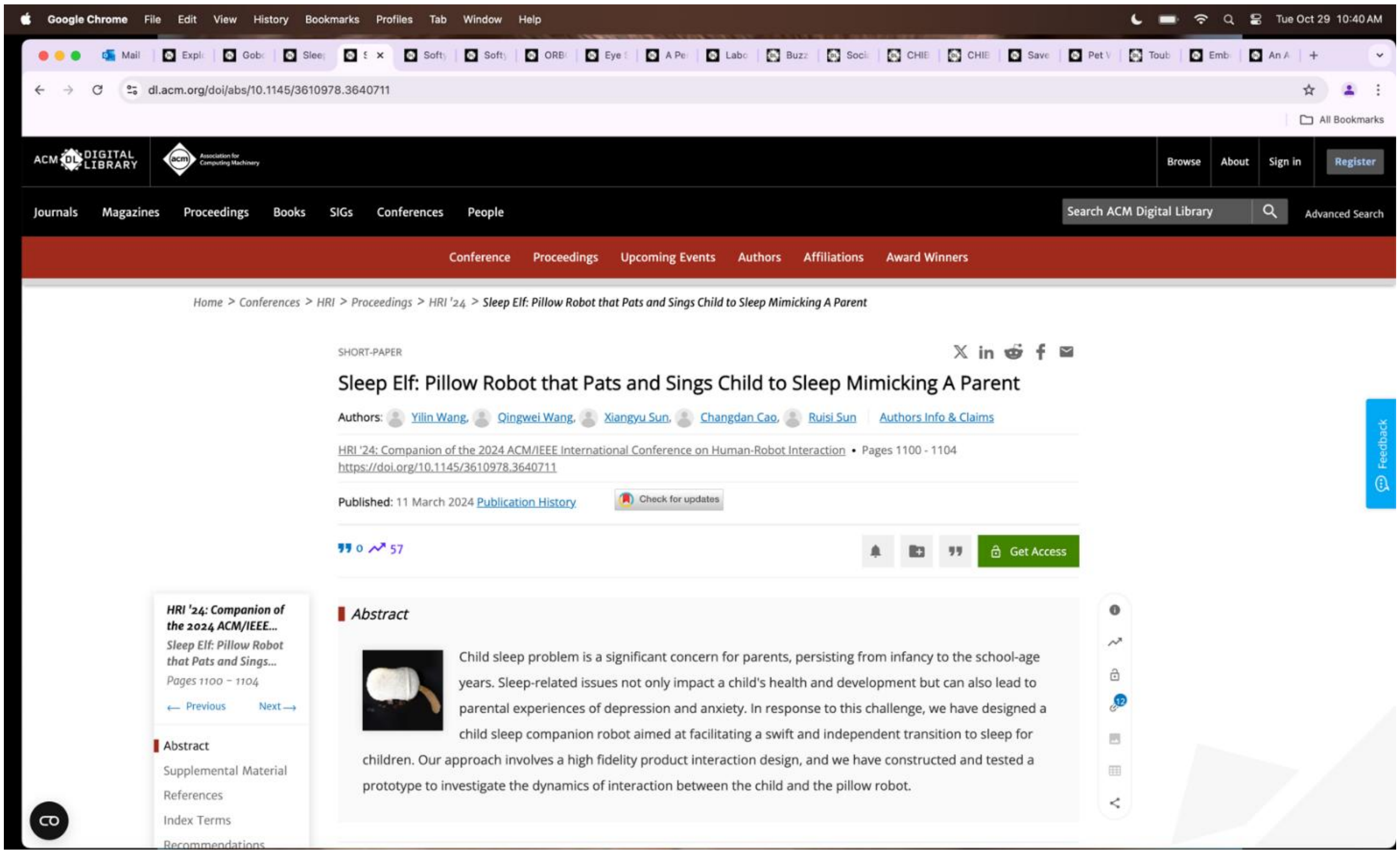
dl.acm.org/doi/abs/10.1145/3610978.3640711
Home > Conferences > HRI > Proceedings > HRI '24 > Sleep Elf: Pillow Robot that Pats and Sings Child to Sleep Mimicking A Parent
SHORT-PAPER
Sleep Elf: Pillow Robot that Pats and Sings Child to Sleep Mimicking A Parent
Authors: Yilin Wang, Qingwei Wang, Xiangyu Sun, Changdan Cao, Ruisi Sun
HRI '24: Companion of the 2024 ACM/IEEE International Conference on Human-Robot Interaction • Pages 1100 - 1104
https://doi.org/10.1145/3610978.3640711
Published: 11 March 2024
Abstract
Child sleep problem is a significant concern for parents, persisting from infancy to the school-age years. Sleep-related issues not only impact a child's health and development but can also lead to parental experiences of depression and anxiety. In response to this challenge, we have designed a child sleep companion robot aimed at facilitating a swift and independent transition to sleep for children. Our approach involves a high fidelity product interaction design, and we have constructed and tested a prototype to investigate the dynamics of interaction between the child and the pillow robot.

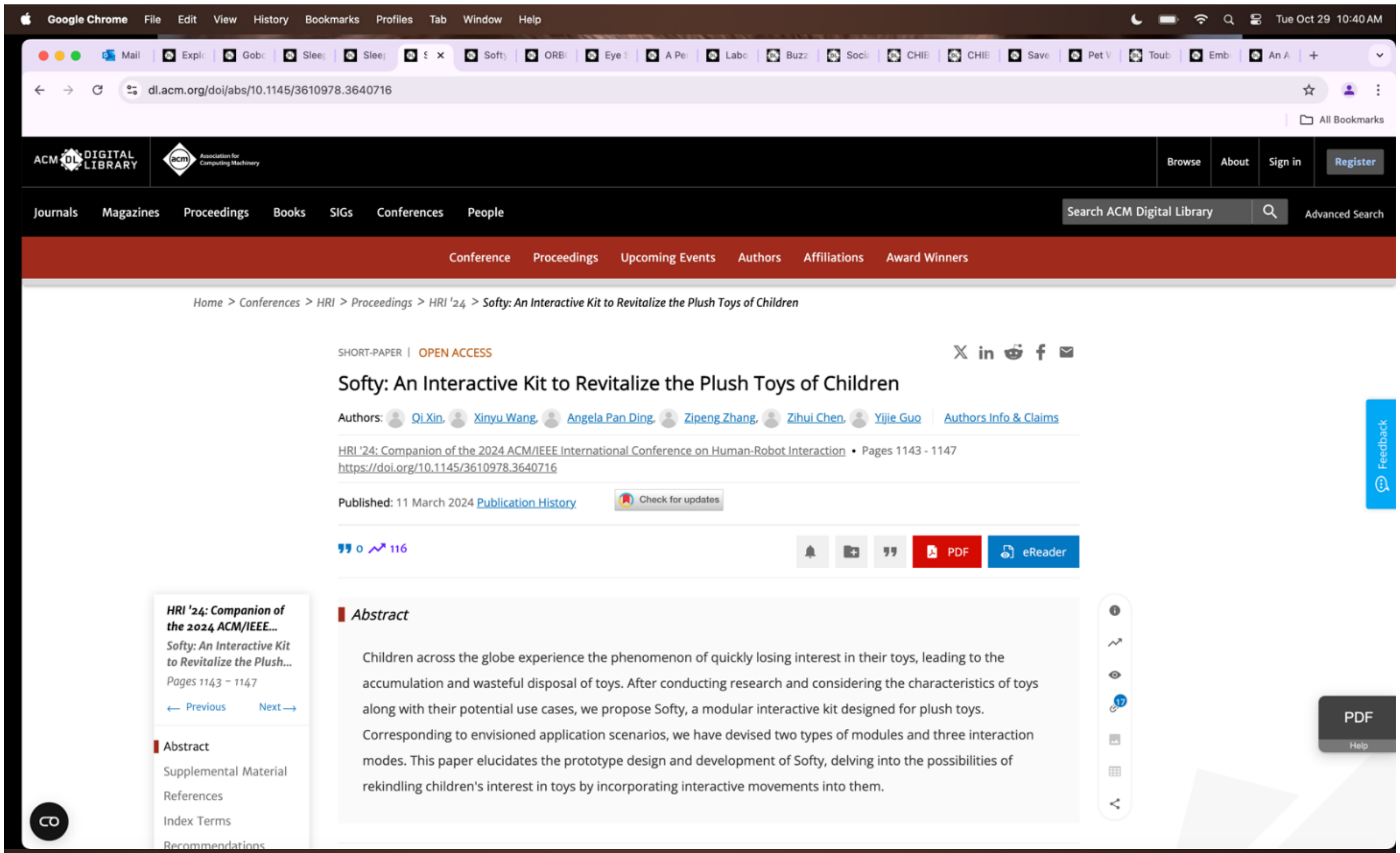
dl.acm.org/doi/abs/10.1145/3610978.3640716
Home > Conferences > HRI > Proceedings > HRI '24 > Softy: An Interactive Kit to Revitalize the Plush Toys of Children
SHORT-PAPER | OPEN ACCESS
Softy: An Interactive Kit to Revitalize the Plush Toys of Children
Authors: Qi Xin, Xinyu Wang, Angela Pan Ding, Zipeng Zhang, Zihui Chen, Yijie Guo Authors Info & Claims
HRI '24: Companion of the 2024 ACM/IEEE International Conference on Human-Robot Interaction • Pages 1143 - 1147
https://doi.org/10.1145/3610978.3640716
Published: 11 March 2024 Publication History
Abstract
Children across the globe experience the phenomenon of quickly losing interest in their toys, leading to the accumulation and wasteful disposal of toys. After conducting research and considering the characteristics of toys along with their potential use cases, we propose Softy, a modular interactive kit designed for plush toys. Corresponding to envisioned application scenarios, we have devised two types of modules and three interaction modes. This paper elucidates the prototype design and development of Softy, delving into the possibilities of rekindling children's interest in toys by incorporating interactive movements into them.

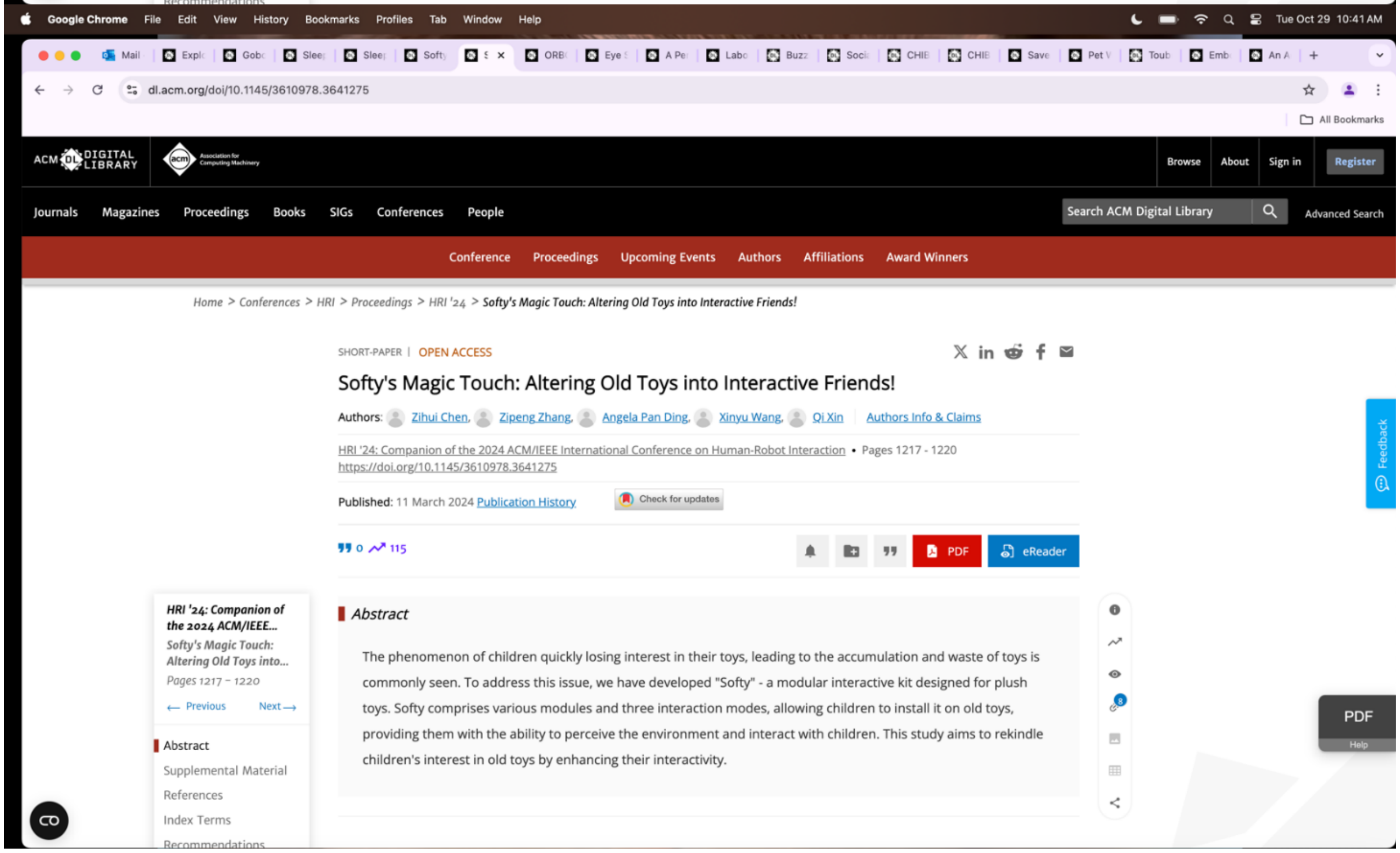
dl.acm.org/doi/10.1145/3610978.3641275
Home > Conferences > HRI > Proceedings > HRI '24 > Softy's Magic Touch: Altering Old Toys into Interactive Friends!
SHORT-PAPER | OPEN ACCESS
Softy's Magic Touch: Altering Old Toys into Interactive Friends!
Authors: Zihui Chen, Zipeng Zhang, Angela Pan Ding, Xinyu Wang, Qi Xin Authors Info & Claims
HRI '24: Companion of the 2024 ACM/IEEE International Conference on Human-Robot Interaction • Pages 1217 - 1220
https://doi.org/10.1145/3610978.3641275
Published: 11 March 2024 Publication History
Abstract
The phenomenon of children quickly losing interest in their toys, leading to the accumulation and waste of toys is commonly seen. To address this issue, we have developed "Softy" - a modular interactive kit designed for plush toys. Softy comprises various modules and three interaction modes, allowing children to install it on old toys, providing them with the ability to perceive the environment and interact with children. This study aims to rekindle children's interest in old toys by enhancing their interactivity.

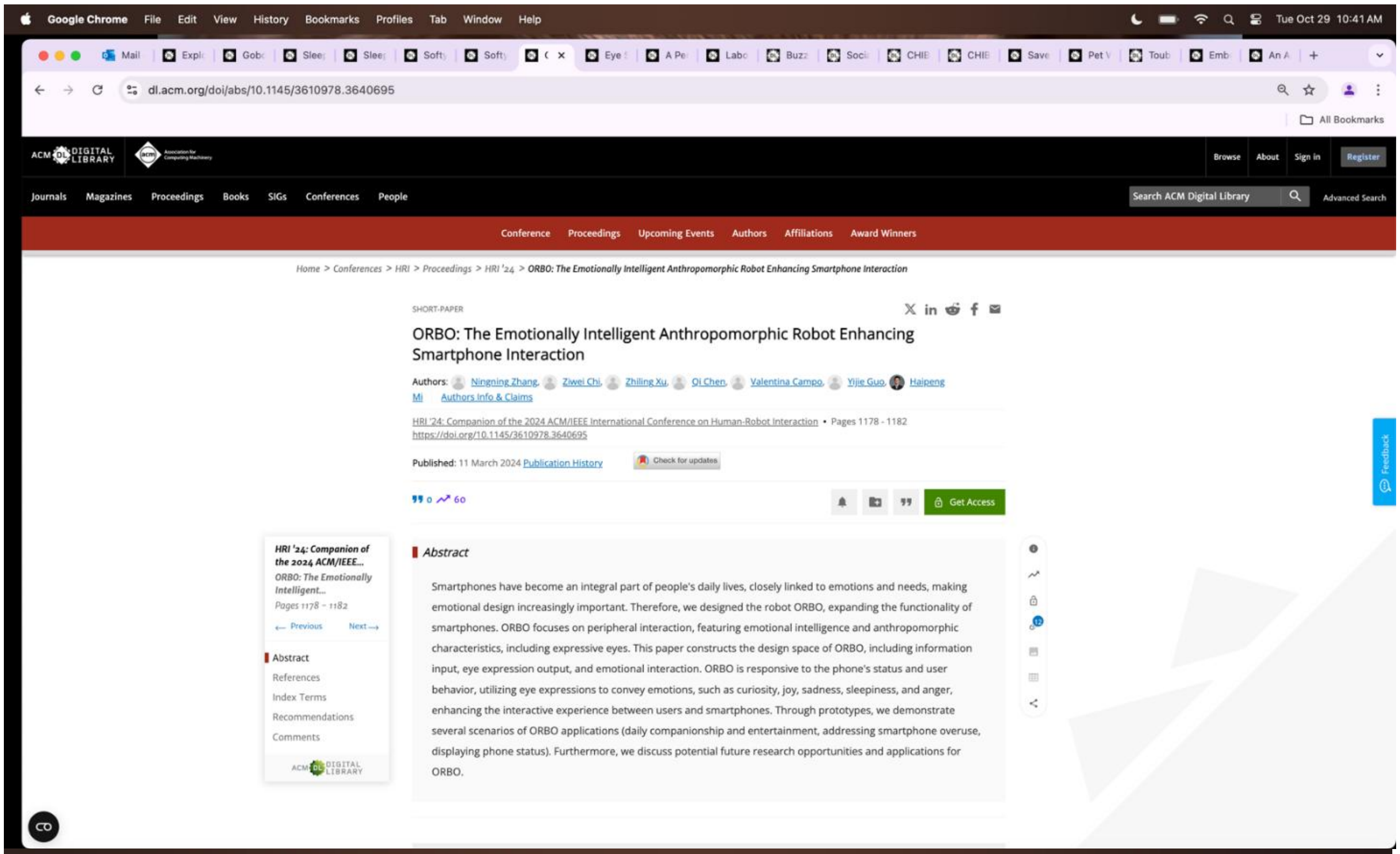
dl.acm.org/doi/abs/10.1145/3610978.3640695
Home > Conferences > HRI > Proceedings > HRI '24 > ORBO: The Emotionally Intelligent Anthropomorphic Robot Enhancing Smartphone Interaction
SHORT-PAPER
ORBO: The Emotionally Intelligent Anthropomorphic Robot Enhancing Smartphone Interaction
Authors: Ningning Zhang, Ziwei Chi, Zhiling Xu, Qi Chen, Valentina Campo, Yijie Guo, Haipeng Mi Authors Info & Claims
HRI '24: Companion of the 2024 ACM/IEEE International Conference on Human-Robot Interaction • Pages 1178 - 1182
https://doi.org/10.1145/3610978.3640695
Published: 11 March 2024 Publication History
Abstract
Smartphones have become an integral part of people's daily lives, closely linked to emotions and needs, making emotional design increasingly important. Therefore, we designed the robot ORBO, expanding the functionality of smartphones. ORBO focuses on peripheral interaction, featuring emotional intelligence and anthropomorphic characteristics, including expressive eyes. This paper constructs the design space of ORBO, including information input, eye expression output, and emotional interaction. ORBO is responsive to the phone's status and user behavior, utilizing eye expressions to convey emotions, such as curiosity, joy, sadness, sleepiness, and anger, enhancing the interactive experience between users and smartphones. Through prototypes, we demonstrate several scenarios of ORBO applications (daily companionship and entertainment, addressing smartphone overuse, displaying phone status). Furthermore, we discuss potential future research opportunities and applications for ORBO.

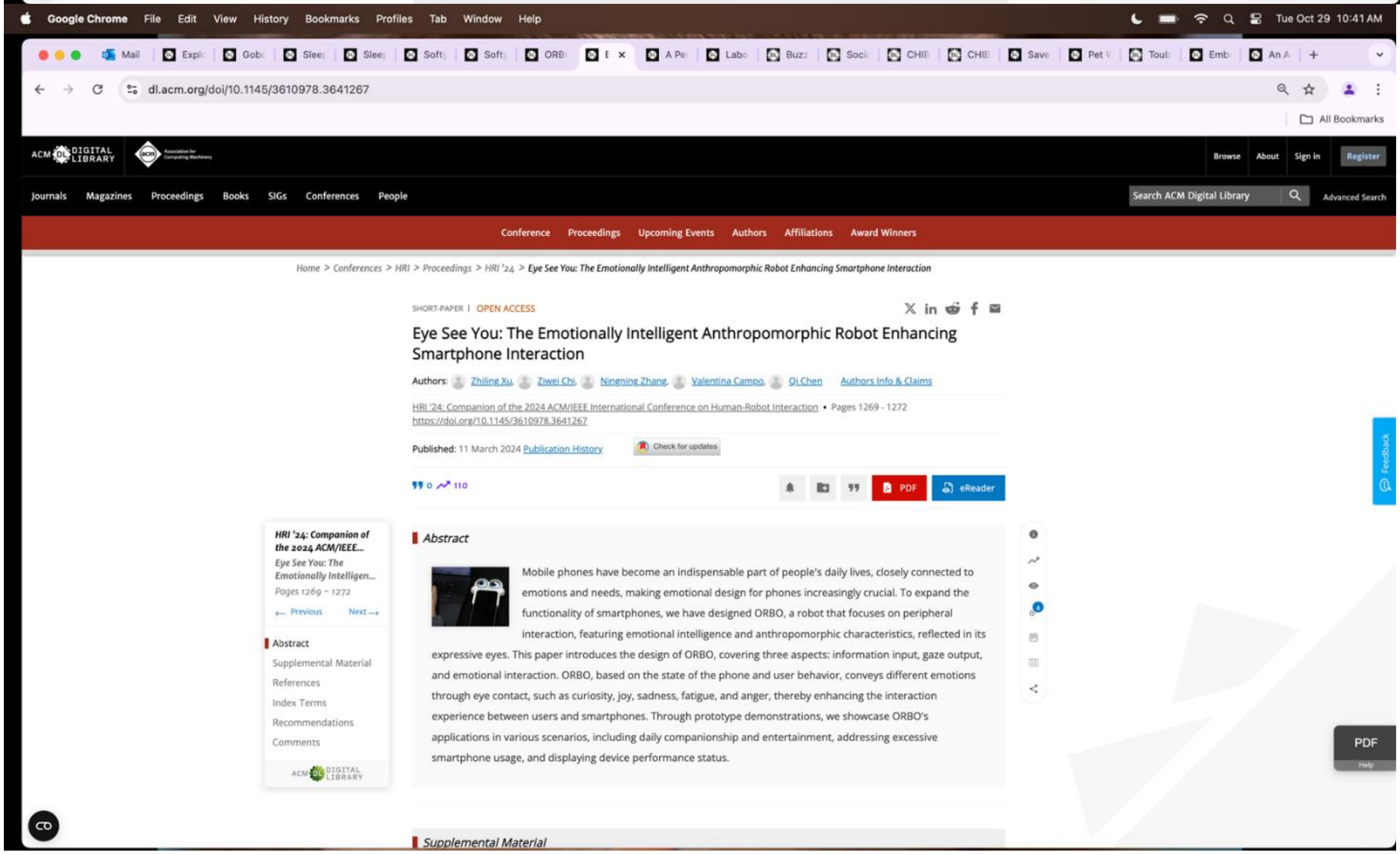
dl.acm.org/doi/10.1145/3610978.3641267
Home > Conferences > HRI > Proceedings > HRI '24 > Eye See You: The Emotionally Intelligent Anthropomorphic Robot Enhancing Smartphone Interaction
SHORT-PAPER | OPEN ACCESS
Eye See You: The Emotionally Intelligent Anthropomorphic Robot Enhancing Smartphone Interaction
Authors: Zhiling Xu, Ziwei Chi, Ningning Zhang, Valentina Campo, Qi Chen Authors Info & Claims
HRI '24: Companion of the 2024 ACM/IEEE International Conference on Human-Robot Interaction • Pages 1269 - 1272
https://doi.org/10.1145/3610978.3641267
Published: 11 March 2024 Publication History
Abstract
Mobile phones have become an indispensable part of people's daily lives, closely connected to emotions and needs, making emotional design for phones increasingly crucial. To expand the functionality of smartphones, we have designed ORBO, a robot that focuses on peripheral interaction, featuring emotional intelligence and anthropomorphic characteristics, reflected in its expressive eyes. This paper introduces the design of ORBO, covering three aspects: information input, gaze output, and emotional interaction. ORBO, based on the state of the phone and user behavior, conveys different emotions through eye contact, such as curiosity, joy, sadness, fatigue, and anger, thereby enhancing the interaction experience between users and smartphones. Through prototype demonstrations, we showcase ORBO's applications in various scenarios, including daily companionship and entertainment, addressing excessive smartphone usage, and displaying device performance status.
Supplemental Material

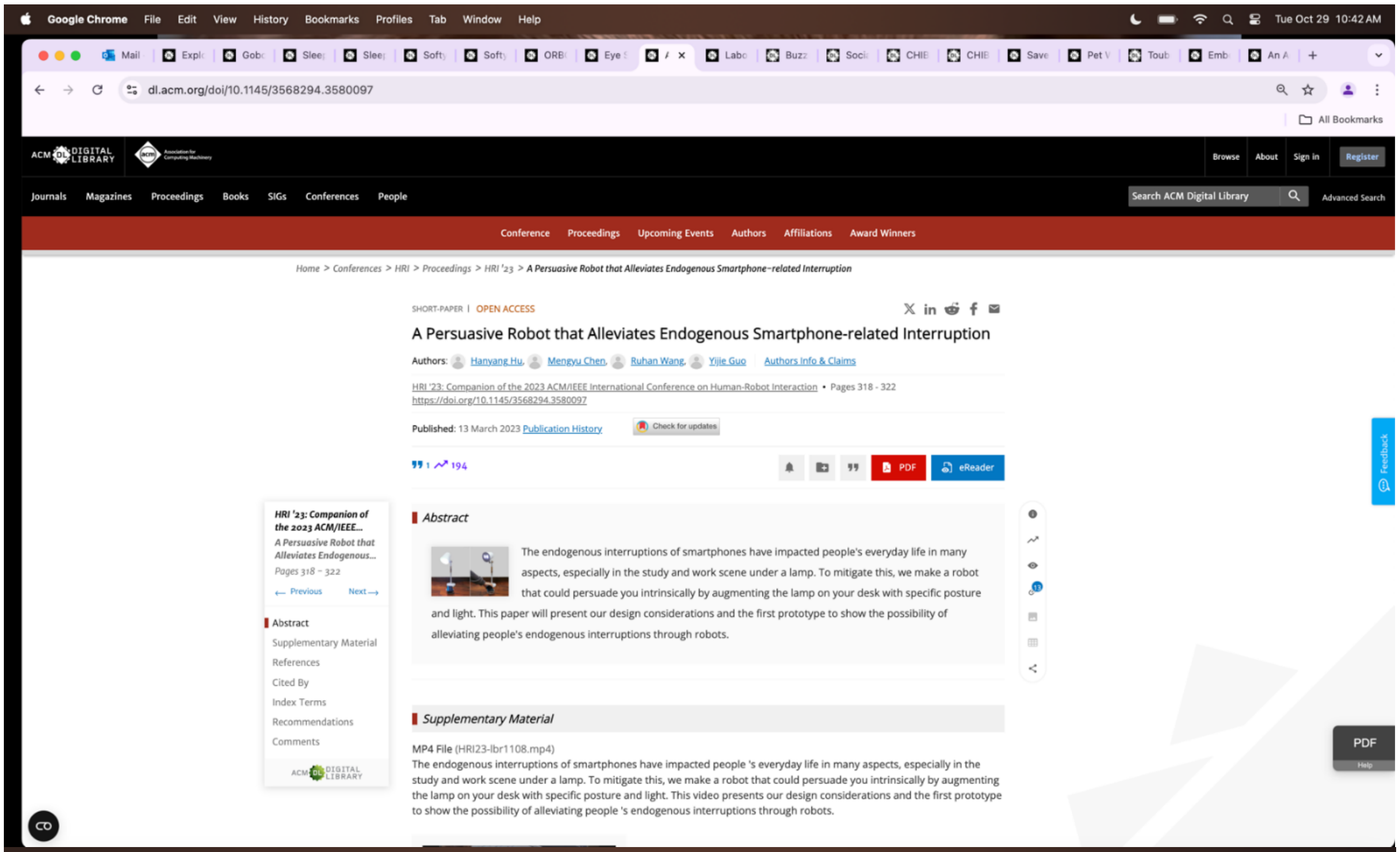

dl.acm.org/doi/10.1145/3568294.3580097

Home > Conferences > HRI > Proceedings > HRI '23 > A Persuasive Robot that Alleviates Endogenous Smartphone-related Interruption

SHORT-PAPER | OPEN ACCESS

## A Persuasive Robot that Alleviates Endogenous Smartphone-related Interruption

Authors: Hanyang Hu, Mengyu Chen, Ruhan Wang, Yijie Guo   Authors Info & Claims

HRI '23: Companion of the 2023 ACM/IEEE International Conference on Human-Robot Interaction • Pages 318 - 322
https://doi.org/10.1145/3568294.3580097

Published: 13 March 2023 Publication History

1   194

### Abstract

The endogenous interruptions of smartphones have impacted people's everyday life in many aspects, especially in the study and work scene under a lamp. To mitigate this, we make a robot that could persuade you intrinsically by augmenting the lamp on your desk with specific posture and light. This paper will present our design considerations and the first prototype to show the possibility of alleviating people's endogenous interruptions through robots.

### Supplementary Material

MP4 File (HRI23-lbr1108.mp4)
The endogenous interruptions of smartphones have impacted people 's everyday life in many aspects, especially in the study and work scene under a lamp. To mitigate this, we make a robot that could persuade you intrinsically by augmenting the lamp on your desk with specific posture and light. This video presents our design considerations and the first prototype to show the possibility of alleviating people 's endogenous interruptions through robots.

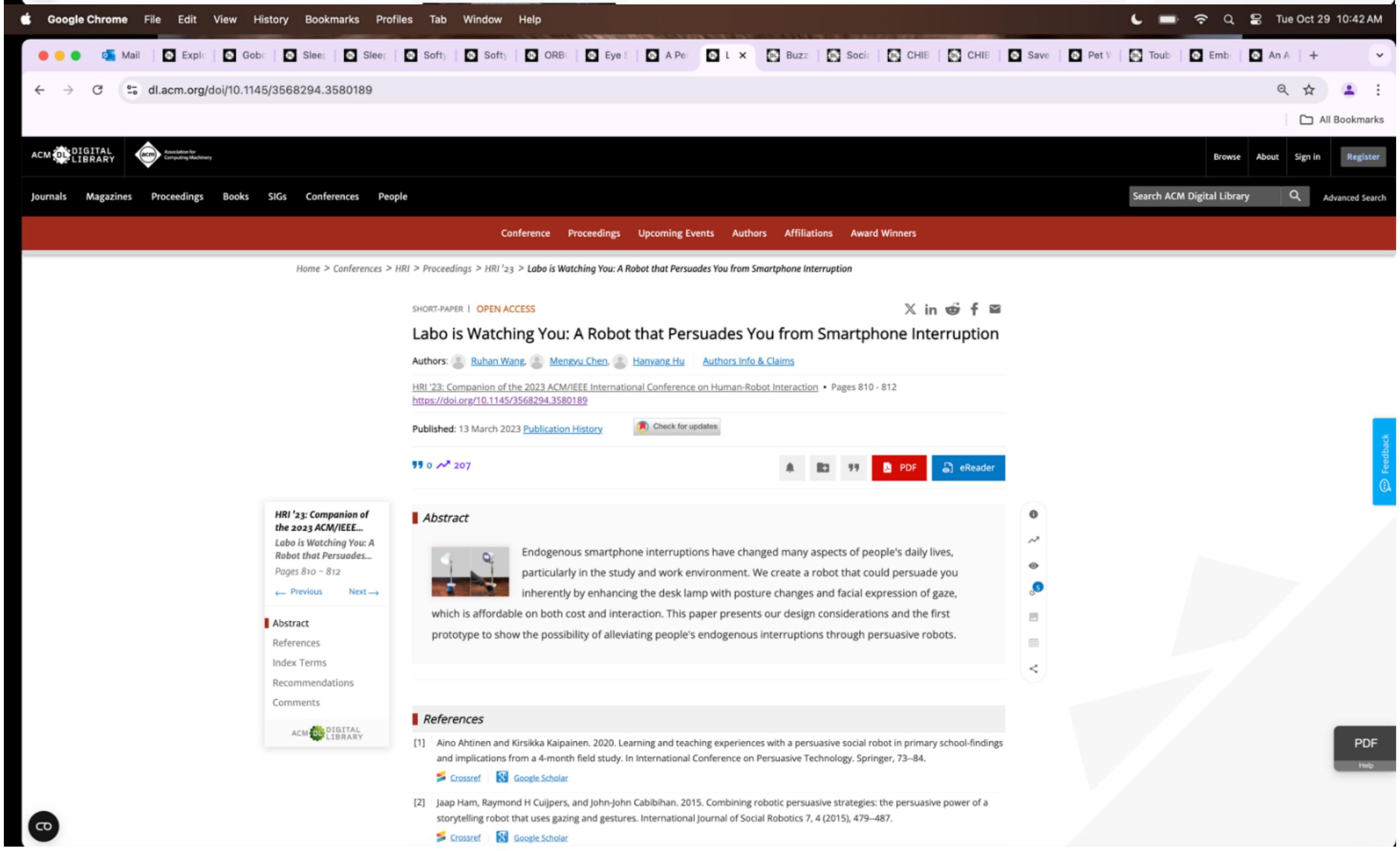

dl.acm.org/doi/10.1145/3568294.3580189

Home > Conferences > HRI > Proceedings > HRI '23 > Labo is Watching You: A Robot that Persuades You from Smartphone Interruption

SHORT-PAPER | OPEN ACCESS

## Labo is Watching You: A Robot that Persuades You from Smartphone Interruption

Authors: Ruhan Wang, Mengyu Chen, Hanyang Hu   Authors Info & Claims

HRI '23: Companion of the 2023 ACM/IEEE International Conference on Human-Robot Interaction • Pages 810 - 812
https://doi.org/10.1145/3568294.3580189

Published: 13 March 2023 Publication History

0   207

### Abstract

Endogenous smartphone interruptions have changed many aspects of people's daily lives, particularly in the study and work environment. We create a robot that could persuade you inherently by enhancing the desk lamp with posture changes and facial expression of gaze, which is affordable on both cost and interaction. This paper presents our design considerations and the first prototype to show the possibility of alleviating people's endogenous interruptions through persuasive robots.

### References

[1] Aino Ahtinen and Kirsikka Kaipainen. 2020. Learning and teaching experiences with a persuasive social robot in primary school-findings and implications from a 4-month field study. In International Conference on Persuasive Technology. Springer, 73–84.

[2] Jaap Ham, Raymond H Cuijpers, and John-John Cabibihan. 2015. Combining robotic persuasive strategies: the persuasive power of a storytelling robot that uses gazing and gestures. International Journal of Social Robotics 7, 4 (2015), 479–487.

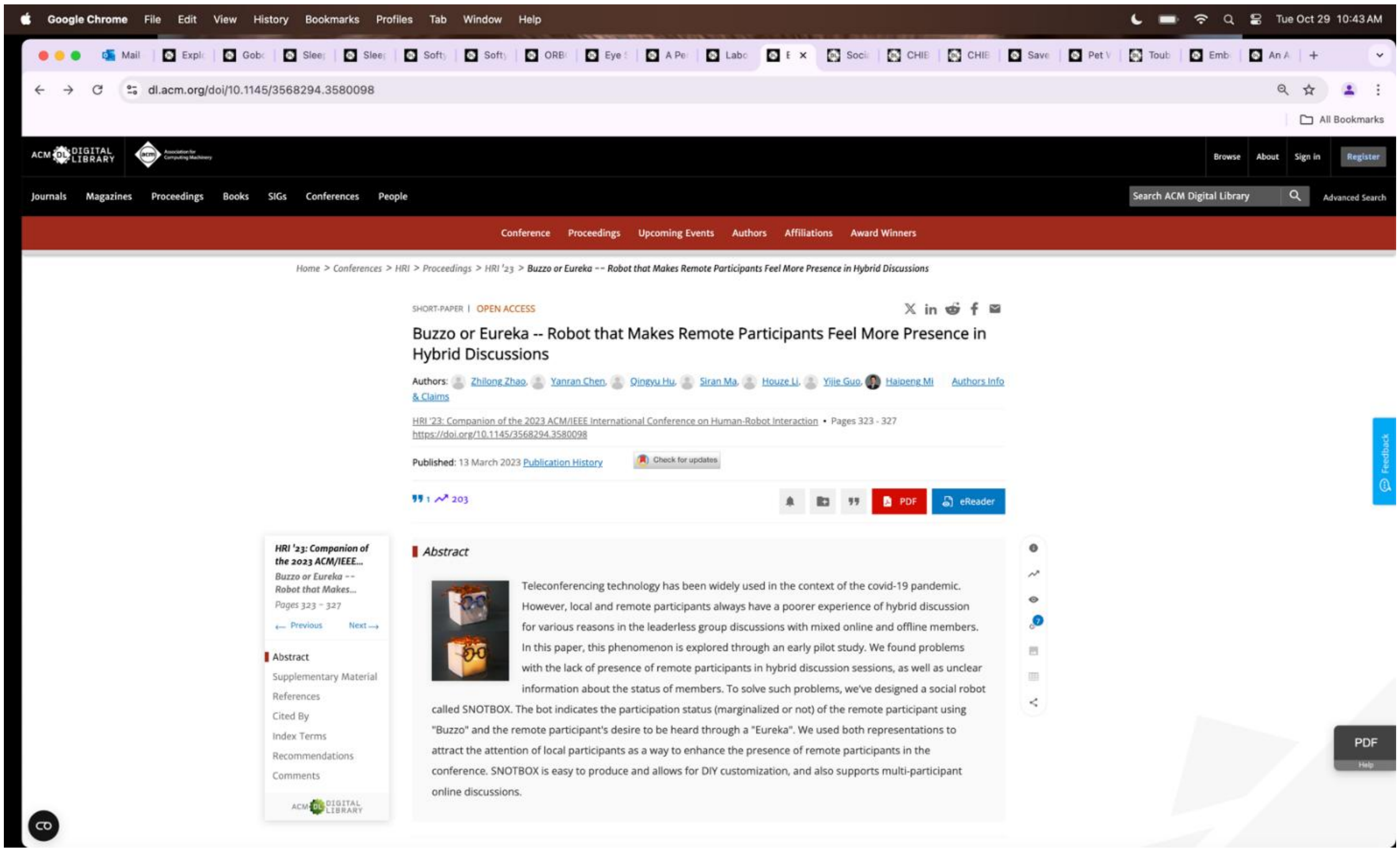

Home > Conferences > HRI > Proceedings > HRI '23 > Buzzo or Eureka -- Robot that Makes Remote Participants Feel More Presence in Hybrid Discussions

SHORT-PAPER | OPEN ACCESS

Buzzo or Eureka -- Robot that Makes Remote Participants Feel More Presence in Hybrid Discussions

Authors: Zhilong Zhao, Yanran Chen, Qingyu Hu, Siran Ma, Houze Li, Yijie Guo, Haipeng Mi  Authors Info & Claims

HRI '23: Companion of the 2023 ACM/IEEE International Conference on Human-Robot Interaction • Pages 323 - 327
https://doi.org/10.1145/3568294.3580098

Published: 13 March 2023 Publication History

Abstract

Teleconferencing technology has been widely used in the context of the covid-19 pandemic. However, local and remote participants always have a poorer experience of hybrid discussion for various reasons in the leaderless group discussions with mixed online and offline members. In this paper, this phenomenon is explored through an early pilot study. We found problems with the lack of presence of remote participants in hybrid discussion sessions, as well as unclear information about the status of members. To solve such problems, we've designed a social robot called SNOTBOX. The bot indicates the participation status (marginalized or not) of the remote participant using "Buzzo" and the remote participant's desire to be heard through a "Eureka". We used both representations to attract the attention of local participants as a way to enhance the presence of remote participants in the conference. SNOTBOX is easy to produce and allows for DIY customization, and also supports multi-participant online discussions.

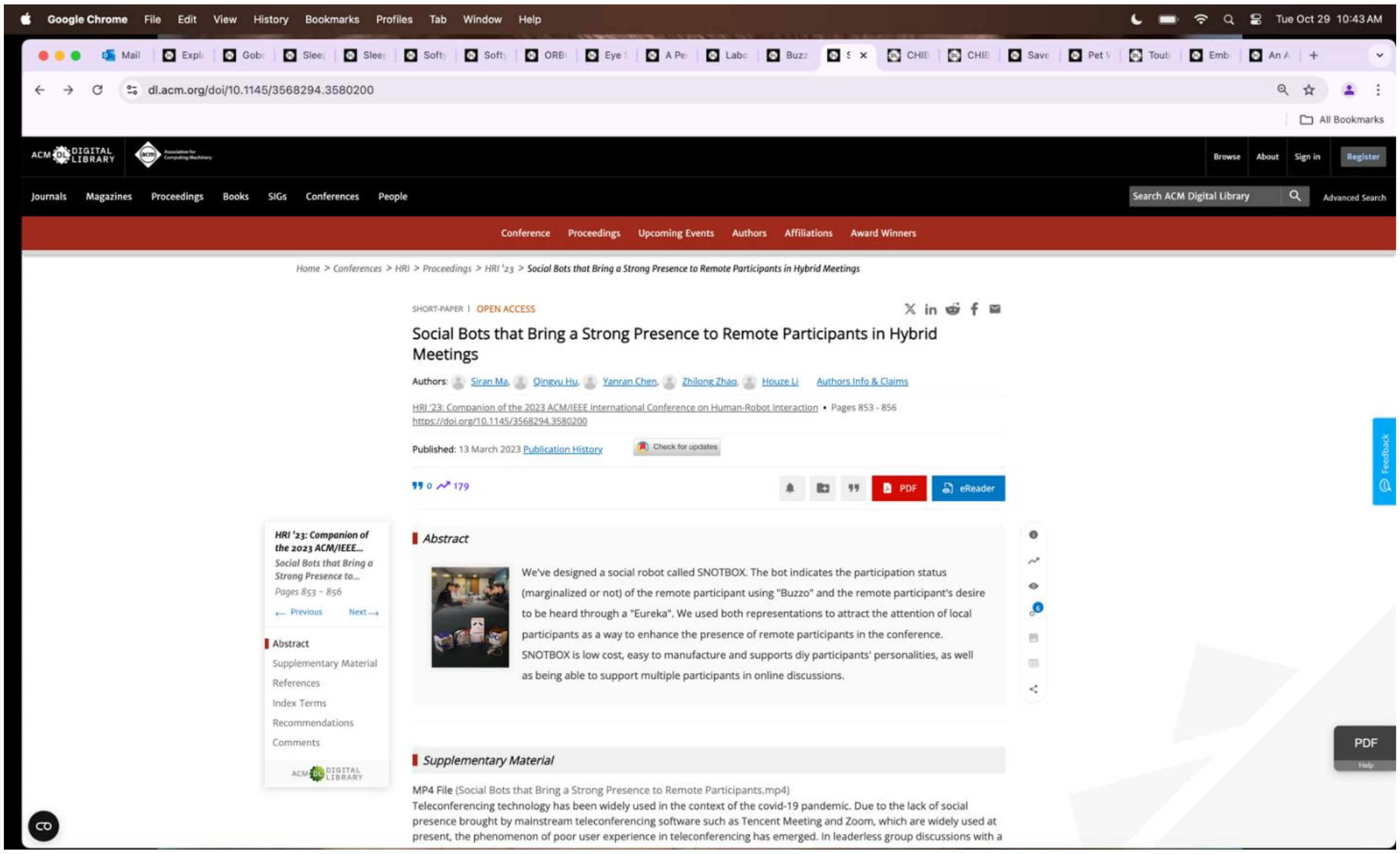

Home > Conferences > HRI > Proceedings > HRI '23 > Social Bots that Bring a Strong Presence to Remote Participants in Hybrid Meetings

SHORT-PAPER | OPEN ACCESS

Social Bots that Bring a Strong Presence to Remote Participants in Hybrid Meetings

Authors: Siran Ma, Qingyu Hu, Yanran Chen, Zhilong Zhao, Houze Li  Authors Info & Claims

HRI '23: Companion of the 2023 ACM/IEEE International Conference on Human-Robot Interaction • Pages 853 - 856
https://doi.org/10.1145/3568294.3580200

Published: 13 March 2023 Publication History

Abstract

We've designed a social robot called SNOTBOX. The bot indicates the participation status (marginalized or not) of the remote participant using "Buzzo" and the remote participant's desire to be heard through a "Eureka". We used both representations to attract the attention of local participants as a way to enhance the presence of remote participants in the conference. SNOTBOX is low cost, easy to manufacture and supports diy participants' personalities, as well as being able to support multiple participants in online discussions.

Supplementary Material

MP4 File (Social Bots that Bring a Strong Presence to Remote Participants.mp4)
Teleconferencing technology has been widely used in the context of the covid-19 pandemic. Due to the lack of social presence brought by mainstream teleconferencing software such as Tencent Meeting and Zoom, which are widely used at present, the phenomenon of poor user experience in teleconferencing has emerged. In leaderless group discussions with a

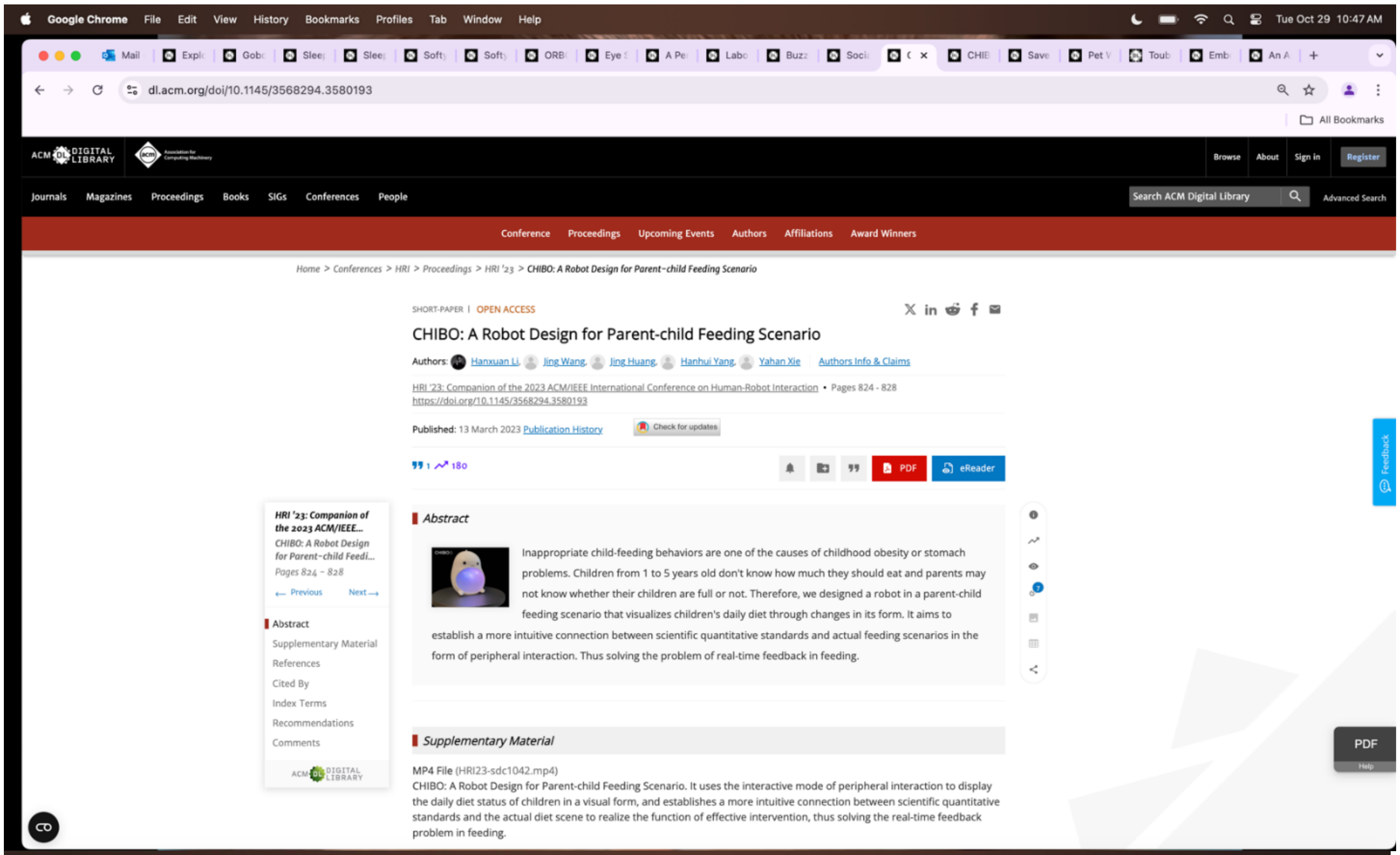
dl.acm.org/doi/10.1145/3568294.3580193
Home > Conferences > HRI > Proceedings > HRI '23 > CHIBO: A Robot Design for Parent-child Feeding Scenario
SHORT-PAPER | OPEN ACCESS
CHIBO: A Robot Design for Parent-child Feeding Scenario
Authors: Hanxuan Li, Jing Wang, Jing Huang, Hanhui Yang, Yahan Xie
HRI '23: Companion of the 2023 ACM/IEEE International Conference on Human-Robot Interaction • Pages 824 - 828
https://doi.org/10.1145/3568294.3580193
Published: 13 March 2023
Abstract
Inappropriate child-feeding behaviors are one of the causes of childhood obesity or stomach problems. Children from 1 to 5 years old don't know how much they should eat and parents may not know whether their children are full or not. Therefore, we designed a robot in a parent-child feeding scenario that visualizes children's daily diet through changes in its form. It aims to establish a more intuitive connection between scientific quantitative standards and actual feeding scenarios in the form of peripheral interaction. Thus solving the problem of real-time feedback in feeding.
Supplementary Material
MP4 File (HRI23-sdc1042.mp4)
CHIBO: A Robot Design for Parent-child Feeding Scenario. It uses the interactive mode of peripheral interaction to display the daily diet status of children in a visual form, and establishes a more intuitive connection between scientific quantitative standards and the actual diet scene to realize the function of effective intervention, thus solving the real-time feedback problem in feeding.

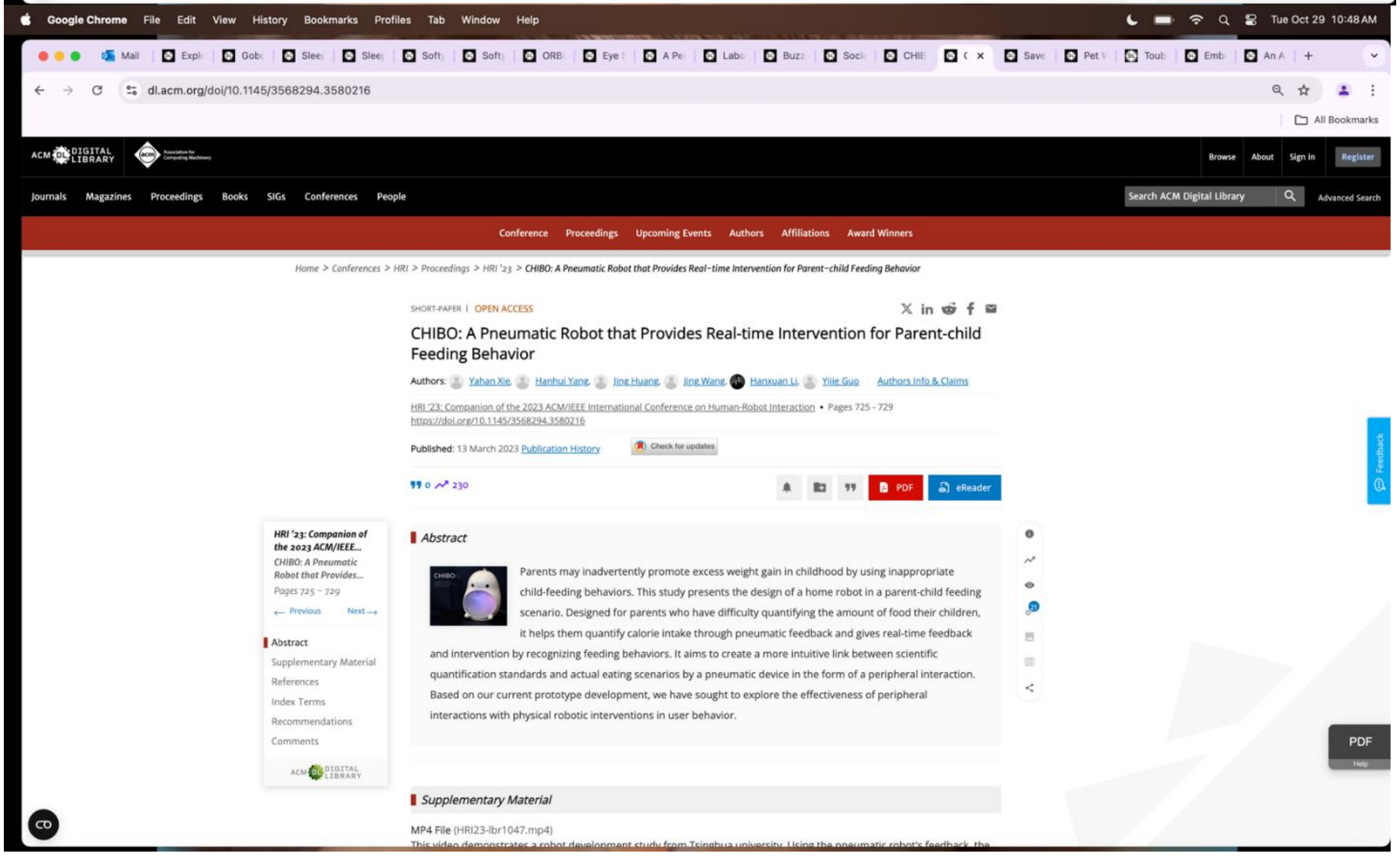
dl.acm.org/doi/10.1145/3568294.3580216
Home > Conferences > HRI > Proceedings > HRI '23 > CHIBO: A Pneumatic Robot that Provides Real-time Intervention for Parent-child Feeding Behavior
SHORT-PAPER | OPEN ACCESS
CHIBO: A Pneumatic Robot that Provides Real-time Intervention for Parent-child Feeding Behavior
Authors: Yahan Xie, Hanhui Yang, Jing Huang, Jing Wang, Hanxuan Li, Yijie Guo
HRI '23: Companion of the 2023 ACM/IEEE International Conference on Human-Robot Interaction • Pages 725 - 729
https://doi.org/10.1145/3568294.3580216
Published: 13 March 2023
Abstract
Parents may inadvertently promote excess weight gain in childhood by using inappropriate child-feeding behaviors. This study presents the design of a home robot in a parent-child feeding scenario. Designed for parents who have difficulty quantifying the amount of food their children, it helps them quantify calorie intake through pneumatic feedback and gives real-time feedback and intervention by recognizing feeding behaviors. It aims to create a more intuitive link between scientific quantification standards and actual eating scenarios by a pneumatic device in the form of a peripheral interaction. Based on our current prototype development, we have sought to explore the effectiveness of peripheral interactions with physical robotic interventions in user behavior.
Supplementary Material
MP4 File (HRI23-lbr1047.mp4)

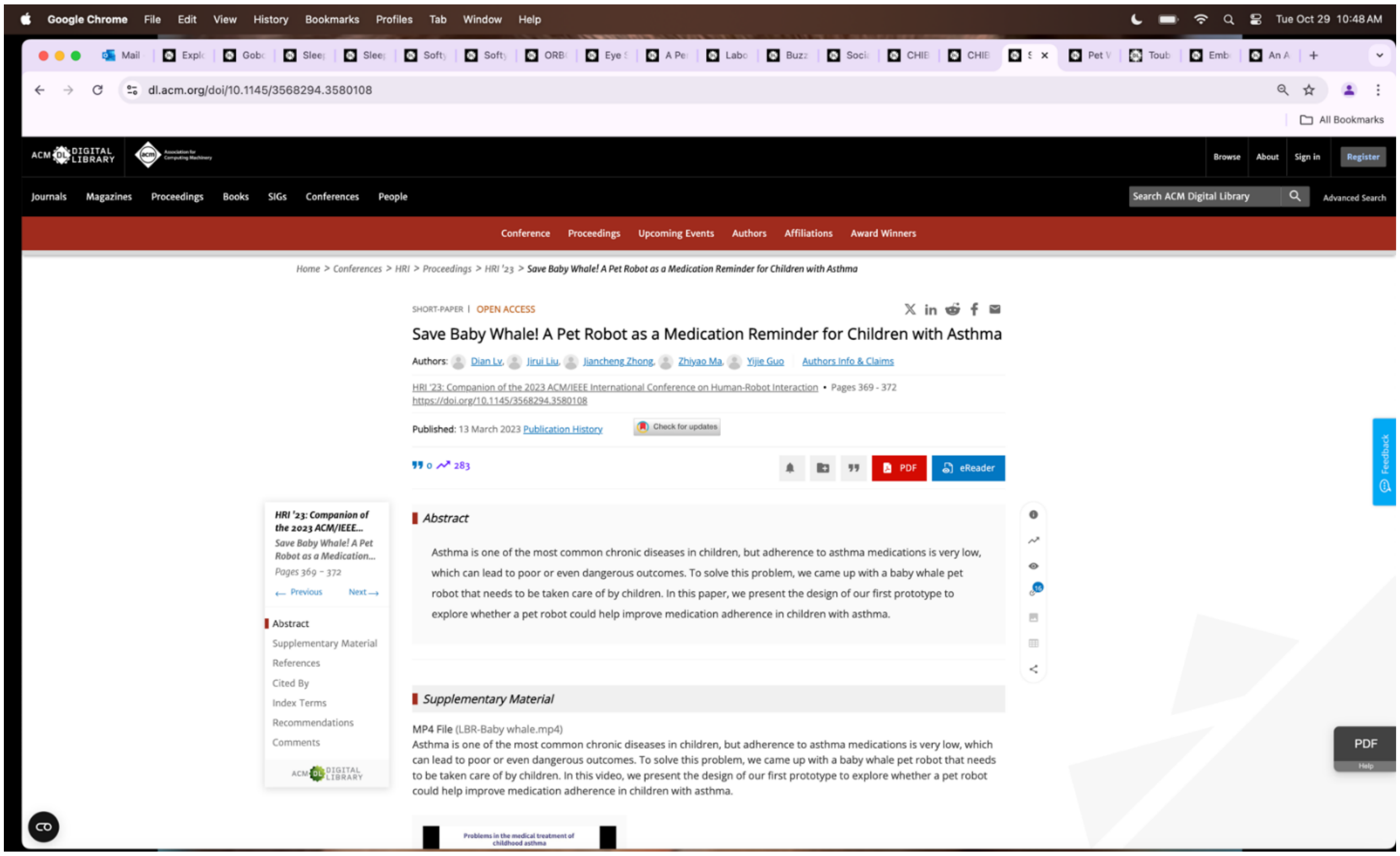
dl.acm.org/doi/10.1145/3568294.3580108
Home > Conferences > HRI > Proceedings > HRI '23 > Save Baby Whale! A Pet Robot as a Medication Reminder for Children with Asthma
SHORT-PAPER | OPEN ACCESS
Save Baby Whale! A Pet Robot as a Medication Reminder for Children with Asthma
Authors: Dian Lv, Jirui Liu, Jiancheng Zhong, Zhiyao Ma, Yijie Guo
HRI '23: Companion of the 2023 ACM/IEEE International Conference on Human-Robot Interaction • Pages 369 - 372
https://doi.org/10.1145/3568294.3580108
Published: 13 March 2023 Publication History
Abstract
Asthma is one of the most common chronic diseases in children, but adherence to asthma medications is very low, which can lead to poor or even dangerous outcomes. To solve this problem, we came up with a baby whale pet robot that needs to be taken care of by children. In this paper, we present the design of our first prototype to explore whether a pet robot could help improve medication adherence in children with asthma.
Supplementary Material
MP4 File (LBR-Baby whale.mp4)
Asthma is one of the most common chronic diseases in children, but adherence to asthma medications is very low, which can lead to poor or even dangerous outcomes. To solve this problem, we came up with a baby whale pet robot that needs to be taken care of by children. In this video, we present the design of our first prototype to explore whether a pet robot could help improve medication adherence in children with asthma.
Tue Oct 29 10:48 AM

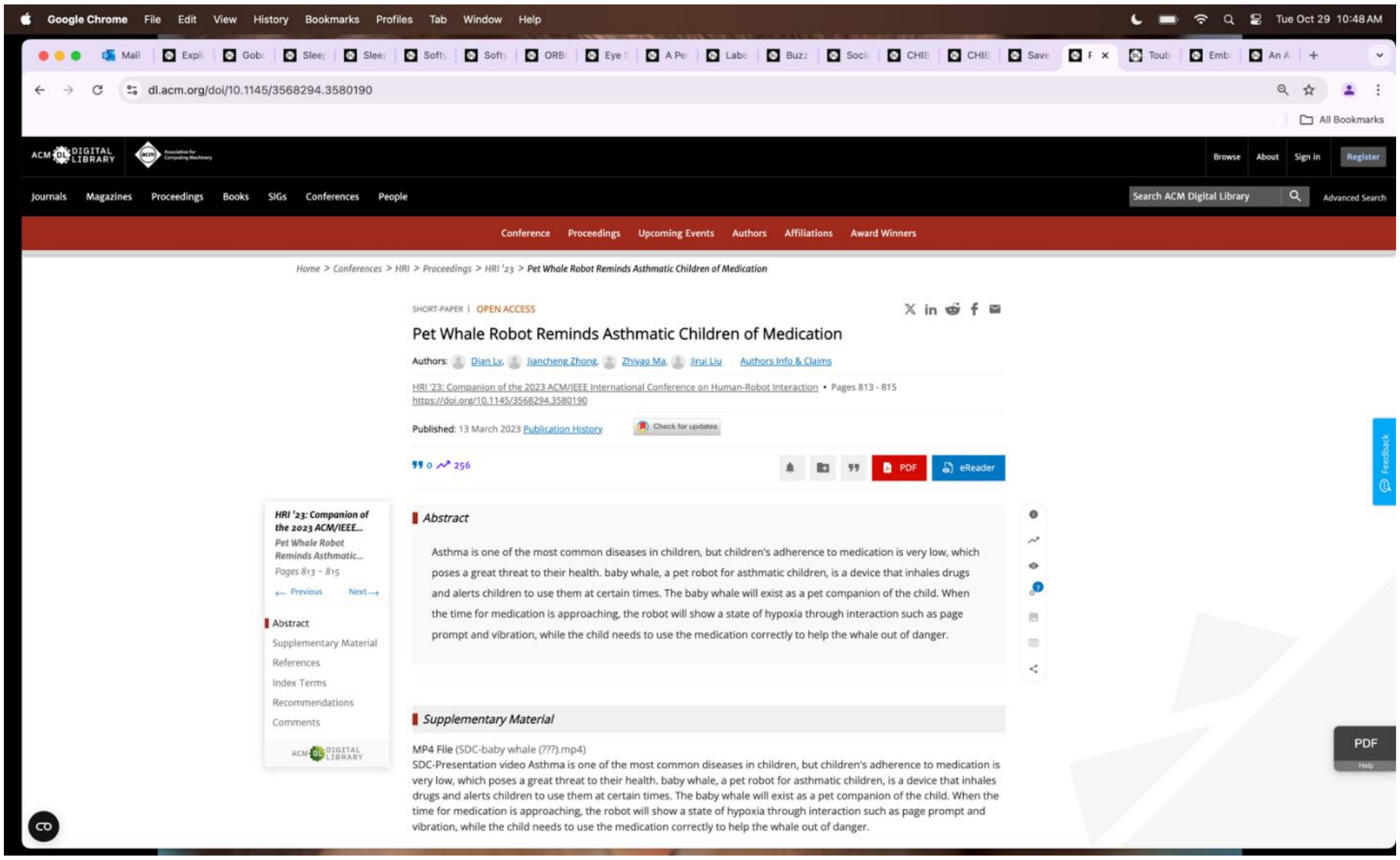
dl.acm.org/doi/10.1145/3568294.3580190
Home > Conferences > HRI > Proceedings > HRI '23 > Pet Whale Robot Reminds Asthmatic Children of Medication
SHORT-PAPER | OPEN ACCESS
Pet Whale Robot Reminds Asthmatic Children of Medication
Authors: Dian Lv, Jiancheng Zhong, Zhiyao Ma, Jirui Liu
HRI '23: Companion of the 2023 ACM/IEEE International Conference on Human-Robot Interaction • Pages 813 - 815
https://doi.org/10.1145/3568294.3580190
Published: 13 March 2023 Publication History
Abstract
Asthma is one of the most common diseases in children, but children's adherence to medication is very low, which poses a great threat to their health. baby whale, a pet robot for asthmatic children, is a device that inhales drugs and alerts children to use them at certain times. The baby whale will exist as a pet companion of the child. When the time for medication is approaching, the robot will show a state of hypoxia through interaction such as page prompt and vibration, while the child needs to use the medication correctly to help the whale out of danger.
Supplementary Material
MP4 File (SDC-baby whale (???).mp4)
SDC-Presentation video Asthma is one of the most common diseases in children, but children's adherence to medication is very low, which poses a great threat to their health. baby whale, a pet robot for asthmatic children, is a device that inhales drugs and alerts children to use them at certain times. The baby whale will exist as a pet companion of the child. When the time for medication is approaching, the robot will show a state of hypoxia through interaction such as page prompt and vibration, while the child needs to use the medication correctly to help the whale out of danger.
Tue Oct 29 10:48 AM

**Appendix D.**

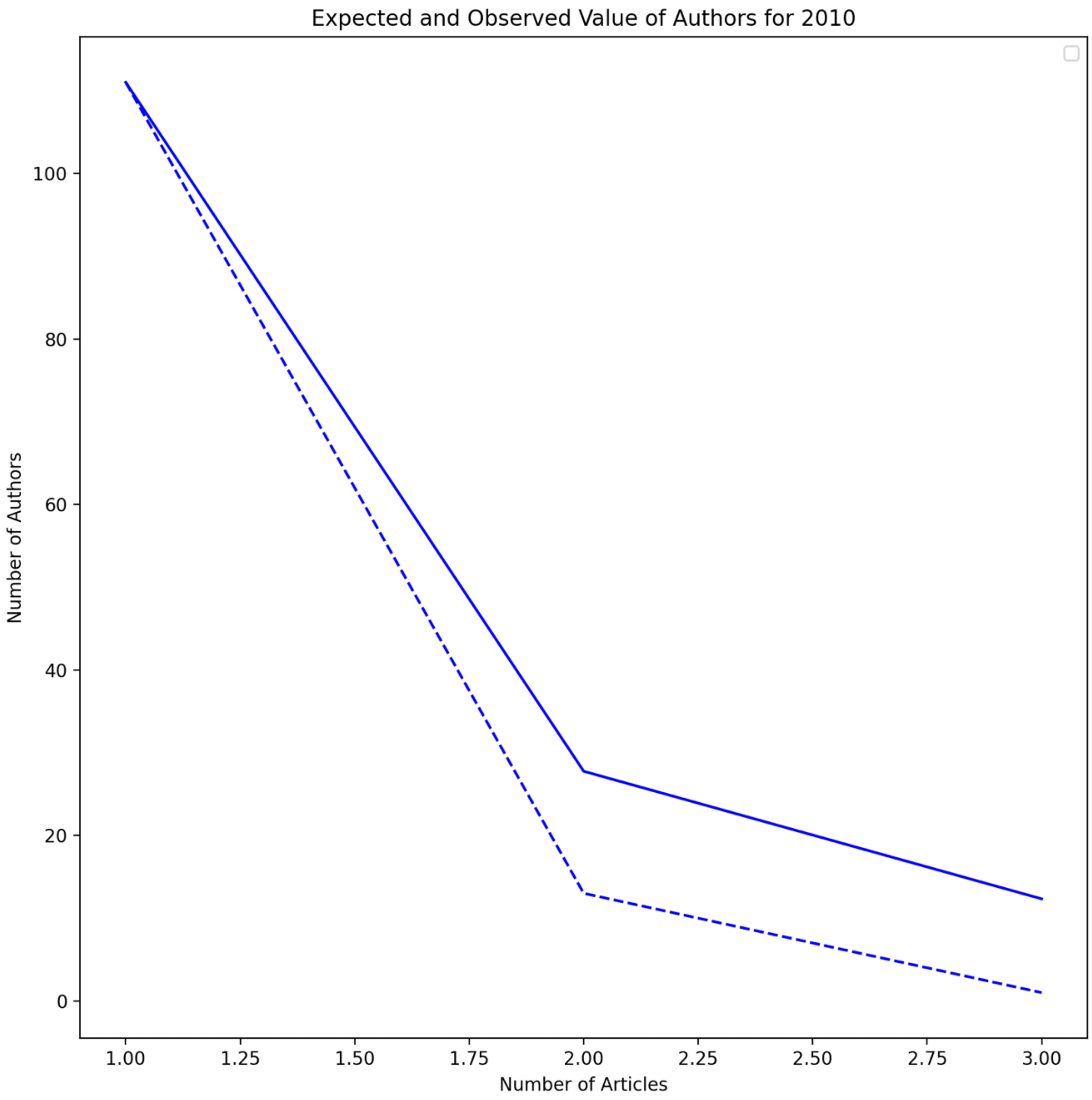

Expected Value of authors for 2010 for publishing 1 paper/s: 111.0
Observed Value of authors for 2010 for publishing 1 paper/s: 111
Expected Value of authors for 2010 for publishing 2 paper/s: 27.75
Observed Value of authors for 2010 for publishing 2 paper/s: 13
Expected Value of authors for 2010 for publishing 3 paper/s: 12.333
Observed Value of authors for 2010 for publishing 3 paper/s: 1

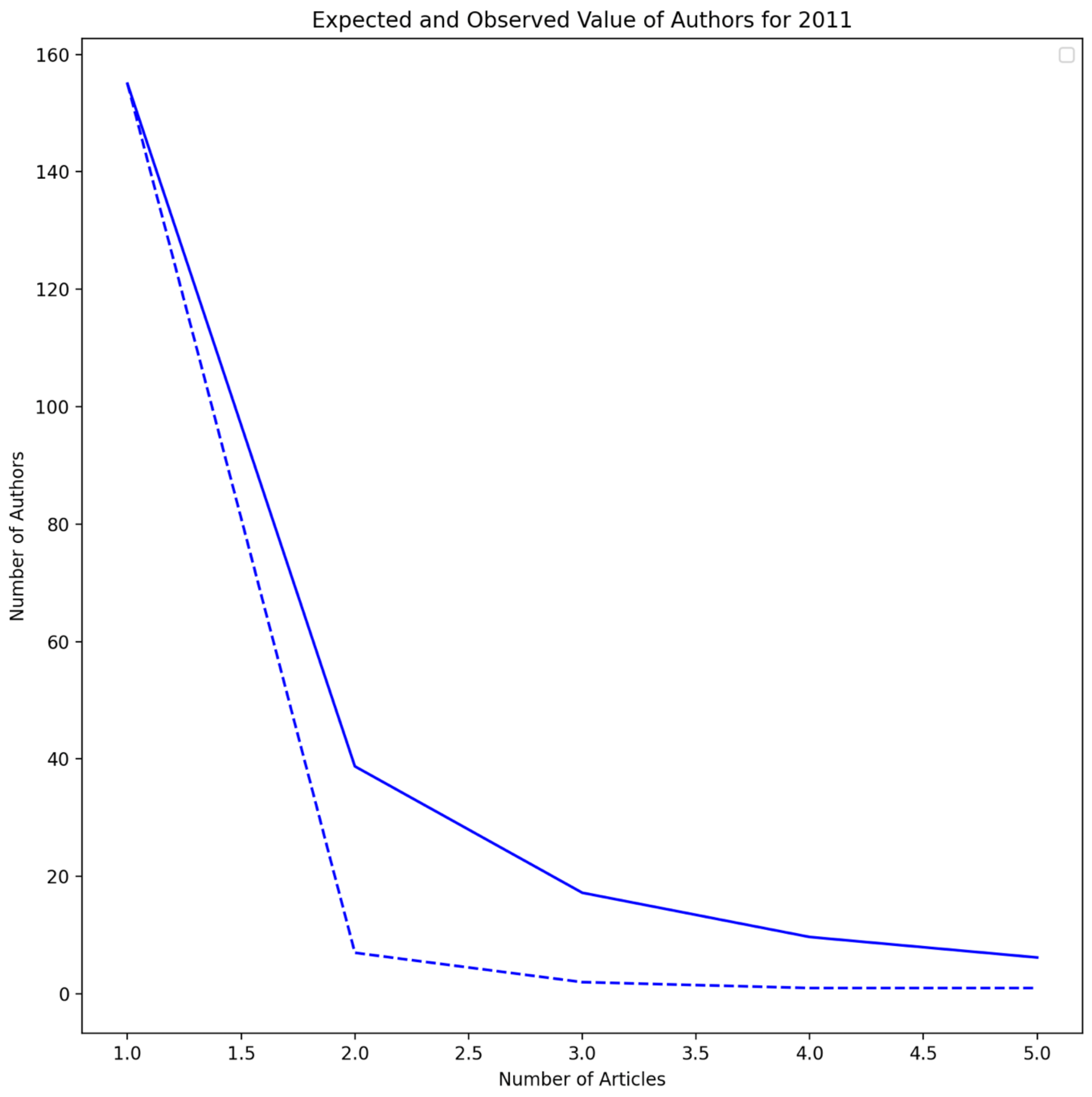

Expected Value of authors for 2011 for publishing 1 paper/s: 155.0
Observed Value of authors for 2011 for publishing 1 paper/s: 155
Expected Value of authors for 2011 for publishing 2 paper/s: 38.75
Observed Value of authors for 2011 for publishing 2 paper/s: 7
Expected Value of authors for 2011 for publishing 3 paper/s: 17.222
Observed Value of authors for 2011 for publishing 3 paper/s: 2
Expected Value of authors for 2011 for publishing 4 paper/s: 9.688
Observed Value of authors for 2011 for publishing 4 paper/s: 1
Expected Value of authors for 2011 for publishing 5 paper/s: 6.2
Observed Value of authors for 2011 for publishing 5 paper/s: 1

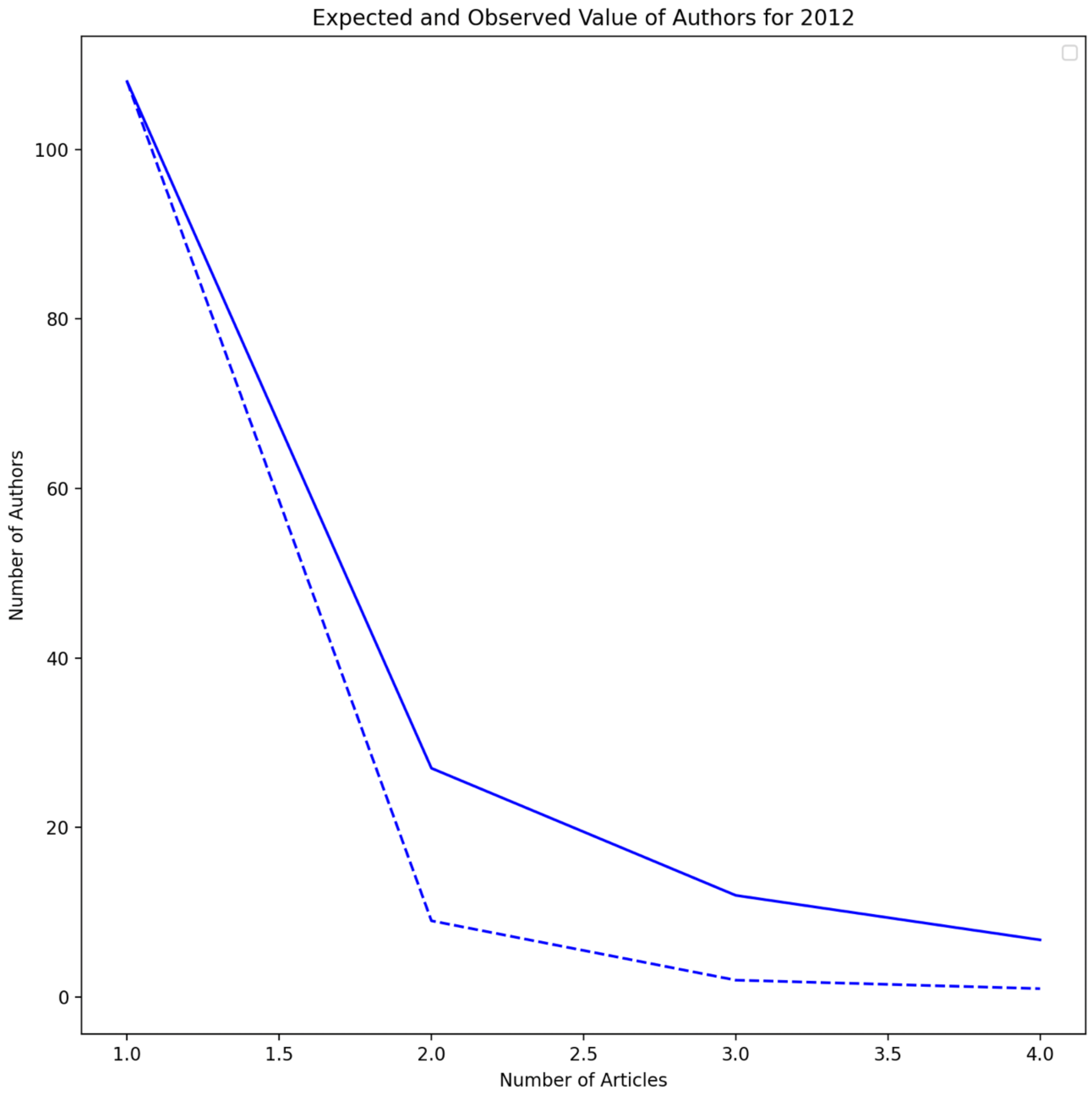

Expected Value of authors for 2012 for publishing 1 paper/s: 108.0
Observed Value of authors for 2012 for publishing 1 paper/s: 108
Expected Value of authors for 2012 for publishing 2 paper/s: 27.0
Observed Value of authors for 2012 for publishing 2 paper/s: 9
Expected Value of authors for 2012 for publishing 3 paper/s: 12.0
Observed Value of authors for 2012 for publishing 3 paper/s: 2
Expected Value of authors for 2012 for publishing 4 paper/s: 6.75
Observed Value of authors for 2012 for publishing 4 paper/s: 1

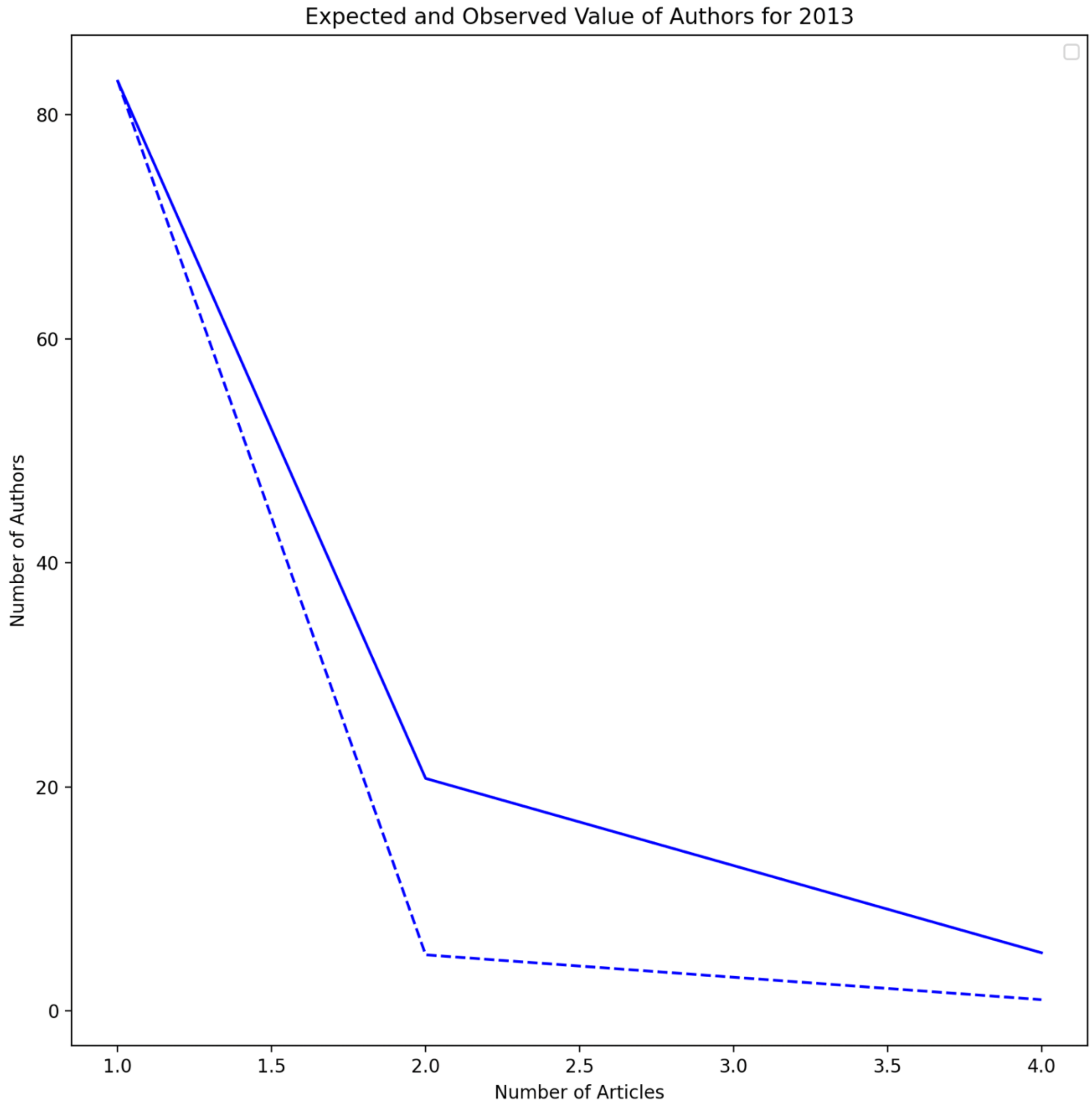

Expected Value of authors for 2013 for publishing 1 paper/s: 83.0
Observed Value of authors for 2013 for publishing 1 paper/s: 83
Expected Value of authors for 2013 for publishing 2 paper/s: 20.75
Observed Value of authors for 2013 for publishing 2 paper/s: 5
Expected Value of authors for 2013 for publishing 4 paper/s: 5.188
Observed Value of authors for 2013 for publishing 4 paper/s: 1

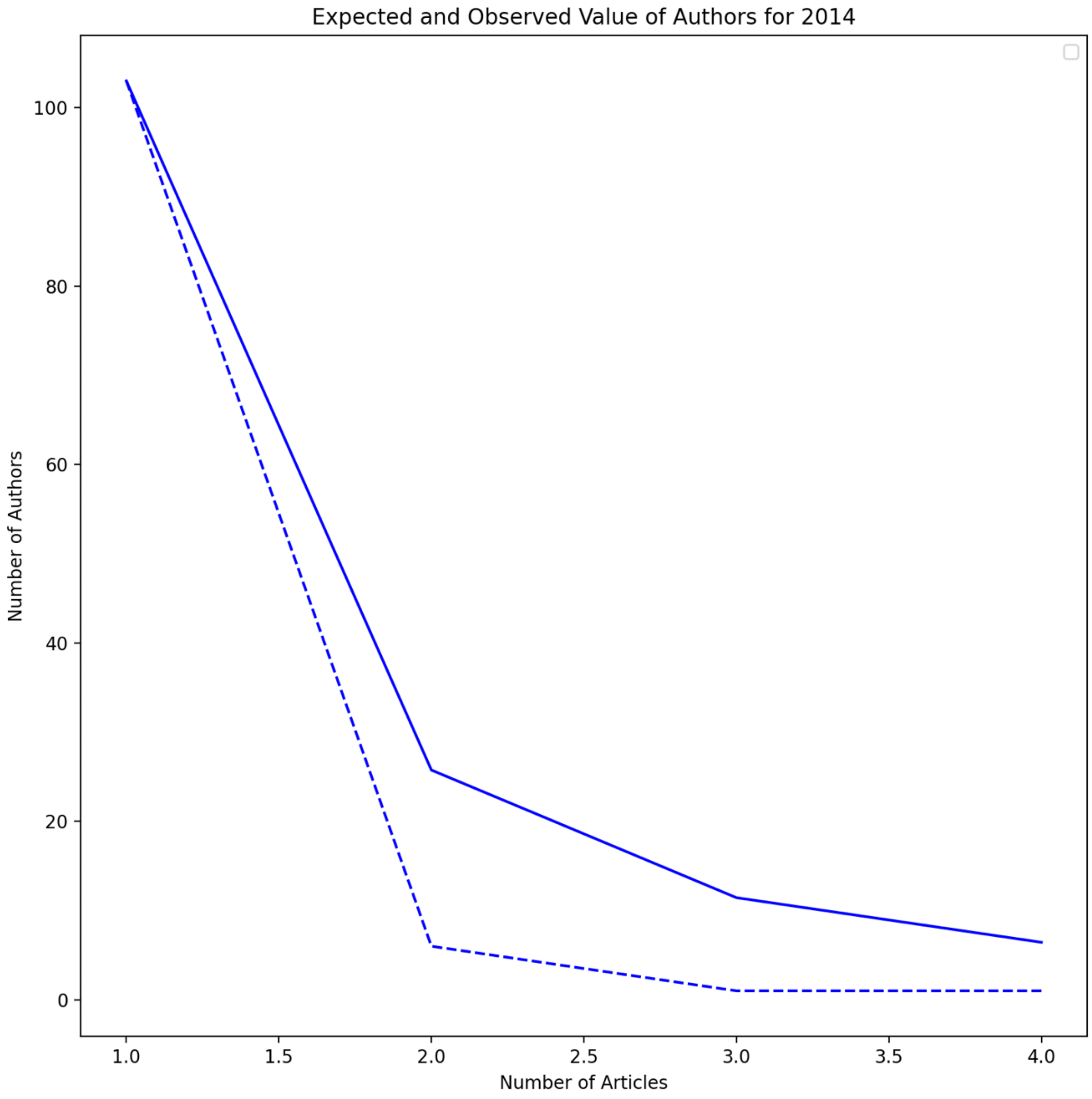

Expected Value of authors for 2014 for publishing 1 paper/s: 103.0
Observed Value of authors for 2014 for publishing 1 paper/s: 103
Expected Value of authors for 2014 for publishing 2 paper/s: 25.75
Observed Value of authors for 2014 for publishing 2 paper/s: 6
Expected Value of authors for 2014 for publishing 3 paper/s: 11.444
Observed Value of authors for 2014 for publishing 3 paper/s: 1
Expected Value of authors for 2014 for publishing 4 paper/s: 6.438
Observed Value of authors for 2014 for publishing 4 paper/s: 1

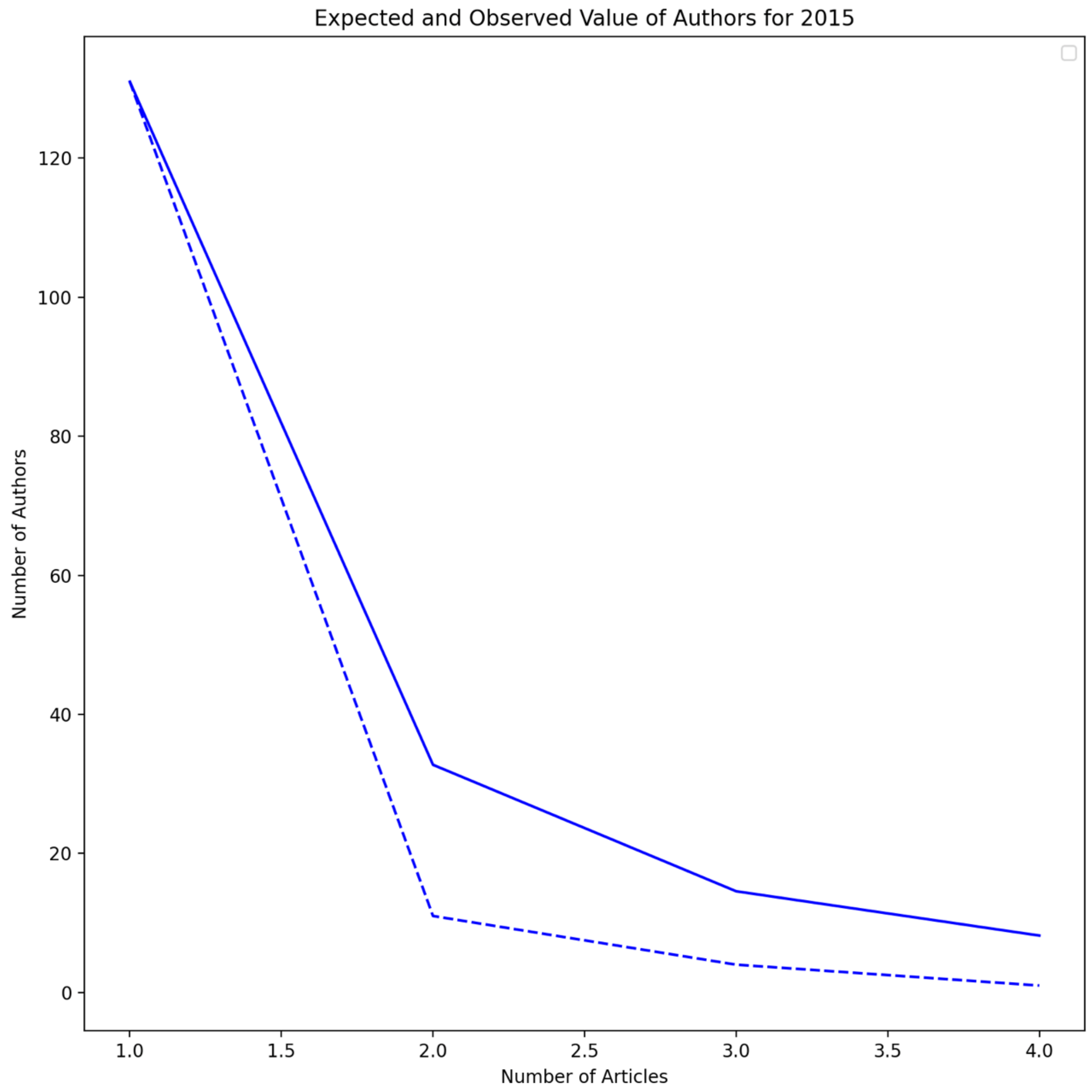

Expected Value of authors for 2015 for publishing 1 paper/s: 131.0
Observed Value of authors for 2015 for publishing 1 paper/s: 131
Expected Value of authors for 2015 for publishing 2 paper/s: 32.75
Observed Value of authors for 2015 for publishing 2 paper/s: 11
Expected Value of authors for 2015 for publishing 3 paper/s: 14.556
Observed Value of authors for 2015 for publishing 3 paper/s: 4
Expected Value of authors for 2015 for publishing 4 paper/s: 8.188
Observed Value of authors for 2015 for publishing 4 paper/s: 1

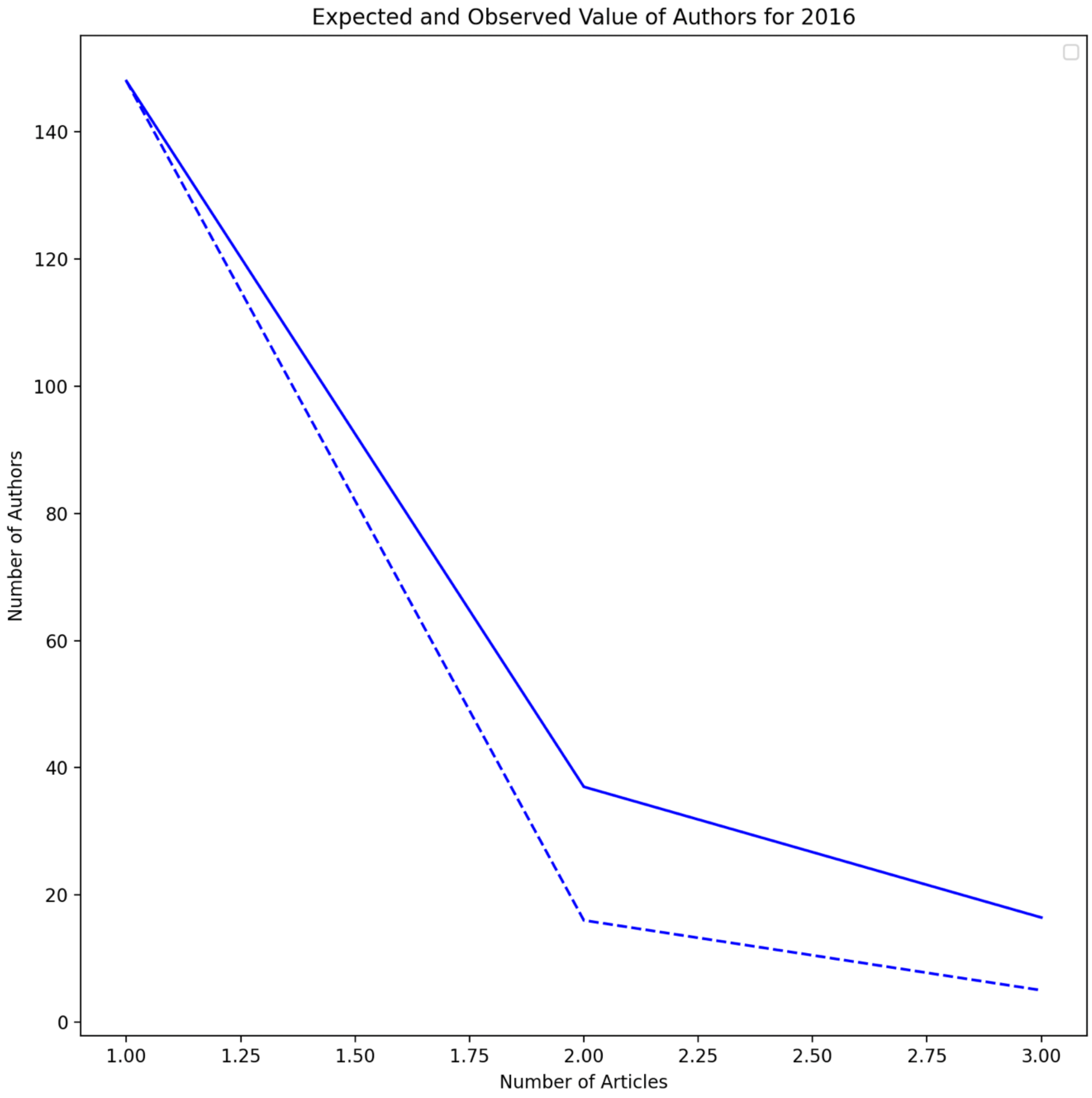

Expected Value of authors for 2016 for publishing 1 paper/s: 148.0
Observed Value of authors for 2016 for publishing 1 paper/s: 148
Expected Value of authors for 2016 for publishing 2 paper/s: 37.0
Observed Value of authors for 2016 for publishing 2 paper/s: 16
Expected Value of authors for 2016 for publishing 3 paper/s: 16.444
Observed Value of authors for 2016 for publishing 3 paper/s: 5

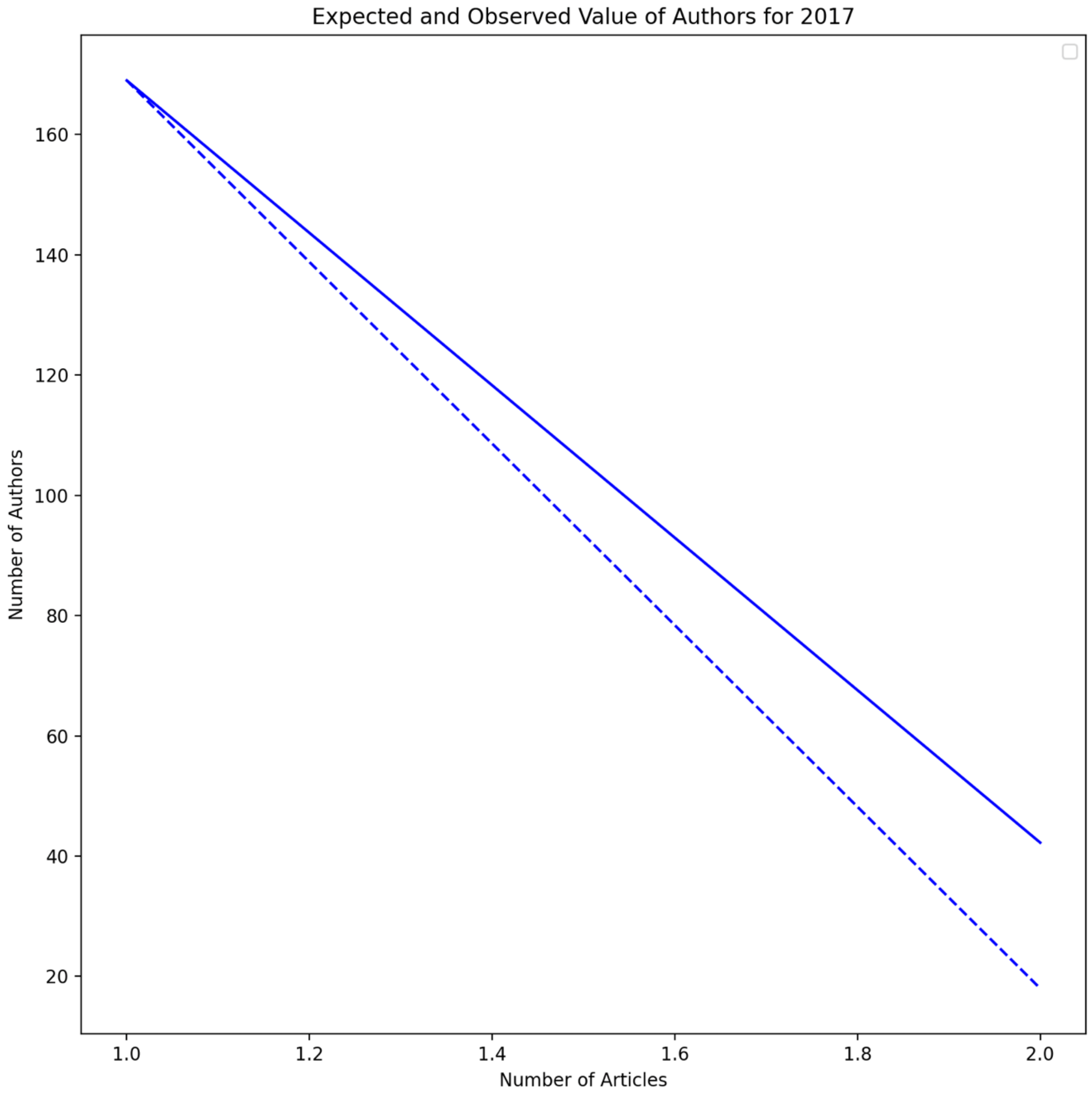

Expected Value of authors for 2017 for publishing 1 paper/s: 169.0
Observed Value of authors for 2017 for publishing 1 paper/s: 169
Expected Value of authors for 2017 for publishing 2 paper/s: 42.25
Observed Value of authors for 2017 for publishing 2 paper/s: 18

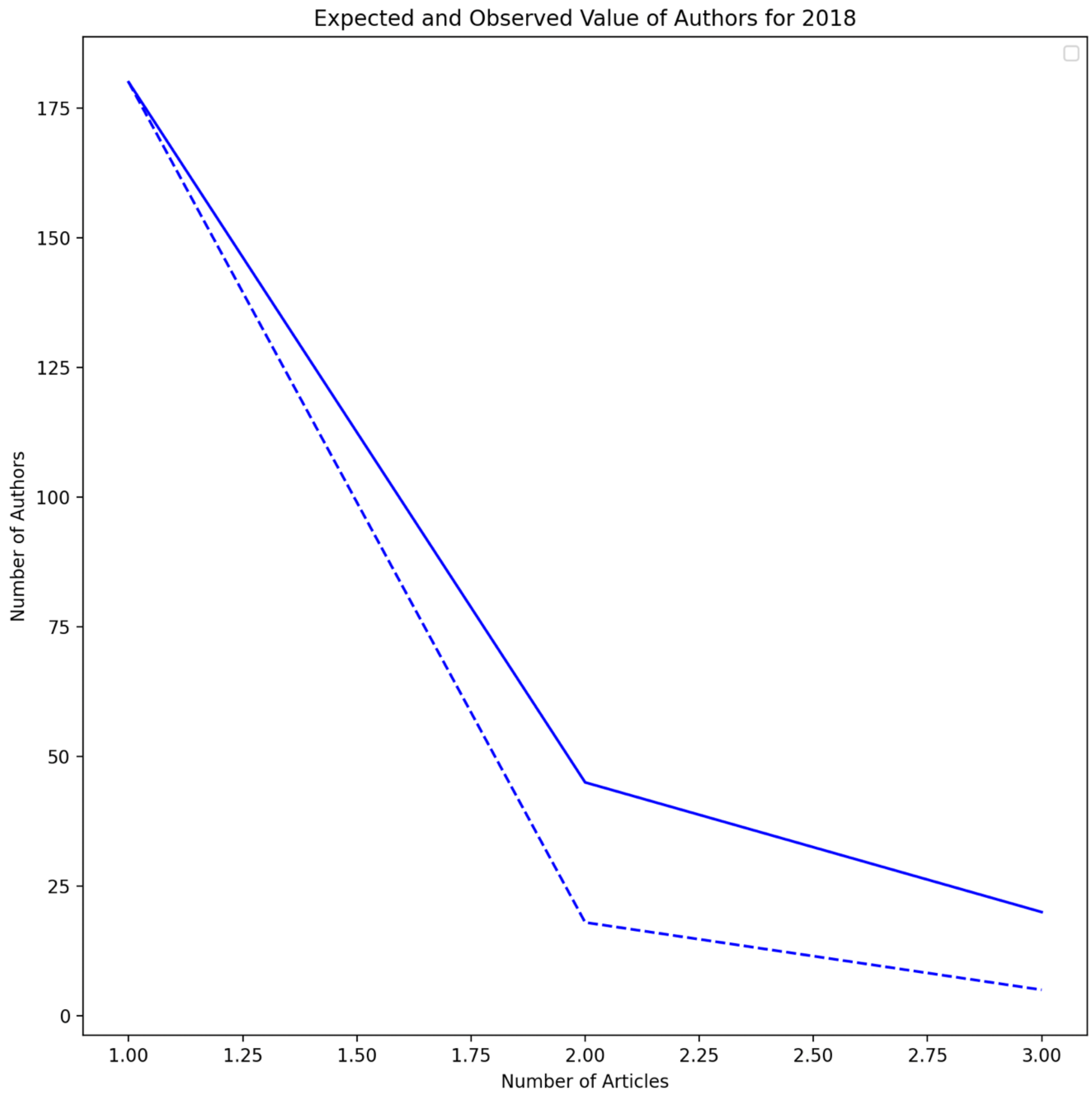

Expected Value of authors for 2018 for publishing 1 paper/s: 180.0
Observed Value of authors for 2018 for publishing 1 paper/s: 180
Expected Value of authors for 2018 for publishing 2 paper/s: 45.0
Observed Value of authors for 2018 for publishing 2 paper/s: 18
Expected Value of authors for 2018 for publishing 3 paper/s: 20.0
Observed Value of authors for 2018 for publishing 3 paper/s: 5

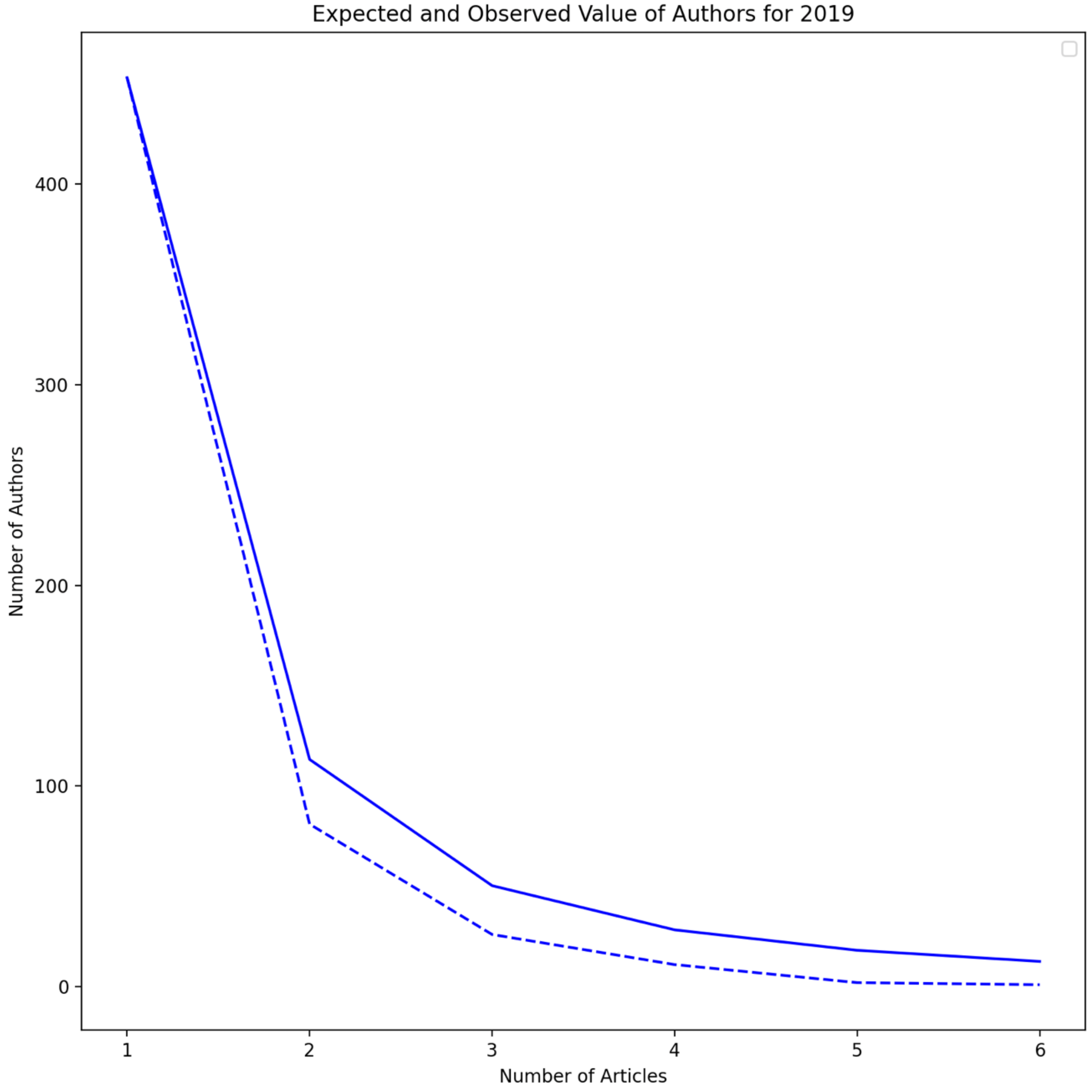

Expected Value of authors for 2019 for publishing 1 paper/s: 453.0
Observed Value of authors for 2019 for publishing 1 paper/s: 453
Expected Value of authors for 2019 for publishing 2 paper/s: 113.25
Observed Value of authors for 2019 for publishing 2 paper/s: 81
Expected Value of authors for 2019 for publishing 3 paper/s: 50.333
Observed Value of authors for 2019 for publishing 3 paper/s: 26
Expected Value of authors for 2019 for publishing 4 paper/s: 28.312
Observed Value of authors for 2019 for publishing 4 paper/s: 11
Expected Value of authors for 2019 for publishing 5 paper/s: 18.12
Observed Value of authors for 2019 for publishing 5 paper/s: 2
Expected Value of authors for 2019 for publishing 6 paper/s: 12.583
Observed Value of authors for 2019 for publishing 6 paper/s: 1

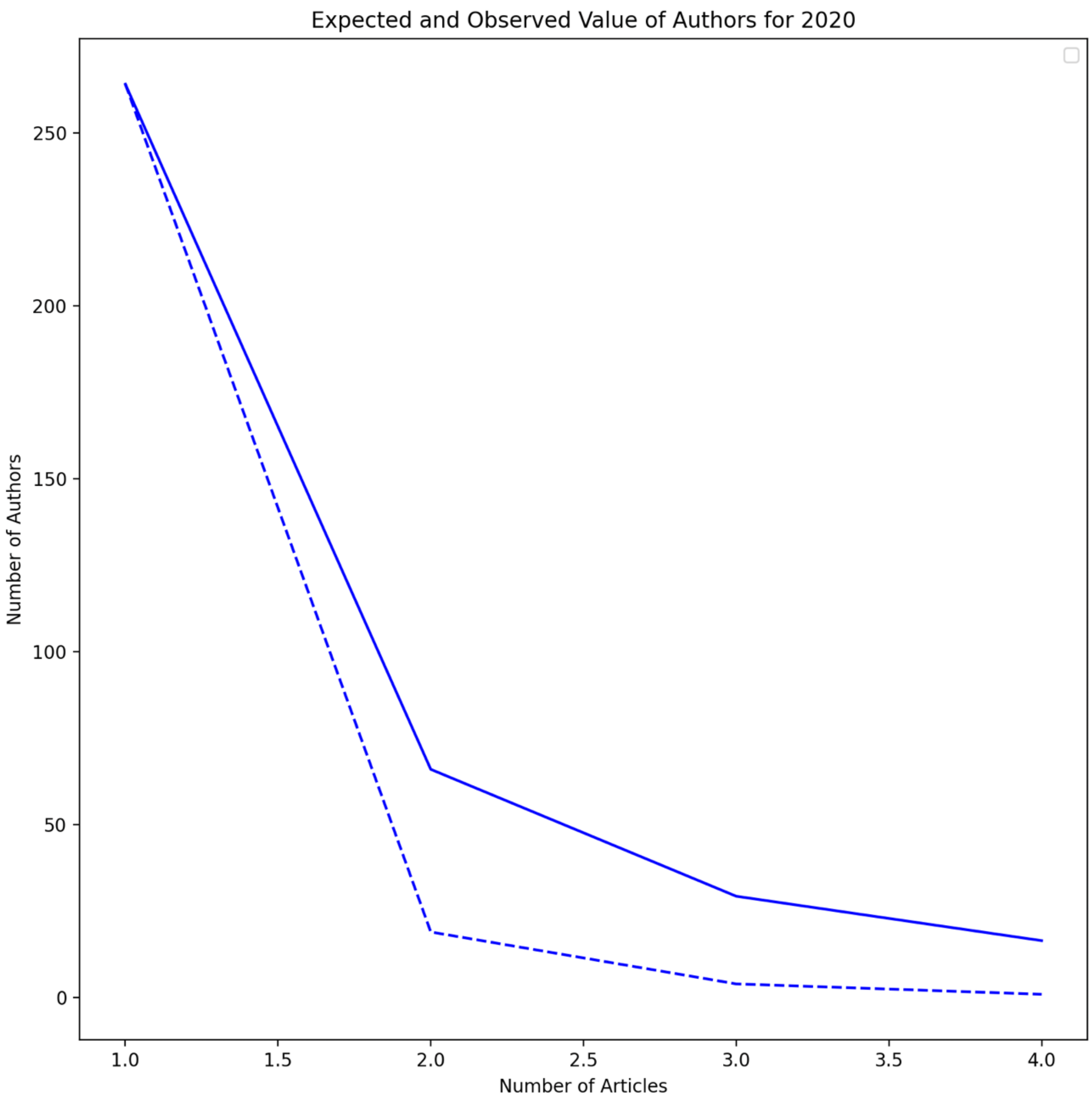

Expected Value of authors for 2020 for publishing 1 paper/s: 264.0
Observed Value of authors for 2020 for publishing 1 paper/s: 264
Expected Value of authors for 2020 for publishing 2 paper/s: 66.0
Observed Value of authors for 2020 for publishing 2 paper/s: 19
Expected Value of authors for 2020 for publishing 3 paper/s: 29.333
Observed Value of authors for 2020 for publishing 3 paper/s: 4
Expected Value of authors for 2020 for publishing 4 paper/s: 16.5
Observed Value of authors for 2020 for publishing 4 paper/s: 1

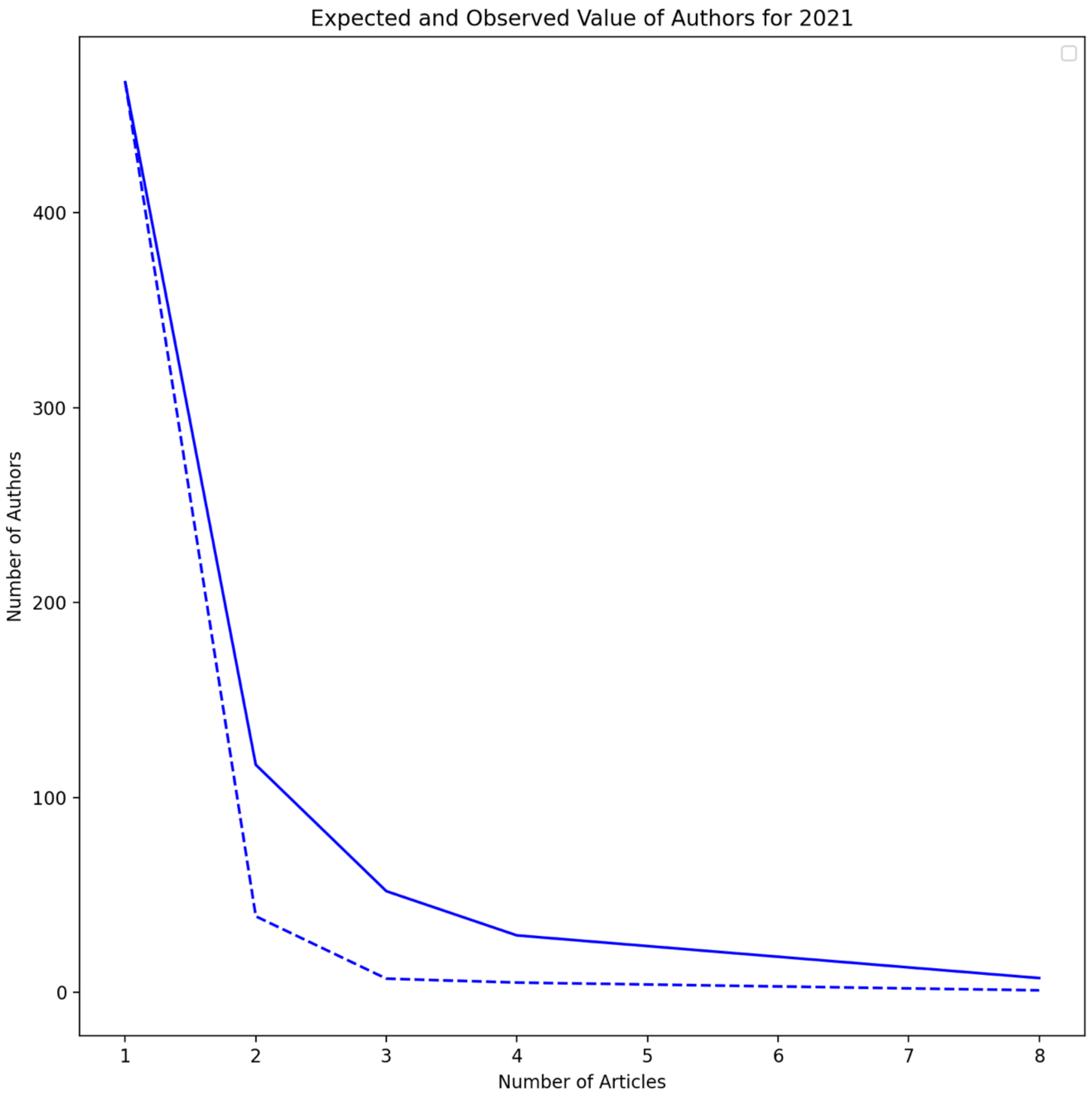

Expected Value of authors for 2021 for publishing 1 paper/s: 467.0
Observed Value of authors for 2021 for publishing 1 paper/s: 467
Expected Value of authors for 2021 for publishing 2 paper/s: 116.75
Observed Value of authors for 2021 for publishing 2 paper/s: 39
Expected Value of authors for 2021 for publishing 3 paper/s: 51.889
Observed Value of authors for 2021 for publishing 3 paper/s: 7
Expected Value of authors for 2021 for publishing 4 paper/s: 29.188
Observed Value of authors for 2021 for publishing 4 paper/s: 5
Expected Value of authors for 2021 for publishing 8 paper/s: 7.297
Observed Value of authors for 2021 for publishing 8 paper/s: 1

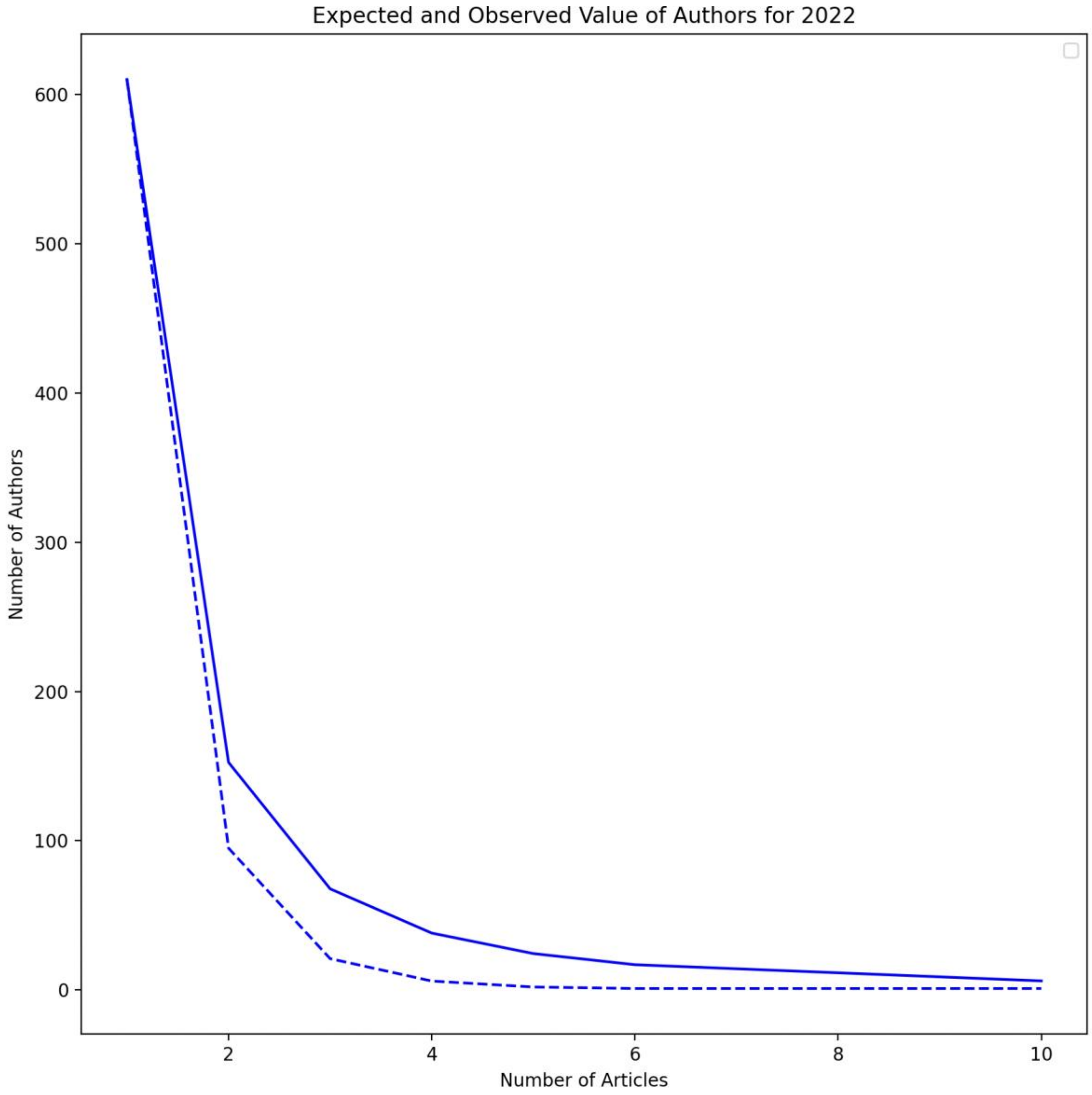

Expected Value of authors for 2022 for publishing 1 paper/s: 610.0
Observed Value of authors for 2022 for publishing 1 paper/s: 610
Expected Value of authors for 2022 for publishing 2 paper/s: 152.5
Observed Value of authors for 2022 for publishing 2 paper/s: 95
Expected Value of authors for 2022 for publishing 3 paper/s: 67.778
Observed Value of authors for 2022 for publishing 3 paper/s: 21
Expected Value of authors for 2022 for publishing 4 paper/s: 38.125
Observed Value of authors for 2022 for publishing 4 paper/s: 6
Expected Value of authors for 2022 for publishing 5 paper/s: 24.4
Observed Value of authors for 2022 for publishing 5 paper/s: 2
Expected Value of authors for 2022 for publishing 6 paper/s: 16.944
Observed Value of authors for 2022 for publishing 6 paper/s: 1
Expected Value of authors for 2022 for publishing 10 paper/s: 6.1
Observed Value of authors for 2022 for publishing 10 paper/s: 1

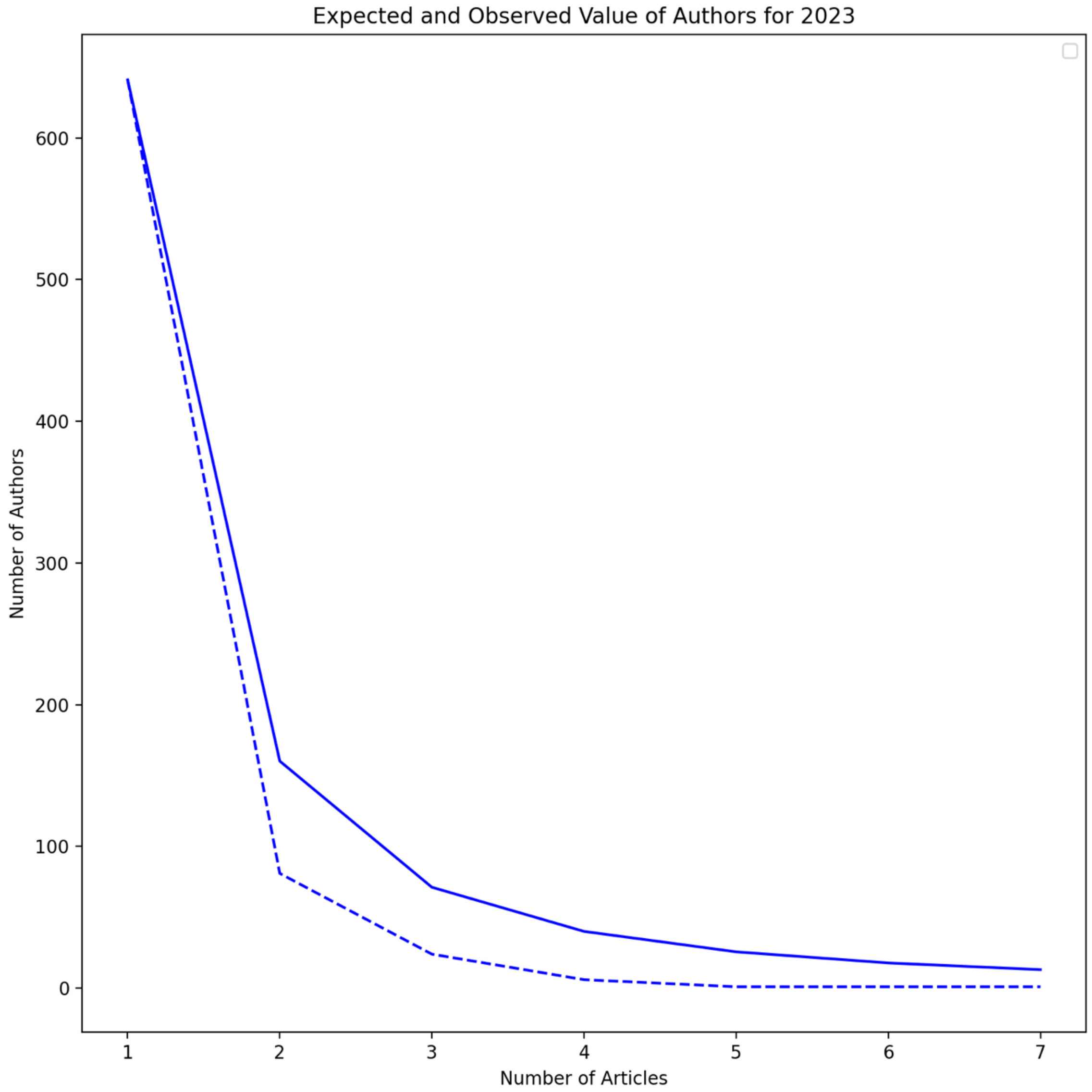

Expected Value of authors for 2023 for publishing 1 paper/s: 641.0
Observed Value of authors for 2023 for publishing 1 paper/s: 641
Expected Value of authors for 2023 for publishing 2 paper/s: 160.25
Observed Value of authors for 2023 for publishing 2 paper/s: 81
Expected Value of authors for 2023 for publishing 3 paper/s: 71.222
Observed Value of authors for 2023 for publishing 3 paper/s: 24
Expected Value of authors for 2023 for publishing 4 paper/s: 40.062
Observed Value of authors for 2023 for publishing 4 paper/s: 6
Expected Value of authors for 2023 for publishing 5 paper/s: 25.64
Observed Value of authors for 2023 for publishing 5 paper/s: 1
Expected Value of authors for 2023 for publishing 6 paper/s: 17.806
Observed Value of authors for 2023 for publishing 6 paper/s: 1
Expected Value of authors for 2023 for publishing 7 paper/s: 13.082
Observed Value of authors for 2023 for publishing 7 paper/s: 1

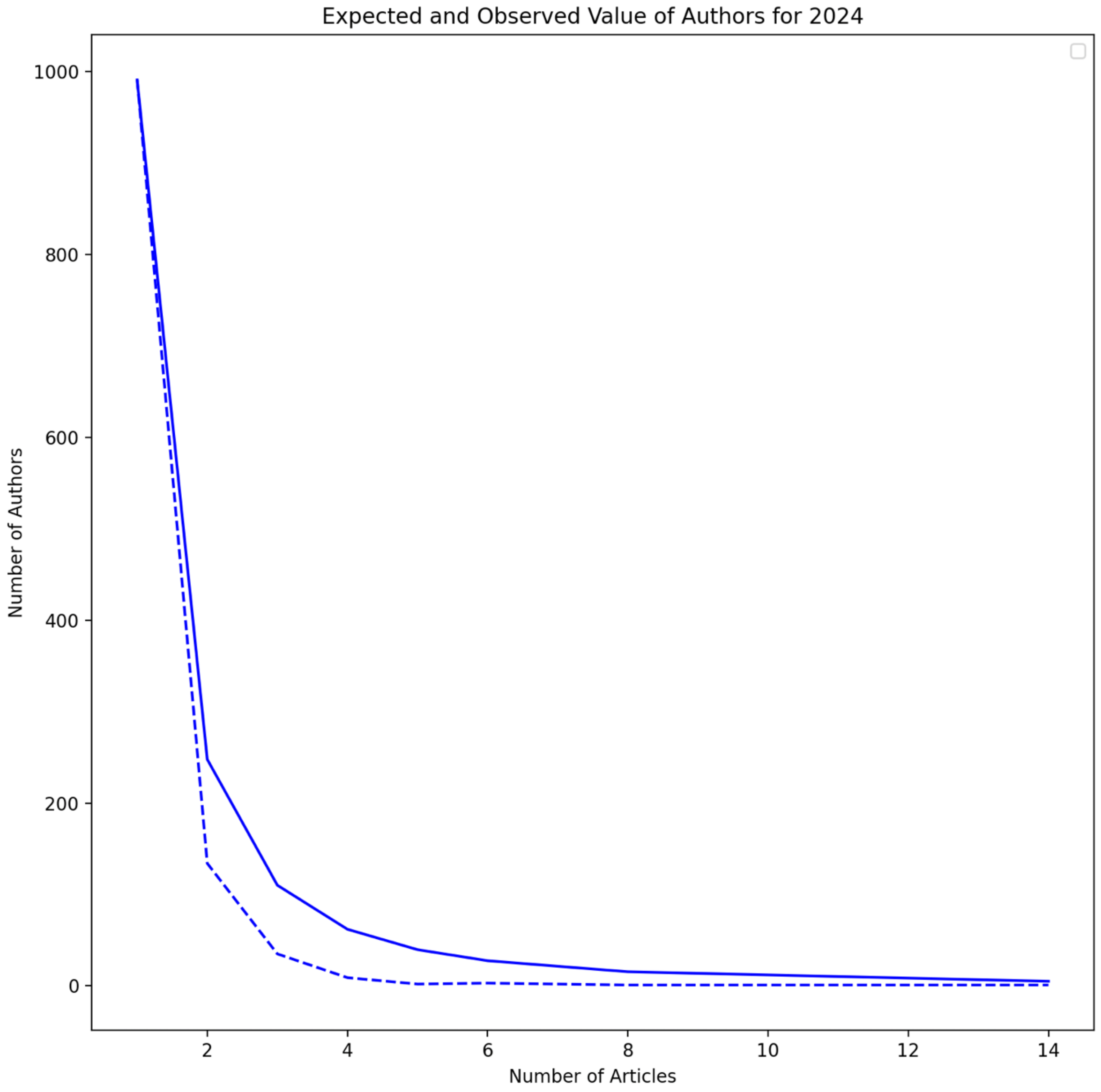

Expected Value of authors for 2024 for publishing 1 paper/s: 991.0
Observed Value of authors for 2024 for publishing 1 paper/s: 991
Expected Value of authors for 2024 for publishing 2 paper/s: 247.75
Observed Value of authors for 2024 for publishing 2 paper/s: 134
Expected Value of authors for 2024 for publishing 3 paper/s: 110.111
Observed Value of authors for 2024 for publishing 3 paper/s: 35
Expected Value of authors for 2024 for publishing 4 paper/s: 61.938
Observed Value of authors for 2024 for publishing 4 paper/s: 9
Expected Value of authors for 2024 for publishing 5 paper/s: 39.64
Observed Value of authors for 2024 for publishing 5 paper/s: 2
Expected Value of authors for 2024 for publishing 6 paper/s: 27.528
Observed Value of authors for 2024 for publishing 6 paper/s: 3

Expected Value of authors for 2024 for publishing 8 paper/s: 15.484
Observed Value of authors for 2024 for publishing 8 paper/s: 1
Expected Value of authors for 2024 for publishing 14 paper/s: 5.056
Observed Value of authors for 2024 for publishing 14 paper/s: 1